\documentclass[12pt]{article}
\usepackage[english]{babel}
\usepackage[latin1]{inputenc}
\usepackage{amsfonts,amssymb,amsmath, epsfig}
\usepackage{color,graphicx,graphics,psfrag}
\usepackage{amsmath,amstext,amssymb,amsfonts, amscd}
\usepackage[lofdepth,lotdepth]{subfig}
\usepackage{bbm}

\usepackage{hyperref}          

\def\Box{\leavevmode\vbox{\hrule
     \hbox{\vrule\kern4pt\vbox{\kern4pt}%
           \vrule}\hrule}}
\def\blackbox{\leavevmode\vrule height 5pt width 4pt depth 0pt\relax}
\def\endproof{\null\hfill {$\blackbox$}\bigskip}

\newcounter{appendix}
\def\appendix{\advance\c@appendix by 1
   \def\thesection{\Alph{section}}
   \ifnum\c@appendix=1 \setcounter{section}{-1} \fi
   \@startsection {section}{1}{\z@}{-3.5ex plus -1ex minus 
   -.2ex}{2.3ex plus .2ex}{\Large\bf}}

\def\paragraph#1{{\bf #1\ }}

\newtheorem{lemma}{Lemma}[section]  

\newtheorem{theorem}[lemma]{Theorem}

\newtheorem{definition}[lemma]{Definition}

\newtheorem{proposition}[lemma]{Proposition}

\newtheorem{remark}{Remark}[section]

\newtheorem{video}{Video}

\title{Topological states and continuum model for swarmalators without force reciprocity} 

\author{P. Degond$^{(1)}$, A. Diez$^{(2,3)}$, A. Walczak$^{(2)}$} 
\date{} 
\begin{document}

\maketitle

\vspace{0.5 cm}

\begin{center}
$^{(1)}$ Institut de Math\'ematiques de Toulouse ; UMR5219 \\
Universit\'e de Toulouse ; CNRS \\
UPS, F-31062 Toulouse Cedex 9, France\\
email: pierre.degond@math.univ-toulouse.fr

\bigskip

$^{(2)}$ Department of Mathematics, Imperial College London, South Kensington Campus \\
London, SW7 2AZ, United Kingdom\\
email: AW: adamvv98@gmail.com

\bigskip

$^{(3)}$ Institute for the Advanced Study of Human Biology (ASHBi), \\
Kyoto University Institute for Advanced Study, \\
Kyoto University,\\ 
Kyoto 606-8315, Japan\\
email: AD: diez.antoinenicolas.4e@kyoto-u.ac.jp

\end{center}

\vspace{0.5 cm}
\begin{abstract}
Swarmalators are systems of agents which are both self-propelled particles and oscillators.
Each particle is endowed with a phase which modulates its interaction force with the other particles. In return, relative positions modulate phase synchronization between interacting particles. In the present model, there is no force reciprocity: when a particle attracts another one, the latter repels the former. This results in a pursuit behavior. In this paper, we derive a hydrodynamic model of this swarmalator system and show that it has explicit doubly-periodic travelling-wave solutions in two space dimensions. These special solutions enjoy non-trivial topology quantified by the index of the phase vector along a period in either dimension. Stability of these solutions is studied by investigating the conditions for hyperbolicity of the model. Numerical solutions of both the particle and hydrodynamic models are shown. They confirm the consistency of the hydrodynamic model with the particle one for small times or large phase-noise but also reveal the emergence of intriguing patterns in the case of small phase-noise. 
\end{abstract}



\medskip
\noindent
{\bf Acknowledgements:} PD holds a visiting professor association with the Department of Mathematics, Imperial College London where part of this research was conducted. Part of this research was conducted when AD was supported by an EPSRC-Roth scholarship cofunded by the Engineering and Physical Sciences Research Council and the Department of Mathematics at Imperial College London.

\medskip
\noindent
{\bf Data statement:} No new data were collected in the course of this research.

\setcounter{equation}{0}
\section{Introduction}
\label{intro}

We investigate a new collective dynamics model based on the swarmalator concept described below. Collective dynamics refers to the coherent behavior of self-propelled agents subject to mutual interactions such as attraction, repulsion and alignment. Examples in nature are provided by birds \cite{lukeman2010inferring}, fish \cite{gautrais2012deciphering, hemelrijk2010emergence}, ants \cite{boissard2013trail, couzin2003self}, bacteria \cite{czirok1996formation}, sperm~\cite{creppy2016symmetry}, colloidal rollers \cite{bricard2015emergent}. How coherence emerges is still the subject of an intense literature in which mathematical models play a central role. In many models, self-propulsion is accounted for by imposing the agents a constant speed, identical for all agents, such as in the Vicsek model \cite{vicsek1995novel} and its many variants \cite{aoki1982simulation, calovi2014swarming, cavagna2015flocking, chate2008collective, costanzo2018spontaneous, couzin2002collective, hemelrijk2012schools, hildenbrandt2010self} (see also the review~\cite{vicsek2012collective}). By contrast, a constant speed is not assumed in the Cucker-Smale model~\cite{cucker2007emergent} and its variants  \cite{dorsogna2006self, ha2008from, ha2009simple, motsch2011new}. The synchronization of oscillator populations shares many similar features with collective dynamics. The paradigmatic model of oscillator synchronization is the Kuramoto model \cite{kuramoto2003chemical} which has stimulated an intense research activity (see e.g. the review \cite{acebron2005kuramoto}). 

Among collective dynamics models, swarmalators have recently attracted increasing attention. Swarmalators are systems of agents that are simultaneously self-propelled particles and oscillators. In addition to position and velocity, they are endowed with a phase which may synchronize with the neighbors' phases. There is a two-way coupling between positions and phases: phase differences between neighboring particles modulate interaction forces and relative positions influence phase synchronization. The concept and terminology was introduced in \cite{o2017oscillators} and further developed in \cite{ha2019emergent, hong2018active, hong2021coupling, jimenez2020oscillatory, lee2021collective, lizarraga2020synchronization, o2022collective, o2018ring} (see also the review~\cite{o2019review}). Swarmalator models have been applied to e.g. the collective swimming of nematode swarms \cite{peshkov2022synchronized} or the intercellular organization of multicellular organisms \cite{japon2021intercellular}. Related earlier models associate phases with rotations of self-propulsion speeds \cite{degond2014hydrodynamics, levis2019activity, liebchen2017collective, paley2007oscillator} but, in these models, there is only a one way coupling of phases on positions.  

Topological states are solutions which have non-trivial topology quantified by some discrete topological index (such as a winding number). Topological states have appeared first in the quantum Hall effect and gradually in other applications such as topological insulators \cite{hasan2010colloquium, qi2011topological}. Non-trivial topology endows the system with increased robustness against perturbations because breaking the topological state requires a jump of the topological index and thus, a finite amount of energy. This is the so-called 'topological protection' effect. In collective dynamics, topological states have been investigated in recent work \cite{degond2021bulk, shankar2017topological, sone2019anomalous, souslov2017topological, zhang2020oscillatory} (see also the review \cite{shankar2020topological}). Swarmalator models also support topological states, such as the ``phase-wave states'' of~\cite{o2017oscillators}.

In the present paper, we propose a new swarmalator model and  demonstrate that it possesses travelling-wave solutions having non-trivial topology. This model differs from the above cited swarmalator models because the phase-modulated force does not enjoy reciprocity: the force acted on a particle by another one is equal to the force acted by the latter on the former, and not the opposite as it should if reciprocity was enforced. The result is a pursuit behavior: when two particles interact, according to their relative phases one particle chases the other one or vice versa (see Fig. \ref{fig:pursuit}). Another difference with \cite{o2017oscillators} is that our model is second order: the self-propulsion component of the velocity obeys a time-continuous version \cite{degond2008continuum} of the Vicsek model \cite{vicsek1995novel}, namely, particles' self-propulsion velocities tend to align with their neighbors up to some noise. Lastly, the model includes noises in both self-propulsion velocity and phase, in contrast to most of the above cited literature. 

Our methodology is based on studying a continuum version of the  swarmalator model. In general, there are three stages of description of particle systems. The finest level of detail is provided by the particle system itself, which, for swarmalators,  consists of a differential system for the positions, velocities and phases of all the particles involved. The next stage is given by the kinetic model where the system is described by the probability distribution of the particles in the space of positions, velocities and phases. The passage between particle and kinetic models requires letting $N \to \infty$ where $N$ is the number of particles. Its mathematical investigation has given rise to a large number of works, (see e.g. \cite{cercignani2013mathematical, degond2004macroscopic, spohn2012large} in classical kinetic theory, \cite{bolley2012mean, briant2020cauchy, diez2019propagation} for the Vicsek model and \cite{ha2021mean} for the swarmalators). Kinetic models of collective dynamics have been proposed in e.g.  \cite{bertin2009hydrodynamic,  peruani2008mean}. For kinetic models derived from the Vicsek model, existence of solutions \cite{figalli2018global, gamba2016global}, phase transitions \cite{degond2013macroscopic, degondfrouvelleliu15, frouvelle2012dynamics}, numerical methods \cite{gamba2015spectral, griette2019kinetic} and relations to models with no velocity normalization \cite{bostan2013asymptotic} have been studied. 

The final stage which gives rise to the coarsest level of details but provides the highest tractability consists of continuum models. These models are obtained from kinetic ones through an asymptotic limit involving a small parameter representing the ratio of the microscopic scale (e.g. the range of particle interactions) and the macroscopic one (typically the size of the observed region). This approach originates from the kinetic theory of gases (see reviews in \cite{cercignani2013mathematical, degond2004macroscopic}). The continuum version of the Viscek model was derived via this asymptotic procedure for the first time in \cite{degond2015macroscopic} (see also \cite{frouvelle2012continuum}) and gave rise to  the new ``Self-Organized Hydrodynamics (SOH)'' model. This derivation provides closed formulas for the parameters of the hydrodynamic model as functions of those of the particle and kinetic model. Later, \cite{jiang2016hydrodynamic} showed the mathematically rigorous validity of the asymptotic limit. Another, more intuitive approach \cite{bertin2006boltzmann, bertin2009hydrodynamic, toner1998flocks} was developed earlier and led to a different model, the ``Toner-Tu'' model. As interesting as it can be, the Toner-Tu model has no mathematically proven connection with the Vicsek particle model, and there are no rigorous formulas relating the parameters of the two models. There have been several extensions of \cite{degond2008continuum}. In relation to the present work, let us mention \cite{degond2015macroscopic} which includes attraction-repulsion forces, and \cite{degond2021body, degondfrouvellemerino17, degondfrouvellemerinotrescases18, degond2018alignment} where alignment of body attitudes (instead of mere self-propulsion velocity) is considered and shown to support topological states \cite{degond2021bulk}. Local existence of solutions for the SOH model was proved in \cite{degond20133hydrodynamic, zhang2017local} and numerical simulations can be found in \cite{degond2015macroscopic, dimarcomotsch16, motsch2011numerical}. 

In this paper, we first derive the continuum version of the swarmalator model under consideration. The derivation strongly relies on \cite{degond2015macroscopic, degond2008continuum, frouvelle2012continuum} and details will be given in an appendix. The resulting model, called ``Swarmalator Hydrodynamics (SH)'' consists of three evolution equations for the particle density, mean self-propulsion velocity and mean phase. We will also consider the case where the phase noise is small which simplifies the system and leads to the ``Noiseless Swarmalator Hydrodynamics (NSH)''. We first analyze the conditions under which the NSH system is hyperbolic. We then specialize the SH system to two spatial dimensions and show the existence of doubly-periodic travelling-wave solutions. These solutions present a non-trivial topology as evidenced by the index of the phase vector field being non-zero. We will then present numerical experiments which have three objectives: (i) the validation of the SH and NSH models as macroscopic descriptions of the swarmalators model; (ii) the numerical verification of the hyperbolicity conditions and (iii) the exploration and comparisons of the patterns obtained by the particle and SH or NSH models in relation with their topology. In forthcoming papers \cite{degond2022topological_bis, degond2022topological}, we will explore other classes of travelling-wave solutions at both the particle and hydrodynamic level, investigate whether they enjoy topological protection and decipher the mechanisms of topological phase transitions when they occur. 

The main innovations of this work are: (i) the introduction of a new swarmalator model involving force non-reciprocity;  (ii) the derivation and hyperbolicity analysis of a system of continuum equations for this model, named the SH system; (iii) the derivation of a class of topologically non-trivial doubly-periodic travelling-wave solutions to the SH system (iv) the numerical validation of the SH model against the particle model and of its hyperbolicity, and the exploration of the patterns appearing with both models in relation with their topology. 

The organization of this paper is as follows. In Section \ref{sec_part_kin}, we introduce the swarmalator system and derive the associated kinetic equations. Some technical points are deferred to Appendix \ref{sec_part_kin_add}. Section \ref{sec_eps_to_0} presents the derivation of the hydrodynamic model, the analysis of its hyperbolicity and the determination of a special class of explicit travelling-wave solutions. Proofs of these results  are collected in Appendices \ref{sec_eps_to_0_proof}, \ref{sec_small_noise_proofs} and \ref{sec_TW_stat}. Numerical experiments are presented in Section \ref{sec:numerics} for the particle model and its hydrodynamic limit. Details on the numerical methods are deferred to Appendix \ref{sec:numericalmethods}. The videos of the outcome of the simulations can be found in the supplementary material. The list and description of the supplementary videos can be found in Appendix \ref{sec:listvideos}.  Finally perspectives are drawn in Section \ref{sec:conclusion}.

\setcounter{equation}{0}
\section{Particle and kinetic models}
\label{sec_part_kin}

In a first subsection, we present the particle swarmalator model on which this study is based. In a second subsection, we provide a kinetic formulation of this model in the limit of a large number of particles.

\subsection{The particle model}
\label{sec_part}

We consider $N$ particles labeled $k = 1, \ldots, N$, having position $X_k \in {\mathbb R}^n$. We suppose these particles are self-propelled with constant and uniform self-propulsion speed $c_0$ and direction of self-propulsion $v_k \in {\mathbb S}^{n-1}$. Additionally we assume that each particle is endowed with a phase $\varphi_k \in {\mathbb R}/(2 \pi {\mathbb Z})$. Neighboring particles interact via alignment of their self-propulsion direction to a local average self-propulsion direction $\bar v_k$ on the one hand and via an attractive-repulsive potential $W$ depending on their positions and phases. The self-propulsion speed is also subject to Brownian noise. The phases of neighboring particle are subject to synchronization to a local average phase $\bar \varphi_k$ and some noise. Finally, the particles are subject to a confinement potential $V$. Specifically, the system for $(X_k, v_k, \varphi_k)_{k=1, \ldots, N}$ is written
\begin{eqnarray}
\frac{d X_k}{dt} &=& c_0 v_k - \gamma \nabla_x W(X_k, \varphi_k), \label{eq:ibm_pos} \\
d v_k &=& P_{v_k^\bot} \circ \big[ \big( \nu \, \bar v_k - \nabla_x V(X_k(t)) \big) \, dt + \sqrt{2D} \, dB_t^k \big], \label{eq:ibm_vel} \\
d \varphi_k &=& - \nu' \, \sin (\varphi_k - \bar \varphi_k) \, dt + \sqrt{2D'} \, dB_t^{'k}, \label{eq:ibm_phas}
\end{eqnarray}
the self propulsion speed $c_0$, the alignment frequency $\nu$, the synchronization frequency $\nu'$, the velocity noise $D$ and the phase noise $D'$ being positive and given. The confinement potential $V$: ${\mathbb R}^d \to [0,\infty)$ with $V \to \infty$ as $|x| \to \infty$ is also given. The terms $dB_t^k$ and $dB_t^{'k}$ describe brownian noises on ${\mathbb R}^n$ and ${\mathbb R}$ respectively. The symbol $\circ$ in \eqref{eq:ibm_vel} means that the stochastic differential equation is interpreted in the Stratonovitch sense, a condition for the vector $v_k$ to remain on ${\mathbb S}^{n-1}$ \cite{hsu2002stochastic}. 

The attractive-repulsive potential $W$: ${\mathbb R}^d \times {\mathbb R}/(2 \pi {\mathbb Z}) \times [0,\infty) \to {\mathbb R}$ takes the form 
\begin{equation}
W(x,\varphi,t) = \frac{1}{N} \sum_{j=1}^N \omega(|x-X_j(t)|) \, \sin (\varphi_j(t) - \varphi), 
\label{eq:attreppot}
\end{equation}
with a given sensing function $\omega$: $[0,\infty) \to {\mathbb R}$ such that $x \mapsto \omega(|x|)$ is  smooth on ${\mathbb R}^n$ and normalized, i.e.  $\int_{{\mathbb R}^n} \omega(|x|) \, dx = 1$. The given constant $\gamma$ can be either positive or negative and specifies the intensity of the attractive-repulsive force. The contribution of the $j$-th particle to the force $- \gamma \nabla_x W(x,\varphi,t)$ depends on the phase difference $\varphi_j(t) - \varphi$. This contribution is in the direction of $- \nabla_x [\omega (x - X_j(t))]$ if $\varphi_j(t)$ is slightly ahead of the phase $\varphi$ and $\gamma >0$ or if $\varphi_j(t)$ is slightly behind the phase $\varphi$ and $\gamma <0$. It is in the opposite direction in the converse cases. Thus, the attractive or repulsive character of this force depends on the relative phases and on the sign of $\gamma$.  

In \eqref{eq:ibm_vel}, for $v \in {\mathbb S}^{n-1}$, $P_{v^\bot}$ stands for the orthogonal projection onto $\{v\}^\bot$ and has expression $P_{v^\bot} = \textrm{Id} - v \otimes v$ where $\textrm{Id}$ is the $n \times n$ identity matrix and $\otimes$ stands for the tensor product. To define $\bar v_k$ we first introduce the current 
\begin{equation}
J_k (x,t) = \frac{1}{N} \sum_{j=1}^N \zeta(|x-X_j(t)|) \, v_j(t), 
\label{eq:def_J}
\end{equation}
with again, a given sensing function $\zeta$: $[0,\infty) \to {\mathbb R}$. Then, we let
\begin{equation}
\bar v_k = \Big( \frac{J_k}{|J_k|} \Big)(X_k(t),t) , 
\label{eq:def_barv}
\end{equation}
assuming that the denominator does not vanish. 

We proceed analogously to define $\bar \varphi_k$. We first introduce 
\begin{equation}
L_k (x,t) = \frac{1}{N} \sum_{j=1}^N \eta (|x-X_j(t)|) \, e^{i \varphi_j(t)}, 
\label{eq:def_L}
\end{equation}
with a given sensing function $\eta$: $[0,\infty) \to {\mathbb R}$. Then, we define $\bar \varphi_k(t) \in {\mathbb R}/(2 \pi {\mathbb Z})$ by
\begin{equation}
e^{i \bar \varphi_k(t)} = \Big( \frac{L_k}{|L_k|} \Big)(X_k(t),t), 
\label{eq:def_barphi}
\end{equation}
again, assuming that the denominator does not vanish. Here again, $dB_t^{'k}$ stands for independent brownian motions in ${\mathbb R}$.

System \eqref{eq:ibm_pos}-\eqref{eq:ibm_phas} is an extension of the time-continuous version of the Vicsek model \cite{vicsek1995novel} proposed in \cite{degond2008continuum}. Indeed, let us temporarily assume that there is neither external potential ($V=0$) nor phase noise ($D'=0$) and that all particle phases are initially equal. Then the attraction-repulsion potential vanishes ($W=0$), the phases remain constant and the position and velocities follow the time-continuous version of the classical Vicsek model \cite{degond2008continuum}:
\begin{eqnarray}
\frac{d X_k}{dt} &=& c_0 v_k, \label{eq:ibm_pos_vic} \\
d v_k &=& P_{v_k^\bot} \circ \big[ \nu \, \bar v_k \, dt + \sqrt{2D} \, dB_t^k \big].  \label{eq:ibm_vel_vic} 
\end{eqnarray}
Eq. \eqref{eq:ibm_pos_vic} describes motion in the direction of $v_k$ at constant speed $c_0$ as a consequence of self-propulsion. The first term at the right-hand side of \eqref{eq:ibm_vel_vic} tends to align $v_k$ with the mean direction of the neighbors $\bar v_k$ computed through \eqref{eq:def_J}, \eqref{eq:def_barv} at rate $\nu$. The second term of \eqref{eq:ibm_vel_vic} generates a brownian motion of $v_k$ on the sphere and models velocity noise with noise intensity $D$ (see e.g. \cite{degond2014hydrodynamics, degond2015macroscopic, degondfrouvellemerino17, degondfrouvellemerinotrescases18, dimarcomotsch16, frouvelle2012continuum} for references on this model and some of its variants).
 
\begin{figure}[ht!]
\centering
\includegraphics[trim={3.5cm 21.cm 2.cm 4.5cm},clip,width=6cm]{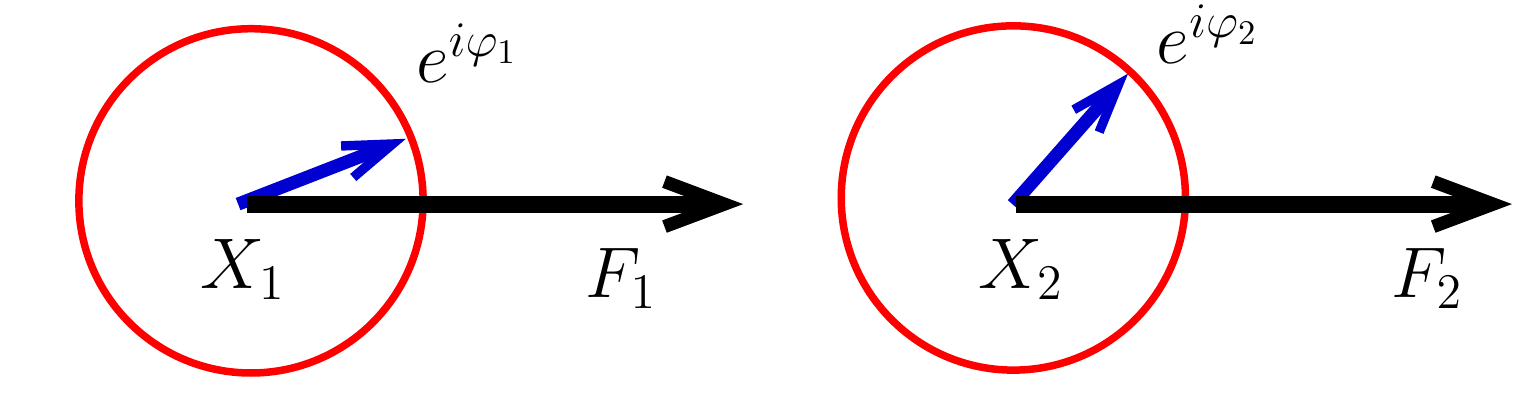}
\caption{Pursuit dynamics between two particles located at $X_1$ and $X_2$. The force $F_1 = - \gamma \nabla_x W (W_1,\varphi_1)$ acting on $X_1$ is equal to the force $F_2 = - \gamma \nabla_x W (W_2,\varphi_2)$ acting on $X_2$. Both are given by $F_1 = F_2 = \frac{\gamma}{2} \omega'(|X_1-X_2|) \frac{X_2-X_1}{|X_2-X_1|} \sin(\varphi_2 - \varphi_1)$ and are depicted by black arrows. The phases of the two particles are symbolized by the unit vectors $e^{i \varphi_k}$, $k=1,\, 2$ in the complex plane which are depicted in blue. The unit circle is drawn in red. The current situation which occurs for $\gamma \omega'(|X_1-X_2|) >0$ diplays the pursuit of $X_2$ by~$X_1$. The converse would occur if $\gamma \omega'(|X_1-X_2|) <0$.}
\label{fig:pursuit}
\end{figure}

When added, the potential $V$ biases the alignment direction in the direction of $-\nabla_x V$, as shown in \eqref{eq:ibm_vel}. This takes into account for instance, external cues in the agents' navigation. With the assumption that $V(x) \to \infty$ as $x \to \infty$, the potential confines the particles in a bounded region of space. Now, if the phases are not constant and/or if phase noise is present ($D' \not = 0$), the attraction-repulsion potential $W$ turns on. It may seem strange that~$W$ appears in \eqref{eq:ibm_pos} and not in \eqref{eq:ibm_vel} like~$V$. We provide a justification of it in Appendix~\ref{subsec:overdamped} through an overdamped limit. Now, $W$ biases particle motion in the direction of $-\gamma \nabla_x W$. It endows the particles with a pursuit behavior illustrated in Fig.~\ref{fig:pursuit} in the case of a pair interaction. The dynamics of the phases follows \eqref{eq:ibm_phas} which is conceptually similar to \eqref{eq:ibm_vel_vic} if phases $\varphi_k$ are associated with unit vectors $e^{i \varphi_k}$ in the complex plane~${\mathbb C}$. Indeed, \eqref{eq:ibm_phas} just states that this vector is subject to alignment with the mean phase vector of the neighbors $e^{i \bar \varphi_k}$ at rate $\nu'$ and to noise with noise intensity~$D'$. 

This system describes agents that swarm (through velocity alignment) and which, at the same time, are oscillators subject to synchronization (through phase alignment). Moreover, swarming and synchronization are coupled through space as motion in space depends on velocity and phase while swarming and synchronisation depend on space through the target velocity and target phase. Thus, this system belongs to the class of ``swarmalators'', a term coined in \cite{o2017oscillators}. Note that in the present system, the natural oscillator frequency is supposed to be zero. There is no conceptual objection to include a non-zero natural frequency but the patterns generated by the system and further explored could be different. This will be investigated in future work. Swarmalators have recently stimulated an intense research, see e.g. \cite{ha2019emergent, hong2018active, hong2021coupling, jimenez2020oscillatory, lee2021collective, lizarraga2020synchronization, o2022collective, o2018ring}.

\subsection{The kinetic model}
\label{sec_kin}

Our goal is now to derive a continuum version of the swarmalator model \eqref{eq:ibm_pos}-\eqref{eq:ibm_phas} in the form of hydrodynamic-type equations. To do so, it is convenient to introduce an intermediate description between the particle and hydrodynamic models, the so-called kinetic model. 

We begin with introducing the empirical measure of the particles in the space ${\mathbb R}^n \times {\mathbb S}^{n-1} \times  {\mathbb R}/(2 \pi {\mathbb Z})$ by
\begin{equation}
\mu^N(x,v,\varphi,t) = \frac{1}{N} \sum_{j=1}^N \delta_{(X_j(t), v_j(t), \varphi_j(t))} (x,v,\varphi), 
\label{eq:emp_meas}
\end{equation}
where $\delta_{(X_j(t), v_j(t), \varphi_j(t))} (x,v,\varphi)$ is the Dirac delta located at $(X_j(t), v_j(t), \varphi_j(t))$. This is a random measure. In the limit $N \to \infty$, under appropriate conditions which we will assume satisfied, $\mu^N$ converges in the weak sense to a deterministic measure $f(x,v,\varphi,t) \, dx \, dv \, d \varphi$ where $f$ satisfies the following kinetic model posed on ${\mathbb R}^n \times {\mathbb S}^{n-1} \times  {\mathbb R}/(2 \pi {\mathbb Z})$:
\begin{eqnarray}
&&\hspace{-1cm}
\partial_t f + \nabla_x \cdot \big[ \big( c_0 v - \gamma \nabla_x W_f(x,\varphi) \big) f \big] + \nabla_v \cdot \big[ P_{v^\bot} \big( \nu \, \bar v_f - \nabla_x V(x) \big) \, f \big] \nonumber \\
&&\hspace{4.5cm}
+ \nu' \, \partial_\varphi \big[ \sin( \bar \varphi_f - \varphi) f \big] = D \Delta_v f + D' \partial_\varphi^2 f, \label{eq:kin}
\end{eqnarray}
with $W_f$ given by
\begin{equation}
W_f(x,\varphi,t) = \int_{{\mathbb R}^n \times {\mathbb S}^{n-1} \times  [0,2\pi]} \omega(|y-x|) \, \sin(\psi - \varphi) \, f(y,w,\psi,t) \, dy \, dw \, d\psi. 
\label{eq:kin_Wf}
\end{equation}
To define $\bar v_f$ and $\bar \varphi_f$, we first define
\begin{eqnarray}
J_f(x,t) &=& \int_{{\mathbb R}^n \times {\mathbb S}^{n-1} \times  [0,2\pi]} \zeta(|y-x|) \, w \, f(y,w,\psi,t) \, dy \, dw \, d\psi, \label{eq:kin_Jf} \\
L_f(x,t) &=& \int_{{\mathbb R}^n \times {\mathbb S}^{n-1} \times  [0,2\pi]} \eta(|y-x|) \, e^{i \psi} \, f(y,w,\psi,t) \, dy \, dw \, d\psi. \label{eq:kin_Lf}
\end{eqnarray}
Then, we let 
\begin{equation}
\bar v_f(x,t) =\Big( \frac{J_f}{|J_f|} \Big)(x,t), \qquad e^{i \bar \varphi_f} (x,t) = \Big( \frac{L_f}{|L_f|} \Big)(x,t) . 
\label{eq:kin_barv_barphi}
\end{equation}
We note that the operators $\nabla_v$, $\nabla_v \cdot$, $\Delta_v$ are respectively the spherical gradient, divergence and Laplace-Beltrami operators. Finally, by the fact that the empirical measure \eqref{eq:emp_meas} is a probability measure, $f$ satisfies the normalization condition 
\begin{equation} 
\int_{{\mathbb R}^n \times {\mathbb S}^{n-1} \times  [0,2\pi]} f(x,v,\varphi,t) \, dx \, dv \, d \varphi = 1. 
\label{eq:f_normaliz}
\end{equation}

A hydrodynamic scaling detailed in Appendix \ref{sec_scaling_appendix}, eventually leads to the following modified kinetic problem, which depends on the scaling parameter $\varepsilon$: 
\begin{eqnarray}
&&\hspace{-1cm}
\partial_t f^\varepsilon + \nabla_x \cdot \big[ \big( v - \gamma \nabla_x U_{f^\varepsilon}(x,\varphi) \big) f^\varepsilon \big] - \nabla_v \cdot \big[ P_{v^\bot} \nabla_x V(x) \, f^\varepsilon \big]\nonumber \\
&&\hspace{-0.2cm}
=  \frac{1}{\varepsilon} \Big\{ D \, \nabla_v \cdot \big[ - k P_{v^\bot} u_{f^\varepsilon}  \, f^\varepsilon +\nabla_v f^\varepsilon  \big] + D' \, \partial_\varphi \big[ - k' \sin( \alpha_{f^\varepsilon} - \varphi) f^\varepsilon + \partial_\varphi f^\varepsilon \big] \Big\}, \label{eq:kin_final} \\
&&\hspace{-1cm}
U_f(x,\varphi,t) = |\ell_f(x,t)| \, \sin(\alpha_f - \varphi), 
\label{eq:kin_U_f} \\
&&\hspace{-1cm}
u_f = \frac{j_f}{|j_f|}, \quad  j_f (x,t) = \int_{{\mathbb S}^{n-1} \times  [0,2\pi]} w \, f(x,w,\psi,t) \, dw \, d\psi, \label{eq:kin_u_f} \\
&&\hspace{-1cm} 
e^{i \alpha_f} = \frac{\ell_f}{|\ell_f|}, \quad  \ell_f (x,t) = \int_{{\mathbb S}^{n-1} \times  [0,2\pi]}  e^{i \psi} \, f(x,w,\psi,t)  \, dw \, d\psi, \label{eq:kin_alpha_f}
\end{eqnarray}
with $k = \frac{\nu}{D}$ and $k' = \frac{\nu'}{D'}$. The parameter $\varepsilon \ll 1$ encodes the ratio of the microscopic scale, i.e. the typical distance or time over which particle response to interactions takes place, and the macroscopic scale, i.e. the typical size or duration of the experiment. The hydrodynamic model is obtained as the $\varepsilon \to 0$ limit of this system. It describes how the macroscopic scale is influenced by the microscopic dynamics on average and is developed in the next section.

\setcounter{equation}{0}
\section{The hydrodynamic model}
\label{sec_eps_to_0}

In a first subsection, we derive the hydrodynamic model by letting $\varepsilon \to 0$ in the kinetic model \eqref{eq:kin_final}-\eqref{eq:kin_alpha_f}. Details are given in Appendix \ref{sec_eps_to_0_proof}. In a second subsection, we simplify the hydrodynamic model by assuming small phase noise, and we study the resulting model.

\subsection{Derivation of the hydrodynamic model}
\label{subsec_general}

We being with introducing the von-Mises distributions: for $u \in {\mathbb S}^{n-1}$ and $\alpha \in {\mathbb R}/(2 \pi {\mathbb Z})$, we let 
\begin{eqnarray}
M_u(v) &=& \frac{1}{Z} \, e^{k v \cdot u}, \quad Z = \int_{{\mathbb S}^{n-1}} e^{k v \cdot u} \, dv, \label{eq:VMv} \\
N_\alpha(\varphi) &=& \frac{1}{Z'} \, e^{k' \cos(\varphi - \alpha)}, \quad Z' = \int_{[0,2\pi]} e^{k' \cos(\varphi - \alpha)} \, d\varphi.  \label{eq:VMphi} 
\end{eqnarray}
We note that $Z$ does not depend on $u$ but only on $k$ and likewise, $Z'$ does not depend on $\alpha$ but only on $k'$. 

Then, we have the formal theorem, whose proof can be found in Appendix \ref{sec_eps_to_0_proof}. 

\begin{theorem}
Suppose that there is a smooth solution $f^\varepsilon$ to the kinetic model \eqref{eq:kin_final} for all $\varepsilon >0$ and that this solution converges smoothly as $\varepsilon \to 0$ to a function $f^0$. Then, 
\begin{equation}
f^0(x,v,\varphi,t) = \rho(x,t) \,  M_{u(x,t)}(v) \, N_{\alpha(x,t)}(\varphi), 
\label{eq:equi}
\end{equation}
where $\rho$, $u$ and $\alpha$ are functions from ${\mathbb R}^n \times [0,\infty)$ to $[0,\infty)$, ${\mathbb S}^{n-1}$ and ${\mathbb R}/(2 \pi {\mathbb Z})$ respectively, which satisfy the following systems of partial differential equations:
\begin{eqnarray}
&&\hspace{-1cm}
\partial_t \rho + \nabla_x \cdot \big[ \rho (c_1 u + b  \rho \nabla_x \alpha) \big] = 0, \label{eq:fl_rho} \\
&&\hspace{-1cm}
\partial_t u + \big[ (c_2 u + b  \rho \nabla_x \alpha) \cdot \nabla_x \big] u + P_{u^\bot} \nabla_x ( \Theta  \log \rho + \kappa V ) = 0, \label{eq:fl_u} \\
&&\hspace{-1cm}
\rho \, \Big( \partial_t \alpha + \big[ (c_1 u + b' \rho \nabla_x \alpha) \cdot \nabla_x \big] \alpha \Big) - \Theta' \, \nabla_x \cdot \big( \rho \nabla_x \rho \big) = 0, \label{eq:fl_al}
\end{eqnarray}
where the coefficients $c_1$, $c_2$, $b$, $b'$, $\Theta$, $\Theta'$ and $\kappa$ are given in Appendix~\ref{sec_eps_to_0_proof} and have the following properties: 
\begin{itemize}
\item $c_1$: $[0,\infty) \to [0,1)$, $k \mapsto c_1(k)$ is an increasing function of $k$ with $c_1(0) = 0$ and $\lim_{k\to \infty} c_1(k) = 1$ \cite{frouvelle2012continuum}. 
\item $c_2$: $[0,\infty) \to (-1,1)$, $k \mapsto c_2(k)$ is a function of $k$ such that $c_2(0) = 0$ and $\lim_{k \to \infty} c_2(k) = 1$. Furthermore, for small and large $k$, we have $0<c_2(k)<c_1(k)$~\cite{frouvelle2012continuum}. 
\item $\Theta = \Theta(k) = k^{-1}$, \, $\kappa = \kappa(k) = \frac{n-1}{k} + c_2(k)$ (see Appendix \ref{sec_eps_to_0_proof}). 
\item $\Theta'$, $b$ and $b'$ are functions of $k'$ and $\gamma$ which have the opposite sign to $\gamma$ (see Appendix~\ref{sec_eps_to_0_proof}). 
\end{itemize}
\label{thm:eps_to_0}
\end{theorem}

\begin{remark}
Numerical simulations in dimension $n=2$ \cite{motsch2011numerical} suggest that $c_2$: $[0,\infty) \to [0,1)$, $k \mapsto c_2(k)$ is an increasing function of $k$ and that $0<c_2(k) <c_1(k)$, $\forall k \in (0,\infty)$. But a rigorous proof of these properties is still lacking. 
\end{remark}  

\begin{remark}
Because of \eqref{eq:f_normaliz}, $\rho$ satisfies the normalization condition
\begin{equation}
\int_{{\mathbb R}^n} \rho(x,t) \, dx = 1. 
\label{eq:rho_normaliz}
\end{equation}
\end{remark}

In the remainder of this paper, we will assume that $k$ is small or large enough so that the property $0<c_2(k)<c_1(k)$ is guaranteed. 

Formula \eqref{eq:equi} relies on the following lemma, whose proof can be found in \cite{degond2008continuum}. Defining 
\begin{equation}
Q(f) =  D \, \nabla_v \cdot \big[ - k P_{v^\bot} u_f  \, f + \nabla_v f  \big] +  D' \, \partial_\varphi \big[ - k' \sin( \alpha_f - \varphi) f + \partial_\varphi f \big], \label{eq:defQ}
\end{equation}
we have :

\begin{lemma}
(i) We can write
\begin{equation}
Q(f) = D \, \nabla_v \cdot \Big[ M_{u_f} \nabla_v \Big( \frac{f}{M_{u_f}} \Big) \Big] +  D' \, \partial_\varphi \Big[ N_{\alpha_f} \partial_\varphi \Big( \frac{f}{N_{\alpha_f}} \Big) \Big]. \label{eq:QVM}
\end{equation}

\noindent
(ii) Defining 
\begin{equation}
{\mathcal D}(f) = \int_{{\mathbb S}^{n-1} \times [0,2\pi]} Q(f) \, \frac{f}{M_{u_f} \, N_{\alpha_f}} \, dv \, d\varphi, \label{eq:defH} 
\end{equation}
we have 
\begin{equation}
{\mathcal D}(f) = - \int_{{\mathbb S}^{n-1} \times [0,2\pi]} M_{u_f} \, N_{\alpha_f} \, 
\Big[ D \Big| \nabla_v \Big( \frac{f}{M_{u_f}\, N_{\alpha_f}} \Big) \Big|^2 
+ D' \Big| \partial_\varphi \Big( \frac{f}{M_{u_f} \, N_{\alpha_f}} \Big) \Big|^2 \Big] \, dv \, d\varphi \leq 0. 
\label{eq:defD}
\end{equation}

\noindent
(iii) Let $f=f(v,\varphi)$. Then, the following assertions are equivalent:

\noindent
\hspace{1cm} (a) $Q(f) = 0$, 

\noindent
\hspace{1cm} (b) ${\mathcal D}(f) = 0$

\noindent
\hspace{1cm} (c) $\exists (\rho, u, \alpha) \in [0,\infty) \times {\mathbb S}^{n-1} \times {\mathbb R}/(2 \pi {\mathbb Z})$ such that 
\begin{equation}
f(v,\varphi) = \rho \, M_u(v) \, N_\alpha(\varphi). 
\label{eq:equi_homo}
\end{equation}

\label{lem:equi}
\end{lemma}

\medskip
We note that Eq. \eqref{eq:kin_final} can be written 
\begin{equation} 
T(f^\varepsilon) = \frac{1}{\varepsilon} Q(f^\varepsilon), 
\label{eq:abstract}
\end{equation}
with 
\begin{equation}
T(f) = \partial_t f + \nabla_x \cdot \big[ \big( v - \gamma \nabla_x U_{f}(x,\varphi) \big) f \big] - \nabla_v \cdot \big[ P_{v^\bot} \nabla_x V(x) \, f \big], 
\label{eq:expressT}
\end{equation}
being the transport operator. Lemma \ref{lem:equi} and Formula \eqref{eq:abstract} suggest that we could modify the collision operator \eqref{eq:QVM} and introduce the relaxation (BGK-type) operator
\begin{equation}
Q_R(f) =  D \big( \rho_f M_{u_f} N_{\alpha_f} - f \big), \label{eq:defQR}
\end{equation}
with $u_f$ and $\alpha_f$ given by \eqref{eq:kin_u_f}, \eqref{eq:kin_alpha_f} and $\rho_f$ by 
\begin{equation}
\rho_f(x,t) = \int_{{\mathbb S}^{n-1} \times  [0,2\pi]} f(x,w,\psi,t) \, dw \, d\psi, \label{eq:defrhof}
\end{equation}
and consider the analogous perturbation problem to \eqref{eq:abstract}, namely 
\begin{equation} 
T(f^\varepsilon) = \frac{1}{\varepsilon} Q_R(f^\varepsilon).
\label{eq:abstractR}
\end{equation}
It can be easily shown that Lemma \ref{lem:equi} (ii) and (iii) still holds with \eqref{eq:defD} replaced by 
\begin{equation}
{\mathcal D}(f) = - D \int_{{\mathbb S}^{n-1} \times [0,2\pi]} M_{u_f} \, N_{\alpha_f} \, 
\Big( \rho_f - \frac{f}{M_{u_f} \, N_{\alpha_f}} \Big)^2 \, dv \, d\varphi \leq 0. 
\label{eq:defD_bgk}
\end{equation}
Then, the following theorem holds 

\begin{theorem}
Theorem \ref{thm:eps_to_0} still holds for Eq. \eqref{eq:abstractR}, but with different expressions of the constants involved. Their expressions can be found in Appendix \ref{sec_eps_to_0_proof}. 
\label{thm:eps_to_0_bgk}
\end{theorem}

Kinetic Eq. \eqref{eq:abstractR} can be given an interpretation in terms of an interacting particle system following a jump process, aka a Piecewise Deterministic Markov Process (PDMP). This interpretation is given in Appendix \ref{sec_particle_bgk}.

We now make comments on the hydrodynamic model \eqref{eq:fl_rho}-\eqref{eq:fl_al}. Again, let us temporarily assume that there is no external potential ($V=0$), no phase noise ($\Theta' = 0$, see Appendix~\ref{sec_small_noise_lemma_proof}) and that the phases are initially constant ($\alpha|_{t=0}$ independent of $x$). Then,~$\alpha$ remains constant in time (and independent of $x$) and the equations for $\rho$ and $u$ reduce to 
\begin{eqnarray}
&&\hspace{-1cm}
\partial_t \rho + \nabla_x \cdot (c_1 \rho u ) = 0, \label{eq:fl_rho_soh} \\
&&\hspace{-1cm}
\partial_t u + c_2 (u \cdot \nabla_x) u +  \frac{\Theta}{\rho} \, P_{u^\bot} \nabla_x \rho  = 0,  \label{eq:fl_u_soh} \\
&&\hspace{-1cm}
|u|=1, \label{eq:fl_u_constraint_soh}
\end{eqnarray}
where we have highlighted in \eqref{eq:fl_u_constraint_soh} the fact that $u$ (the average self-propulsion direction) is a normalized vector. This model is the continuum version of the Vicsek model derived in \cite{degond2008continuum}, referred to as the Self-Organized Hydrodynamics (SOH). Eq. \eqref{eq:fl_rho_soh} is the mass conservation (or continuity) equation for the fluid density $\rho$. It shows that the fluid velocity $c_1 u$ establishes along the average self-propulsion direction $u$ and has norm $c_1$. Hence, the fluid speed is less than the particle speed (whose value is $1$ after scaling) and this is because the direction of particle velocities is spread around $u$ according to the von Mises distribution \eqref{eq:VMv}. Eq. \eqref{eq:fl_u_soh} is akin to the momentum conservation equation in the isothermal Euler equation of gas dynamics. Indeed, the first two terms correspond to the material derivative of $u$ and are balanced by a pressure force $- \Theta \nabla_x \rho$. However, we can spot several differences. The first one is the normalization condition \eqref{eq:fl_u_constraint_soh} which has no counterpart in the Euler equation and which gives rise to the projection operator $P_{u^\bot}$ in factor of the pressure term. Indeed, this projection is needed to ensure consistency with the constraint \eqref{eq:fl_u_constraint_soh}. Another difference is that the material derivative does not involve the fluid velocity $c_1 u$ but a different velocity $c_2 u$ since $c_2 \not = c_1$. This feature makes the model non Galilean-invariant, but this is no surprise because the particle model itself is not Galilean-invariant: there is a preferred frame where the particle speed is $1$. The quantity $c_2$ is the speed at which information propagates among the agents for them to update their velocity in response to density gradients. We noted that, fairly generally, we have $c_2 < c_1$. The fact that $c_2 \not = c_1$ will be key to the existence of topological travelling-wave solutions discussed in the next section. 

Compared with the SOH model, the full model involves the additional equation \eqref{eq:fl_al} for the average phase $\alpha(x,t)$. Naturally, only spatial gradients of $\alpha$ influence the dynamics of $\rho$ and $u$ (as only differences of phases influence the particle dynamics) and their influence is the same in Eqs. \eqref{eq:fl_rho} and \eqref{eq:fl_u}: it adds the same term $b \rho \nabla_x \alpha$ to both the fluid velocity~$c_1 u$  and the information velocity $c_2 u$. This reflects the influence of the attraction-repulsion potential which, at the particle level, adds to the self-propulsion velocity a vector whose orientation depends on the differences (i.e. gradients) of the phases and on the sign of~$\gamma$ (which controls that of $b$ as stated in the theorem). The phase equation \eqref{eq:fl_al} is a balance equation similar to the velocity equation \eqref{eq:fl_u}. The material derivative of $\alpha$ combines $c_1 u$ and the contribution of $\rho \nabla_x \alpha$ as in \eqref{eq:fl_rho} but intriguingly, this contribution is weighted by a different coefficient $b'$ compared to Eqs. \eqref{eq:fl_rho}, \eqref{eq:fl_u}. Finally the material derivative is balanced by a diffusive term of the density $\rho$. However, note that $\Theta'$ may have either signs as well as $\gamma$. Finally, the external potential appears in the velocity equation~\eqref{eq:fl_u} and biases the mean alignment direction $u$ (or its material derivative) consistently with what it does in the particle model. The full model 
\eqref{eq:fl_rho}-\eqref{eq:fl_al} will be referred to as ``Swarmalator Hydrodynamics'' (SH).

\subsection{Small noise limit in the phase equation}
\label{subsec_small_noise}

In this section, we neglect the noise in the phase equation, i.e., we make $k' \to \infty$. It has been numerically observed that the most interesting patterns are obtained in this regime. Indeed, phase gradients increase coherence between the particles through the pursuit mechanism (see Fig.~\ref{fig:pursuit}). On the opposite, phase noise, which contributes to equalizing the phases, destroys this coherence. 

We have the following lemma, the proof of which can be found in Appendix \ref{sec_small_noise_lemma_proof}. 

\begin{lemma}
In the limit $k' \to \infty$, we have 
\begin{equation}
 b \to - \gamma, \qquad b' \to - \gamma, \qquad \Theta' \to 0.  
\label{eq:small_noise_coef}
\end{equation}
In this limit, the macroscopic system becomes (assuming $\rho \not = 0$): 
\begin{eqnarray}
&&\hspace{-1cm}
\partial_t \rho + \nabla_x \cdot \big[ \rho (c_1 u + b  \rho \nabla_x \alpha) \big] = 0, \label{eq:fl_rho_sm} \\
&&\hspace{-1cm}
\partial_t u + \big[ (c_2 u  + b  \rho \nabla_x \alpha) \cdot \nabla_x \big] u + P_{u^\bot}  \nabla_x ( \Theta \log \rho  + \kappa V ) = 0, \label{eq:fl_u_sm} \\
&&\hspace{-1cm}
\partial_t \alpha + \big[ (c_1 u  + b \rho \nabla_x \alpha) \cdot \nabla_x \big] \alpha = 0, \label{eq:fl_al_sm}
\end{eqnarray}
\label{lem:small_noise_limit}
\end{lemma}

We note that the contributions of $\rho \nabla_x \alpha$ are now weighted by the same coefficient $b$ in all three material derivatives (there was a different coefficient $b'$ for the phase equation in the general SH \eqref{eq:fl_rho} - \eqref{eq:fl_al}). Consequently, in \eqref{eq:fl_al_sm}, the phase is transported by the fluid velocity as defined from \eqref{eq:fl_rho_sm}, i.e. $c_1 u  + b \rho \nabla_x \alpha$. 

In the remainder of this paper, we will focus on this system, further referred to as the ``Noiseless Swarmalator Hydrodynamics'' (NSH). 

Taking the gradient of \eqref{lem:small_noise_limit} and introducing $z=\nabla_x \alpha$, we find that $(\rho,u,z)$ satisfies the following system: 
\begin{eqnarray}
&&\hspace{-1cm}
\partial_t \rho + \nabla_x \cdot \big[ \rho (c_1 u  + b  \rho z) \big] = 0, \label{eq:fl_rho_nl} \\
&&\hspace{-1cm}
\partial_t u + \big[ (c_2 u + b  \rho z) \cdot \nabla_x \big] u + P_{u^\bot}  \nabla_x ( \Theta \log \rho  + \kappa V ) = 0, \label{eq:fl_u_nl} \\
&&\hspace{-1cm}
\partial_t z + \nabla_x \big[ (c_1 u + b \rho z) \cdot z \big] = 0, \label{eq:fl_al_nl} \\
&&\hspace{-1cm}
\nabla_x \wedge z = 0, \label{eq:fl_const_nl}
\end{eqnarray}
where $\nabla_x \wedge z$ is the exterior derivative of $z$ if $z$ is identified to a differential form, i.e. $\nabla_x \wedge z$ is the antisymmetric matrix with entries $(\nabla_x \wedge z)_{ij} = \partial_{x_i} z_j - \partial_{x_j} z_i$. Eq. \eqref{eq:fl_const_nl} is a structural constraint that says that $z$ is a gradient, namely that there exists $\alpha$ such that $z = \nabla_x \alpha$. Eq. \eqref{eq:fl_al_nl} preserves this structural constraint as the exterior derivative of a gradient is always zero. Now, the relation between the NSH System \eqref{eq:fl_rho_sm}-\eqref{eq:fl_al_sm} and System \eqref{eq:fl_rho_nl}-\eqref{eq:fl_const_nl} is expressed in the following lemma, the proof of which is obvious. 

\begin{lemma}
(i) Let $(\rho, u, \alpha)$ be a solution of the NSH System \eqref{eq:fl_rho_sm}-\eqref{eq:fl_al_sm}. Then, $(\rho, u, z)$ with $z = \nabla_x \alpha$ is a solution of System \eqref{eq:fl_rho_nl}-\eqref{eq:fl_const_nl}. 

\noindent
(ii) Conversely, let $(\rho, u, z)$ be a solution of System \eqref{eq:fl_rho_nl}-\eqref{eq:fl_const_nl}. Suppose $z_0= z|_{t=0}$ satisfies $\nabla_x \wedge z_0 = 0$. Let $\alpha_0$ be a solution of $\nabla_x \alpha_0 = z_0$ and let $\alpha$ be a solution of 
$$ \partial_t \alpha + (c_1 u + b \rho z) \cdot z  = 0, \qquad \alpha|_{t=0} = \alpha_0. $$
Then, $(\rho, u, \alpha)$ is a solution of the NSH System \eqref{eq:fl_rho_sm}-\eqref{eq:fl_al_sm}.
\label{lem:equiv}
\end{lemma}

\begin{figure}[ht!]
\centering
\includegraphics[trim={3.5cm 13.5cm 4.cm 4.5cm},clip,width=6cm]{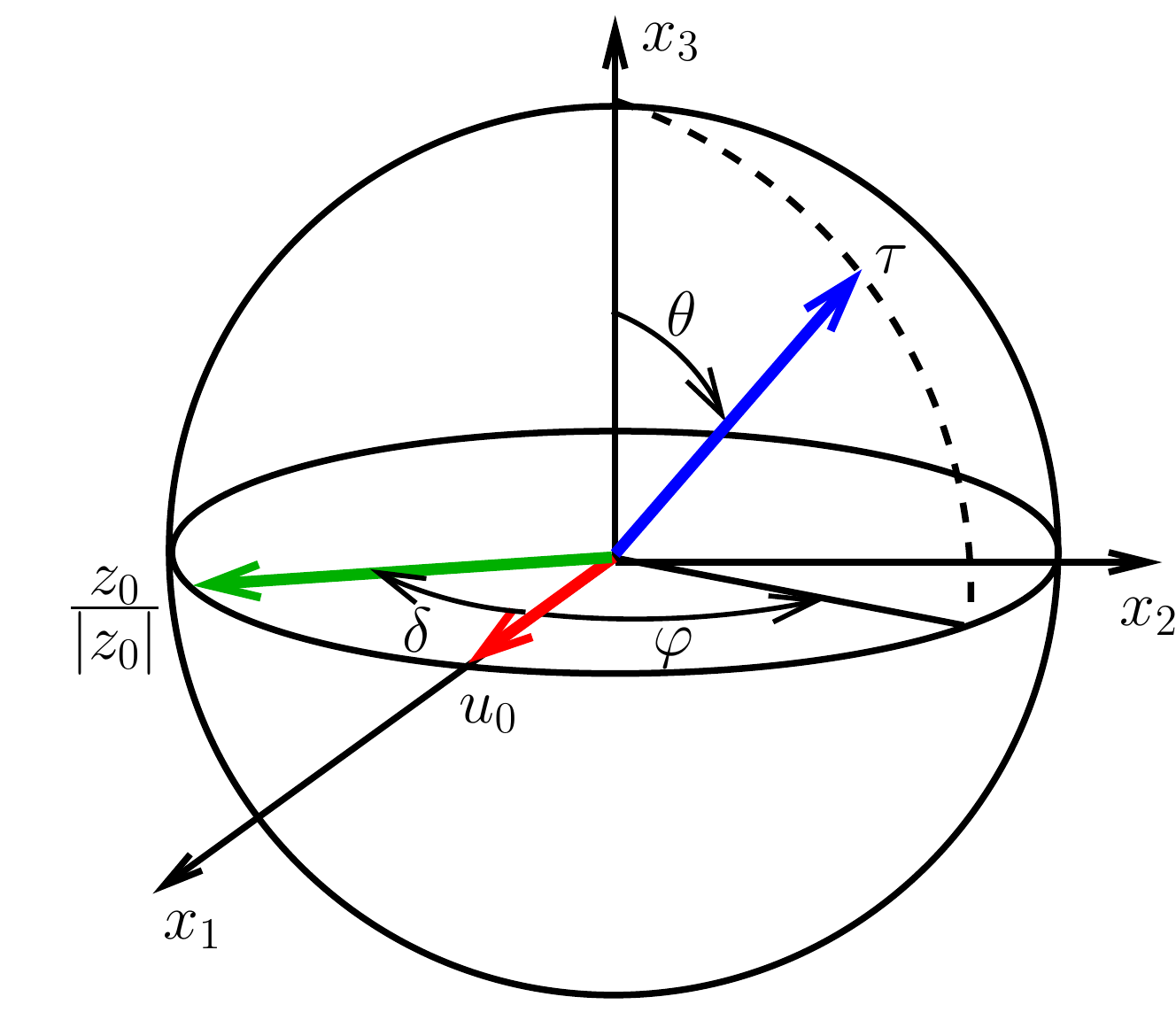}
\caption{Geometric setting of the hyperbolicity analysis}
\label{fig:angles}
\end{figure}

We investigate the hyperbolicity of System \eqref{eq:fl_rho_nl}-\eqref{eq:fl_const_nl} in the case $V=0$. Consider a uniform steady state of density $\rho_0$, velocity $u_0$ and phase gradient $z_0$ (note that for this study, we ignore the normalization condition \eqref{eq:rho_normaliz}). In view of Lemma \ref{lem:equiv},  such a steady-state corresponds to a solution $(\rho_0, u_0, \alpha_0)$ of the NSH System \eqref{eq:fl_rho_sm}-\eqref{eq:fl_al_sm} with 
\begin{equation}
\alpha_0 = \alpha_0(x,t) = z_0 \cdot x - (c_1 u_0 + b \rho_0 z_0) \cdot z_0 \, t. 
\label{eq:alpha0}
\end{equation}
In other words, a uniform steady-state solution for System \eqref{eq:fl_rho_nl}-\eqref{eq:fl_const_nl} is a travelling-wave solution of the NSH System  \eqref{eq:fl_rho_sm}-\eqref{eq:fl_al_sm}. We take the spatial Fourier transform of the linearized system to \eqref{eq:fl_rho_nl}-\eqref{eq:fl_const_nl} and denote by $\xi$ the Fourier variable and $\tau = \xi/|\xi|$ its direction. Using frame indifference, we introduce a reference frame $(e_1, \ldots, e_n)$ such that $u_0 \in \textrm{Span} \{e_1\}$, $z_0 \in \textrm{Span} \{e_1,e_2\}$ and $\xi \in \textrm{Span} \{e_1,e_2,e_3\}$. Then, we introduce the angles $\delta \in [-\pi,\pi)$, $\theta \in [0,\pi]$, $\phi \in [-\pi,\pi)$ such that 
\begin{equation}
u_0 = e_1, \, \, z_0 = |z_0| \, (\cos \delta \, e_1 + \sin \delta \, e_2), \, \,  \tau = \sin \theta \, \cos \phi \, e_1 + \sin \theta \, \sin \phi \, e_2 + \cos \theta \, e_3.
\label{eq:frame}
\end{equation}
In other words, $\delta$ is the polar angle of $z_0$ in the frame $(e_1,e_2)$ and $(\theta,\phi)$ are the spherical angles of $\tau$ in $(e_1,e_2,e_3)$. This geometric setting is illustrated in Figure \ref{fig:angles}. Now, we have

\begin{lemma}
In this Lemma, we make $V=0$. 

\noindent
(i) If $z_0=0$ or $z_0 \parallel u_0$ and if $c_1-c_2 \not = 2 |b| \rho_0 |z_0|$, System \eqref{eq:fl_rho_nl}-\eqref{eq:fl_const_nl} is hyperbolic about $(\rho_0,u_0,z_0)$. 
\begin{itemize}
\item If $|z_0| \not = 0$, the eigenvalue (propagation speed) 
$$\lambda_1 = (c_2 + b \rho_0 |z_0|) \sin \theta \, \cos \phi,$$
is of multiplicity at least $n-2$. The other three eigenvalues are simple and are the real roots of a cubic equation. In some special cases, one of these roots can coincide with $\lambda_1$, in which case it has multiplicity $n-1$. In the general case, $\lambda_1$ has multiplicity equal to $n-2$. 
\item If $|z_0| = 0$, w.l.o.g. we can choose $\theta = \pi/2$. Then, the eigenvalues are (generically): 
\begin{itemize}
\item[$*$] $\lambda_1 = c_2 \cos \phi$ of multiplicity $n-2$ corresponding to propagation of the components of the velocity normal to both $u_0$ and $\xi$, 
\item[$*$] $\lambda_2 = c_1 \cos \phi$ of multiplicity $1$ corresponding to propagation of phase perturbations, 
\item[$*$] a pair of simple eigenvalues
$$\lambda_\pm = \frac{1}{2} \Big( (c_1 + c_2) \cos \phi \pm \sqrt{ (c_1 - c_2)^2  \cos^2 \phi + 4 c_1 \Theta \sin^2 \phi} \Big), $$
corresponding to the intertwining of the density perturbation and the component of the velocity perturbation along $P_{u_0^\bot} \xi$. 
\end{itemize}
\end{itemize}

\noindent
(ii) If $z_0 \not =0$ and $z_0 \not \parallel u_0$, there are two constants $C_1$ and $C_2$  with $0 < C_1 < C_2$, depending on $c_1$, $c_2$ and $\Theta$, such that for all $\rho_0 |b| |z_0| \in (0,C_1) \cup (C_2,\infty)$, there exists values of the angle $\delta$ between $u_0$ and $z_0$ such that System \eqref{eq:fl_rho_nl}-\eqref{eq:fl_const_nl} is \textbf{not} hyperbolic about $(\rho_0,u_0,z_0)$. 
\label{lem:hyperbolic}
\end{lemma}

The proof of this Lemma can be found in Appendix \ref{sec_hyperbolic_proof}. 

\begin{remark}
(i) In the case $z_0=0$, the eigenvalues $\lambda_1$ and $\lambda_\pm$ are the eigenvalues of the SOH model \eqref{eq:fl_rho_soh}, \eqref{eq:fl_u_soh} \cite{degond20133hydrodynamic}. The phase equation only adds the extra eigenvalue $\lambda_2$. In this case, no eigenmode depends on the coupling constant $b$. 

\medskip
\noindent
(ii) The condition that $c_1-c_2 \not = 2 |b| \rho_0 |z_0|$ is purely technical. In this case, there is a double real eigenvalue which is not explicit. Thus, the dimension of the corresponding eigenspace is not known and we cannot guarantee hyperbolicity. However, this is an isolated value of $|b| \rho_0 |z_0|$  and so, we may state that generically, the problem is hyperbolic when $z_0 \parallel u_0$.

\medskip
\noindent
(iii) As soon as $z_0 \not \parallel u_0$ and $\rho_0 |b| |z_0|$ ie either small or large, there are configurations $(\rho_0,u_0,z_0)$ that lead to hyperbolicity loss. The proof in Appendix \ref{sec_hyperbolic_proof} shows that this loss happens when the angle between $u_0$ and $z_0$ is close to $\pi/2$. So, there is no contradiction with the hyperbolicity result when $z_0 \parallel u_0$. Although we do not know if  hyperbolicity is lost at intermediate values of $\rho_0 |b| |z_0|$, we may expect to observe instabilities as soon as the angle between $u_0$ and $z_0$ is sizeable. 
\end{remark}

In the next section, we will consider the NSH model and derive several classes of explicit solutions in dimension $n=2$.

\subsection{Doubly periodic travelling-wave solutions in dimension $n=2$}
\label{subsec:explicit}

In this section, we restrict ourselves to dimension $n=2$. We let $(x_1,x_2)$ be the cartesian coordinates of a point $x \in {\mathbb R}^2$ and $(e_1,e_2)$ be the cartesian coordinate basis. We denote by $(u_1,u_2)$ the two coordinates of the self-propulsion velocity $u$ in this basis. We recall that $u$ is a normalized vector, i.e. 
\begin{equation}
u_1^2 + u_2^2 = 1. 
\label{eq:str_const}
\end{equation}
We assume a spatial domain $\Omega = (0,1)^2$ with periodic boundary conditions. Then, we have the following

\begin{proposition}
(i) Travelling-wave solutions: let $(m,p) \in {\mathbb Z}^2$ and let $U \in {\mathbb S}^1$ be arbitrary. Then, the following is a periodic travelling-wave solution of the NSH system \eqref{eq:fl_rho_sm}-\eqref{eq:fl_al_sm} (with $V=0$) in $\Omega$ satisfying the normalization condition \eqref{eq:rho_normaliz}: 
\begin{eqnarray}
\rho &=& 1, \label{eq:rho_per_TW} \\
u &=& U, \label{eq:u_per_TW} \\
\alpha &=& 2 \pi \, (p x_1 + m x_2) - \lambda t + \alpha_0, \label{eq:alp_per_TW} \\
\lambda &=& 2 \pi c_1 (p U_1 + m U_2) + 4 \pi^2 b (p^2 + m^2), \label{eq:lambda_per_TW}
\end{eqnarray}
where $U=U_1 e_1 + U_2 e_2 $ and $\alpha_0 \in {\mathbb R}$ is an arbitrary constant.

\noindent
(ii) Stationary solutions with non-constant phase:  let $(p,m) \in {\mathbb Z}^2 \setminus \{(0,0)\}$ be such that 
\begin{equation}
\sqrt{p^2 + m^2} \leq \frac{c_1}{2 \pi |b|}. 
\label{eq:TW_constraint} 
\end{equation}
Then there exist two vectors (if the inequality in \eqref{eq:TW_constraint} is strict) or a unique vector (if there is equality in \eqref{eq:TW_constraint}) $U \in {\mathbb S}^1$ such that 
\begin{equation}
p U_1 + m U_2 = - \frac{2 \pi b}{c_1} (p^2 + m^2). 
\label{eq:TW_constraint2} 
\end{equation}
For these two choices of $U$ (respectively unique choice of $U$), then $\lambda = 0$ and the solution \eqref{eq:rho_per_TW}-\eqref{eq:alp_per_TW} is a stationary solution. 

\noindent 
(iii) Stationary solutions with constant phase: If $(p,m)=(0,0)$, then $\lambda = 0$ and $\alpha = \alpha_0$ is a constant. Then the solution \eqref{eq:rho_per_TW}-\eqref{eq:alp_per_TW} is a stationary solution for any choice of $U \in {\mathbb S}^1$. 

\label{prop_per}
\end{proposition}

The proof of this theorem is given in Appendix \ref{sec_TW_stat}. These periodic solutions can easily be generalized to arbitrary dimensions. They are also solutions to the SH model \eqref{eq:fl_rho}-\eqref{eq:fl_al} if in Prop. \ref{prop_per}, we change $b$ into $b'$. 

\begin{figure}[ht!]
\centering
\includegraphics[trim={3.5cm 10cm 1.cm 4.5cm},clip,width=7cm]{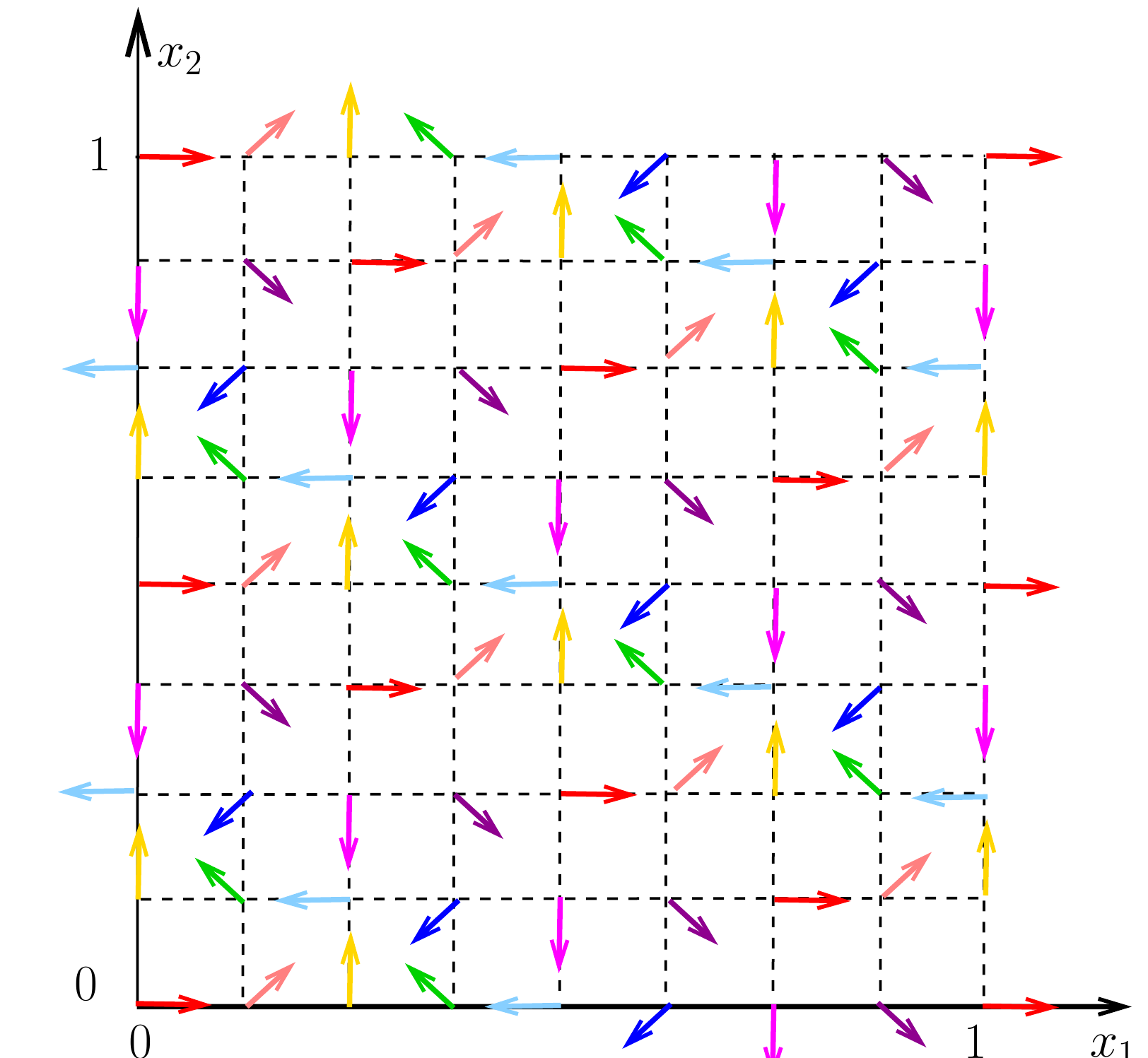}
\caption{Plot of the phase field $\alpha(x,0)$ (when $\alpha_0 = 0$) given by \eqref{eq:alp_per_TW} in the case $(p,m) = (1,2)$. The vectors $e^{i \alpha(x,0)}$ at the points $(x_1,x_2) = (\frac{k}{8}, \frac{\ell}{8})$, for $k$, $\ell$ in $\{0, 1, \ldots , 8 \}$ are shown. The color code corresponds to the angle $\alpha$ with red, pink, yellow, green, light blue, blue, magenta, purple corresponding to $\alpha = 0, \, \frac{\pi}{4}, \frac{\pi}{2}, \frac{3 \pi}{4}, \pi, \frac{5\pi}{4}, \frac{3\pi}{2}, \frac{7\pi}{4}$ respectively. The indices of this vector field about the origin are $1$ and $2$ when one moves along the $x_1$- and $x_2$-axes respectively.}
\label{fig:periodic}
\end{figure}

We note that when $x_1$ or $x_2$ increases by one (i.e. one period), the phase $\alpha$ is increased by an integer multiple of $2 \pi$. So, as an element of ${\mathbb R}/(2 \pi {\mathbb Z})$, $\alpha$ is ${\mathbb Z}$-periodic, as $\rho$ and $u$ are. An alternate view consists of introducing the unit vector field $e^{i \alpha(x,t)}$ where the plane is identified with~${\mathbb C}$. In this representation, the integers $p$ and $m$ are the indices of this unit vector field when, at a given time $t$, one moves along the $x_1$- and $x_2$-axes by one period respectively (see Fig. \ref{fig:periodic}). Thus, the pair $(p,m)$ is a  topological index of this solution and obviously, remains unchanged with time. 

These travelling-wave solutions are nothing but the restrictions to a periodic spatial domain of the uniform solutions of System \eqref{eq:fl_rho_nl}-\eqref{eq:fl_const_nl}, with the additional constraint that the phase itself must be periodic (hence the restriction to a discrete set of values of $z_0$, namely $z_0 \in 2 \pi {\mathbb Z}^2 \setminus \{(0,0)\}$). The stability of these solutions has been studied in Section \ref{subsec_small_noise}. Thus, we may apply Lemma \ref{lem:hyperbolic} and conclude that, as long as $U = (p,m)/\sqrt{p^2 + m^2}$, these solutions are stable. Also, we may state that otherwise, if  $|b| \sqrt{p^2+m^2}$ is either small or large, these solutions are unstable. In particular, the stationary solutions which are guaranteed to be stable are those for which we have equality in \eqref{eq:TW_constraint}. In the unstable case, one may wonder what happens to the topological index $(p,m)$. Since, it can only change by integer values, only two possibilities may occur: either the solution transitions to another solution with the same topological index, or, a phase discontinuity occurs somewhere. To distinguish between these two scenarios, numerical simulations are needed and will be developed in future work.

\setcounter{equation}{0}
\section{Numerical experiments}
\label{sec:numerics}

In this section, we present numerical simulations of both the particle model \eqref{eq:ibm_pos}-\eqref{eq:ibm_phas} and the hydrodynamic SH model \eqref{eq:fl_rho}-\eqref{eq:fl_al}.  The goals and main conclusions of this section are summarized below. 

\begin{itemize}
\item In Section \ref{sec:validationsh}, we provide a validation of the SH model as a macroscopic description of the particle model by showing that the particle model follows the behavior predicted by the SH model. For this, we use the explicit doubly periodic solutions of Section~\ref{subsec:explicit} as a baseline and we provide a quantitative analysis of the convergence of the particle model to the SH model as the number of particles goes to infinity. 

\item In Section \ref{sec:segregationparticles} we study the long-time behavior of the particle system and the stability of the doubly-periodic travelling-wave solutions. We investigate the role of the noise in the phase equation (i.e. of the parameter~$k'$) and of the direction of the initial velocity $u_0$. In particular, we demonstrate that the doubly-periodic solution is all the more stable than the level of phase-noise is high (i.e. than the value of $k'$ is low). For low levels of noise in the phase equation (i.e for large values of $k'$), the behavior of the particle system after 
a certain simulation time is characterized by the emergence of a strong segregation phenomenon between regions of constant phase separated by thin boundaries of very-low density.  By contrast, for large levels of noise in the phase equation (i.e. for $k'$ small), the doubly-periodic travelling wave solution is very stable starting from any configuration. Moreover, in this case, topologically non-trivial states can emerge even starting from an initially disordered state. Finally, these experiments also give a numerical verification of the hyperbolicity condition stated in Lemma \ref{lem:hyperbolic}. In particular we observe that when $u_0$ and $\nabla_x \alpha_0$ are normal and for low phase-noise levels, the destabilization of the corresponding travelling-wave solution is faster than when $u_0$ and $\nabla_x \alpha_0$ are aligned. In addition, we also observe that the solution is more stable when $u_0$ and $\nabla_x \alpha_0$ are negatively aligned (i.e. in opposite directions) than when they are positively aligned (i.e. in the same direction). 

\item In Section \ref{sec:simulationsh}, we present simulations of the SH model. The segregation behavior observed in the particle simulations is not clearly observed in the simulation of the SH system although we still observe the formation of thin regions of low-density for low levels of noise in the phase equation. Finally, the particle and SH simulations both show that the doubly-periodic travelling wave solution is more stable for large levels of noise in the phase equation and when the initial velocity is negatively aligned with the phase gradient. 

\end{itemize}

\subsection{Validation of the hydrodynamic limit}\label{sec:validationsh}

In order to numerically validate the derivation of the Swarmalator Hydrodynamics model \eqref{eq:fl_rho}-\eqref{eq:fl_al}, we simulate the particle system \eqref{eq:ibm_pos}-\eqref{eq:ibm_phas} with a set of parameters chosen accordingly to the scaling conditions presented in Appendix \ref{sec_scaling_appendix} and under which the mean-field and hydrodynamic limits are taken. Then we confront statistical quantities measured at the particle level with their prediction given by the hydrodynamic model.

\subsubsection{Scaling}\label{sec:scaling_numerics}

Following the notations of Appendix \ref{sec_scaling_appendix}, we first fix the macroscopic space and time scales $x_0=1$ and $t_0=1$. Then we choose a large number of particles $N$ and a small radius of interaction $R$. In a spatially homogeneous setting, each particle interact in average with $N_\mathrm{neigh} = \pi R^2 N$ neighboring particles. As a rule of thumb, the mean-field regime is attained when a particle typically interacts with at least a few tens of other particles. We will therefore choose the parameters $N$ and $R$ such that $N_\mathrm{neigh} \sim 10^2$. 

Then, as explained in Appendix \ref{sec_scaling_appendix}, in order to take the hydrodynamic limit, the radius of interaction $R$ is the only scaling parameter from which all the other parameters of the particle simulations can be defined. Namely, we choose a linear potential for the phase attraction-repulsion force 
\[\omega(r) = \frac{x_0^{4}}{t_0 R^2}\tilde{\omega}\left(\frac{r}{R}\right),\] 
where $\tilde{\omega}(\tilde{r}) = \frac{3}{\pi}(1-\tilde{r})\mathbbm{1}_{\tilde{r}\leq 1}$ and $\tilde{r} = r/R$. Note that with $x_0=1$ and $t_0=1$ then it holds that $\int_{\mathbb{R}^2} \omega(|x|)\, dx = 1$ and for $y\in \mathbb{R}^2$, 
\begin{equation}\label{eq:gradomega}\nabla_x \,\omega(|y-x|) = C\frac{y-x}{|y-x|} \,\mathbbm{1}_{|y-x|\leq R},\end{equation}
with $C = \frac{x_0^4}{t_0 R^3}$. 
Moreover, we choose 
\[\zeta(r) = \eta(r) =  \frac{x_0^2}{R^2}\mathbbm{1}_{r\leq R}.\]
The other parameters are also chosen depending on $R$ as  
\[D =\frac{x_0}{t_0 R} \tilde{D},\quad \nu =\frac{x_0}{t_0 R} \tilde{\nu},\quad  D' =\frac{x_0}{t_0 R} \tilde{D}',\quad \nu' =\frac{x_0}{t_0 R} \tilde{\nu}',\]
so that $\tilde{D}, \tilde{\nu}, \tilde{D}', \tilde{\nu}'$ are dimensionless parameters. Note that the parameter $\gamma$ is already a dimensionless parameter.

\subsubsection{Setting of the experiment}

In order to check the behavior of the particle system as $N\to+\infty$ and $R\to0$, we use as a test case the doubly periodic travelling wave solution derived in Section \ref{subsec:explicit}. Given arbitrary alignment parameters $\tilde{D}, \tilde{\nu}, \tilde{D}', \tilde{\nu}'$, we initialize the particle system by drawing $N$ particles uniformly on the torus $(0,1)^2$ and the initial velocities with respect to the von Mises distribution $M_{u_0}$ with $u_0=(0,-1)^\mathrm{T}$. The phase of a particle at position $(x_1,x_2)$ is sampled from the von Mises distribution $N_{2\pi x_2}$ (corresponding to $p=0$ and $m=1$ in Proposition \ref{prop_per}). Then, we choose the intensity of the phase attraction-repulsion force such that $b'=\frac{1}{2\pi}(1+c_1)$, i.e. 
\[\gamma = -\frac{c_1+1}{2\pi c_1'(\frac{1}{k'}+c_2')} ,\]
so that the theoretical travelling wave speed is equal to 
\[\lambda = -2\pi c_1 + 4\pi^2 b' = 2\pi,\]
where we recall that $b'$ is given by \eqref{eq:express_b'}. We recall that with the potential \eqref{eq:gradomega}, a particle chases the particles which are slightly ahead of phase when $\gamma<0$ and flees them when $\gamma>0$. In the present case, with $\gamma<0$, $p=0$ and $m=1$, it means that the phase attraction-repulsion force is opposite to the mean self-propulsion velocity of the particles (see e.g. the term inside the divergence operator in \eqref{eq:fl_rho}) and $\lambda=2\pi$ means that the travelling-wave is moving upward and is 1-periodic. 

At the particle level, in order to observe a travelling wave and measure its speed, we need to define a suitable statistical indicator. First, we define the ``mass" of a particle with phase $\varphi$ as
\[m(\varphi) = 1 + \cos\varphi.\]
Then we consider the $x_2$-coordinate of the ``center of mass'' (in the torus) at time $t$ of the particle system, given by the formula:
\begin{equation}\label{eq:centermass}\overline{x}^N_2(t) = \frac{1}{2\pi}\mathrm{arg}\left(\frac{1}{N}\sum_{k=1}^N m(\varphi_k) e^{2i\pi X_{k,2}(t)}\right),\end{equation}
where $X_{k,2}$ is the $x_2$-coordinate of the position of particle $k$. The formal mean-field limit when $N\to+\infty$ leads to
\[\frac{1}{N}\sum_{k=1}^N m(\varphi_k) e^{2 i \pi X_{k,2}(t)} \underset{N\to+\infty}{\longrightarrow} \int_0^1 \int_0^{2\pi} (1 + \cos\varphi) \, e^{2i\pi x_2}\, N_{\alpha(t,x_2)}(\varphi) \,dx_2 \, d\varphi,\]
where we recall that $N_\alpha$ denotes the von Mises distribution \eqref{eq:VMphi} with the parameter $\alpha$ given by the macroscopic model  \eqref{eq:alp_per_TW}. Namely in this experiment $\alpha(t,x_2) = 2\pi x_2 - \lambda t + \alpha_0$ where we chose $\alpha_0 = \pi$. A direct computation shows that 
\begin{equation}\label{eq:predictioncentermass}\overline{x}^N_2(t) \to \frac{1}{2} +  t \,\,\, \mathrm{mod} \,\,\,1.\end{equation}

\subsubsection{Results}

We run the particle simulation with the parameters described above and we measure the quantity \eqref{eq:centermass}. 

\begin{figure}[ht!]
\centering
\subfloat[$N=10^4$ $(N_\mathrm{neigh}\sim 0.8)$]{\includegraphics[width=0.31\textwidth]{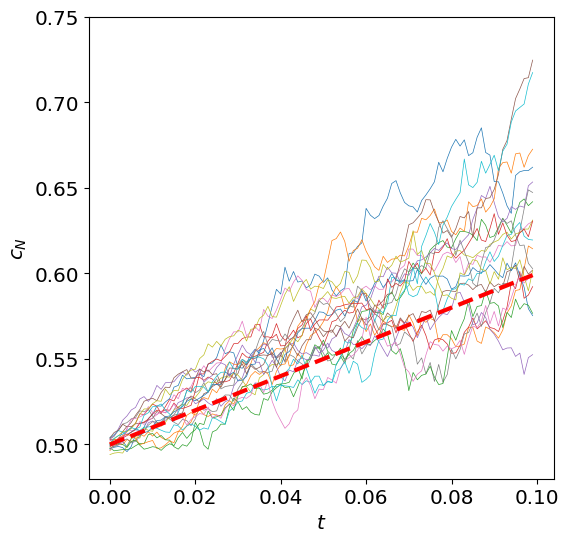}}\hspace{0.01\textwidth}
\subfloat[$N=123015$ $(N_\mathrm{neigh}\sim 10)$]{\includegraphics[width=0.31\textwidth]{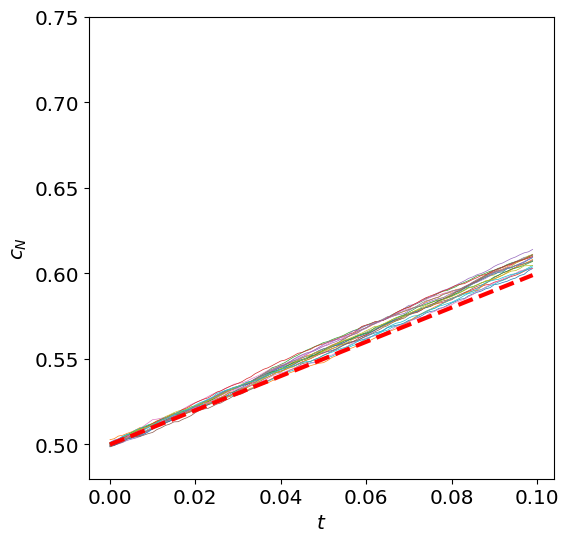}}\hspace{0.01\textwidth}
\subfloat[$N=1999999$ $(N_\mathrm{neigh}\sim 157)$]{\includegraphics[width=0.31\textwidth]{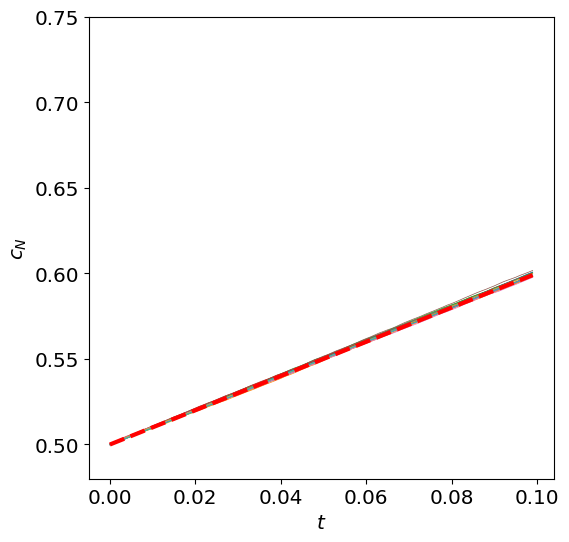}}
\caption{Twenty independent measurements of the $x_2$-coordinate of the center of mass of the particle system over 0.1 units of time for three different values of $N$. The thick dashed red line is the theoretical prediction derived from the macroscopic model. For each value of $N$, twenty independent experiments are displayed in colored plain thin lines. For each value of $N$, the number $N_\mathrm{neigh} = \pi R^2 N$ is the average number of neighbours in a ball of radius $R$. Parameters: $R=0.005$, $\tilde{\nu} = 3$, $\tilde{D} = 1$, $\tilde{\nu}'=5$, $\tilde{D}'=1$.  }
\label{fig:convergenceN_threeN}
\end{figure}

First, Figure \ref{fig:convergenceN_threeN} illustrates the behavior of the particle simulation when $N$ goes to infinity with a fixed small interaction radius. We measure the quantity $\overline{x}^N_2(t)$ for various values of $N$ and we compare it to the theoretical prediction \eqref{eq:predictioncentermass} when $N\to+\infty$. For small values of $N$, in a regime where the particles have only a few neighbors to interact with ($N_\mathrm{neigh}\sim1$), only a noisy behavior is observed although the center of mass is clearly biased to move in the expected direction. As $N$ grows and since $R$ is fixed, each particle interacts in average with a larger number of other particles. When the number of neighbours crosses $\sim 100$, the system can be considered as being in a mean-field regime and we observe a very good qualitative agreement with the macroscopic prediction \eqref{eq:predictioncentermass}. 

\begin{figure}[ht!]
\centering
\subfloat[Speed for various $N$]{\includegraphics[height=0.42\textwidth]{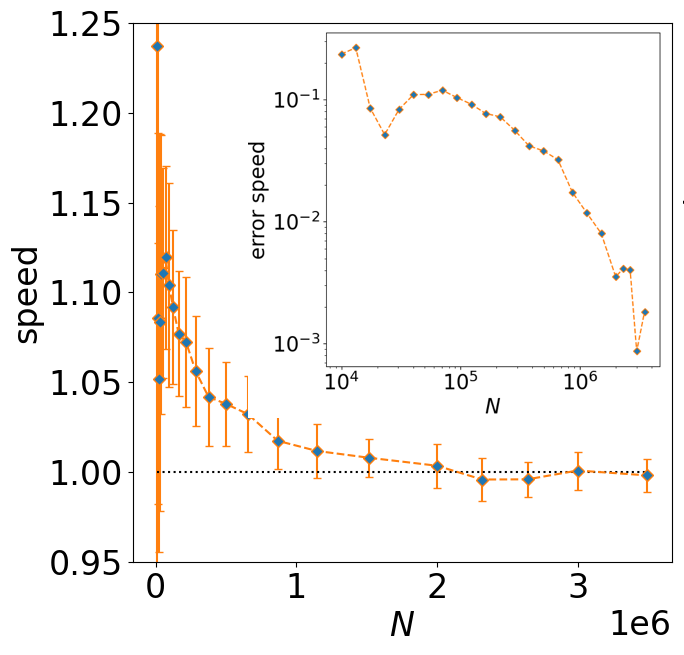}\label{subfig:Nconvergence}}\hspace{0.01\textwidth}
\subfloat[Standard deviation around the mean value]{\includegraphics[height=0.42\textwidth]{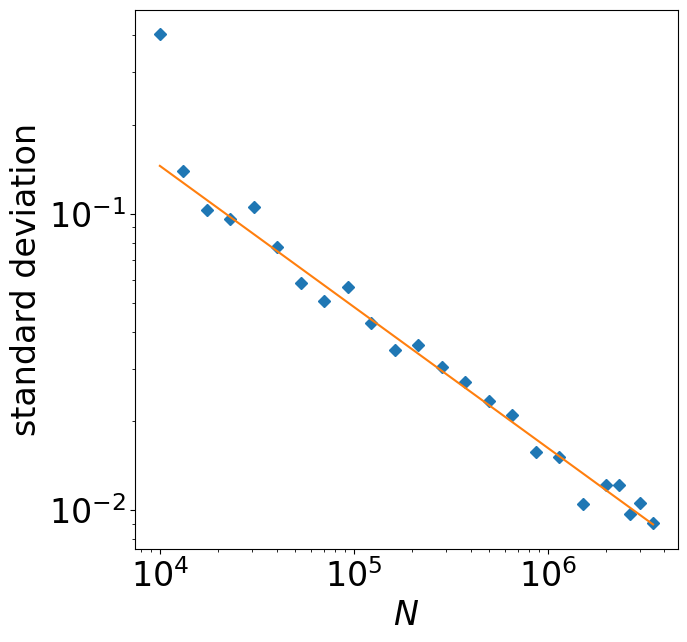}\label{subfig:convergenceNsd}}
\caption{Convergence of the particle scheme as $N\to+\infty$. (a) For different values of $N$, twenty independent experiments are run and for each of them the measured travelling-wave speed is computed as the slope coefficient of $\overline{x}_2^N(t)$ using a standard linear regression. The blue points correspond to the mean value over the twenty experiments for each value of $N$ and the error bar in orange shows the standard deviation. The theoretical speed is equal to 1 as indicated by the dotted black horizontal line. The inset shows the absolute value of the difference between the mean value of the measured speed averaged over the 20 simulations and the theoretical speed in log-scale. (b) The blue points show the standard deviation around the mean value of the measured travelling-wave speed for each value of $N$. The orange line is the regression line (excluding the first point) and has a slope coefficient equal to $0.48$ which is close to the value $0.5$ predicted by the central limit theorem. Parameters: $R=0.005$, $\tilde{\nu} = 3$, $\tilde{D} = 1$, $\tilde{\nu}'=5$, $\tilde{D}'=1$.}
\label{fig:convergenceN}
\end{figure}

Having shown a qualitative agreement between the measured and predicted behaviors, we then perform a quantitative analysis of the convergence of the particle scheme when $N\to+\infty$. Since the quantity $\overline{x}^N_2(t)$ qualitatively and theoretically converges as $N\to+\infty$ towards a straight line, we compute the slope of this line (using a standard linear regression method) and compare it to the predicted travelling wave speed, theoretically equal to 1. The results are shown in Figure \ref{fig:convergenceN}.

In Figure \ref{subfig:Nconvergence}, we observe an excellent quantitative agreement between the measured and predicted travelling wave speeds when $N$ is large. Namely, when $N>3\cdot10^6$, the absolute value of the difference between the two quantities is, in average, of order $10^{-3}$ with a standard deviation of the order $10^{-2}$. 
Moreover, Figure \ref{subfig:Nconvergence} gives an indication on the behavior of the particle scheme for small and moderate values of $N$. In particular, it should be noted that, when $N$ is small, the measured speed is actually larger than the predicted one. To understand this phenomenon, let us first point out that with the chosen parameters, there are two competing effects: the phase gradient produces a positive force in the $(0,1)^\mathrm{T}$ direction (i.e. pointing upward) while the self-propulsion velocity is in the direction $(0,-1)^\mathrm{T}$ (i.e. pointing downward). However there is a difference in the nature of these two forces. The self-propulsion velocity is subject to noise and different particles have independent noises. On the contrary, the force exerted by the phase gradient is just computed by taking an average over all (neighboring) particles and is thus less sensitive to the individual noises affecting the particles self-propulsion velocities and phases. Even when $N$ is small, we can infer that the phase attraction-repulsion force has a smaller variance than the self-propulsion velocity. Since the norm of the self-propulsion velocity is constant equal to 1, a larger variance means that in average, the self-propulsion velocity opposite to the phase attraction-repulsion force has a norm smaller than 1.  The phase attraction-repulsion force is thus winning over the two, which can explain the positive bias observed. Note also that when the self-propulsion velocity and the phase attraction-repulsion forces are positively aligned, we have observed (not shown here) that the measured speed is slower than the one predicted, which is expected since due to the noise, the two forces are not perfectly aligned and the norm of their sum is smaller than the theoretical value obtained when they are perfectly aligned. 

A second important observation that we have made, but which is not directly shown in Figure \ref{fig:convergenceN}, is that for larger values of $R$ ($R\sim 0.01$) and for very large values of $N$ ($N>10^6$), there is a perceptible negative bias in the measured travelling-wave speed, meaning that it is slower than expected (although it is still larger for $N$ small). For $R=0.01$ and for all values of $N>10^6$, the measured speed is about $0.97$ instead of $1$. Since the theoretical speed is proportional to $c_1(k')$ this can be explained by the fact that when taking the hydrodynamic limit (see Section \ref{sec_eps_to_0_proof}), this latter quantity appears as the limit when $R\to0$ of  
\[|L^R| = \Big|\int_{\mathbb{R}^n\times[0,2\pi]} \eta^{R}(|y-x|) e^{i\psi}N_{\alpha(y,t)}(\psi)\,dy\,d\psi\Big| \\
= c_1(k') \Big|\int_{\mathbb{R}^n} \eta^{R}(|y-x|) e^{i\alpha(y,t)}\,dy\Big|\]
under the assumption that the particle distribution is equal to the equilibrium distribution $f(y,w,\psi,t)=\rho(y,t) M_{u(y,t)}(w)N_{\alpha(y,t)}(\psi)$ with $\rho(y,t)\equiv1$ and $u,\alpha$ are given by the macroscopic model. The kernel $\eta^{R} \equiv \eta^{R}(|x|) := R^{-n}\eta(|x|/R)$ is non negative, has integral one and tends to a Dirac delta at 0 when $R\to0$. Consequently, since $\eta^R$ has integral one, by the triangle inequality, the last term on the right-hand side is smaller than $c_1(k')$ for any $R>0$. However, this effect is not perceptible for the value $R= 0.005$ as shown on Figure~\ref{fig:convergenceN}. 

Finally, Figure \ref{subfig:convergenceNsd} shows that the standard deviation around the average measured travelling-wave speed decreases as an inverse power law as $N$ is increasing. The exponent of this inverse power law is computed using a standard linear regression and is equal to approximately $0.48$. This experimental value has a simple theoretical explanation. In the mean-field limit, the particles can be shown to behave as independent and identically distributed random variables (thanks to the so-called propagation of chaos property). Consequently, by the central limit theorem, the standard deviation around the average~\eqref{eq:centermass} should behave like $N^{-1/2}$, which is consistent with the observations.

\subsection{Segregation phenomena}\label{sec:segregationparticles}

In a macroscopic regime, the doubly periodic travelling wave observed in the particle simulations is stable during a certain time but may eventually transition towards another state. This destabilization phenomenon may be understood as a natural consequence of the numerical noise induced by the scheme (which combines both the finite size effect and the inherent stochasticity of the particles). It can also be seen as a practical way to study the stability of the different solutions of the macroscopic system \eqref{eq:fl_rho}-\eqref{eq:fl_al}. In this section we investigate how the particle system departs from the doubly periodic travelling wave configuration depending on the initial direction of the velocity and on the value of the parameter $k'$. This will give a numerical confirmation of the results of Lemma~\ref{lem:hyperbolic}. In addition, we show that for low levels of noise in the phase equation (i.e. for large values of $k'$), the particle system transitions towards configurations characterized by a strong segregation between populations of particles with constant phase separated by thin boundaries of low density.

\subsubsection{Parameters and setting of the experiments}\label{sec:parameters_particles}

All the simulations in this section take place in an intermediate regime where the interaction radius is sizeable compared to the dimension of the domain. Consequently, this setting is slightly farther from the hydrodynamic limit (which requires $R \to 0$) than that of Section \ref{sec:validationsh} and we note a small but perceivable departure of the observed travelling-wave speed from the value predicted by the hydrodynamic model. We choose this setting because simulations are quicker to run without qualitatively altering the results. All the simulations presented below use the following parameters: 
\[N=10^6,\quad R=0.01,\quad \tilde{\nu}=5,\quad\tilde{D}=1,\quad\gamma=0.2.\]
The particles' positions are sampled independently uniformly in the torus. The initial phase of a particle at position $(x_1,x_2)$ is sampled from the von Mises distribution $N_{2\pi x_2}$. With this choice, the phase attraction-repulsion force drags the particles downward (i.e. in the direction $(0,-1)^\mathrm{T}$). Given an initial velocity $v_0\in \mathbb{S}^{2}$, the particles' velocities are sampled according to the von Mises distribution $M_{v_0}$.

In order to test the influence of the phase attraction-repulsion interaction, we vary the parameters $\tilde{\nu}'$ and $\tilde{D}'$ and the choice of the initial velocity $v_0$.  First, we consider four choices for the parameters $\tilde{\nu}'$ and $\tilde{D}'$ which correspond to four different levels of noise in the phase equation. 

\begin{enumerate} 
\item Very-low noise: $k'\equiv k'_\mathrm{hi+} = 10000$ ($\tilde{\nu}'=10$, $\tilde{D}'=0.001$).
\item Low noise: $k'\equiv k'_\mathrm{hi} = 200$ ($\tilde{\nu}'=10$, $\tilde{D}'=0.05$). 
\item Medium noise: $k'\equiv k'_\mathrm{med} = 10$ ($\tilde{\nu}'=10$, $\tilde{D}'=1$).
\item Large noise: $k'\equiv k'_\mathrm{lo} = 3$ ($\tilde{\nu}'=3$, $\tilde{D}'=1$).
\end{enumerate} 

For smaller values of $k'$, only a noisy behavior is observed as it can be expected. Then, for each value of $k'$, we will consider three different choices for the initial velocity~$v_0$.
\begin{enumerate}
\item When $v_0 = (0,-1)^\mathrm{T}$ we say that the velocity and the phase attraction repulsion force are Positively Aligned (PA). 
\item When $v_0 = (0,1)^\mathrm{T}$ we say that the velocity and the phase attraction repulsion force are Negatively Aligned (NA).
\item When $v_0 = (1,0)^\mathrm{T}$ we say that the velocity and the phase attraction repulsion force are Orthogonal (OT). 
\end{enumerate}

As a control system, for each value of $k'$, we also consider the system where the positions, velocities and phases of the particles are initially sampled independently uniformly respectively in $(0,1)^2$, $\mathbb{S}^1$ and $[0,2\pi]$. This configuration is referred as UF in the following.

\subsubsection{Results}\label{sec:results_particles}

The results for all the simulations are shown in Videos \ref{video:verylownoise_NA} to \ref{video:largenoise_UF3} in Appendix \ref{sec:listvideos_particles}. The main observations are summarized below. 

\begin{figure}[ht!]
\centering
\subfloat[Density $k'_\mathrm{hi+}$\label{subfig:finalNA_density_verylownoise}]{\includegraphics[width=0.25\textwidth]{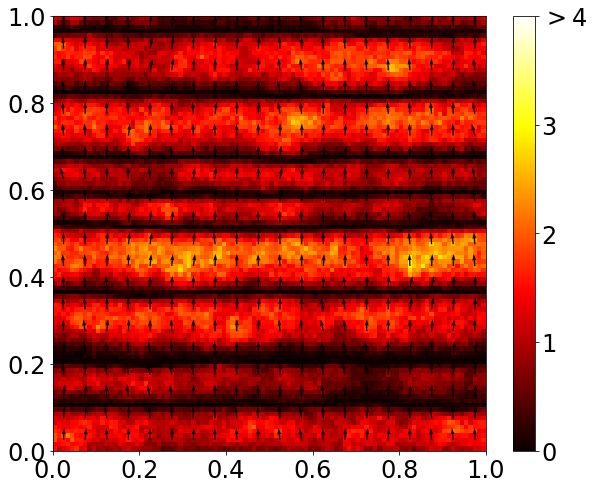}}\hspace{-0.02\textwidth}
\subfloat[Phase $k'_\mathrm{hi+}$\label{subfig:finalNA_phase_verylownoise}]{\includegraphics[width=0.25\textwidth]{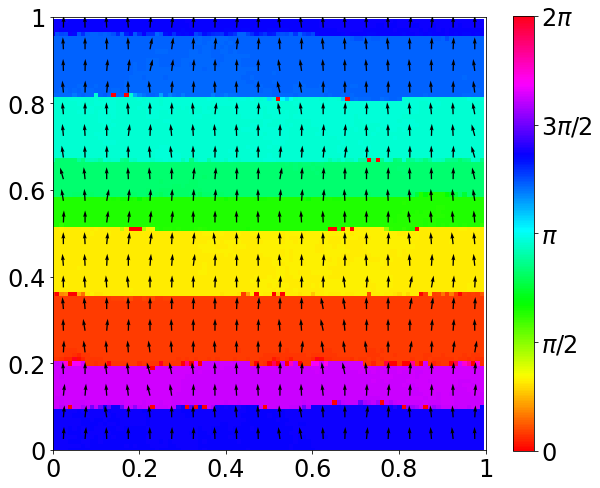}}\hspace{0.0\textwidth}
\subfloat[Density $k'_\mathrm{hi}$\label{subfig:finalNA_density_lownoise}]{\includegraphics[width=0.25\textwidth]{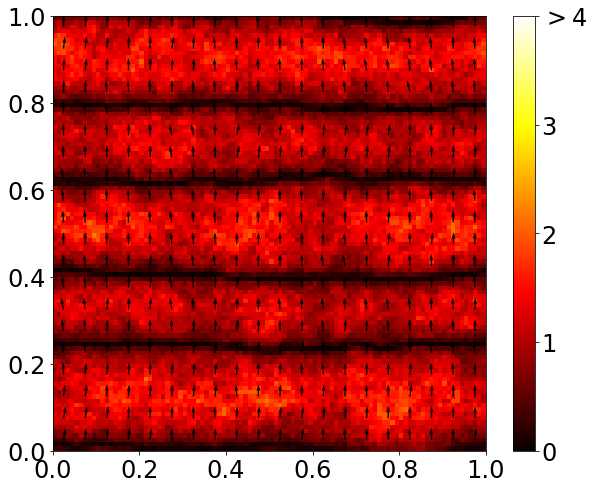}}\hspace{-0.02\textwidth}
\subfloat[Phase $k'_\mathrm{hi}$\label{subfig:finalNA_phase_lownoise}]{\includegraphics[width=0.25\textwidth]{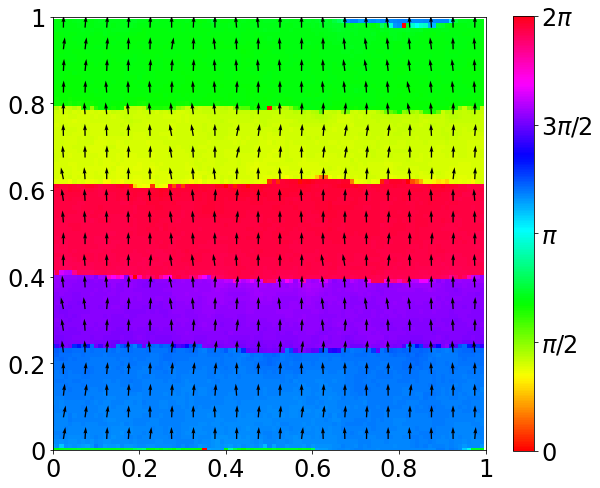}}\\
\subfloat[Density $k'_\mathrm{med}$\label{subfig:finalNA_density_mediumnoise}]{\includegraphics[width=0.25\textwidth]{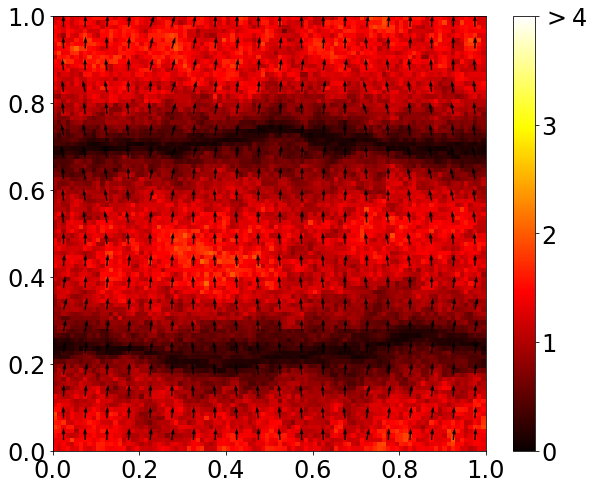}}\hspace{-0.02\textwidth}
\subfloat[Phase $k'_\mathrm{med}$\label{subfig:finalNA_phase_mediumnoise}]{\includegraphics[width=0.25\textwidth]{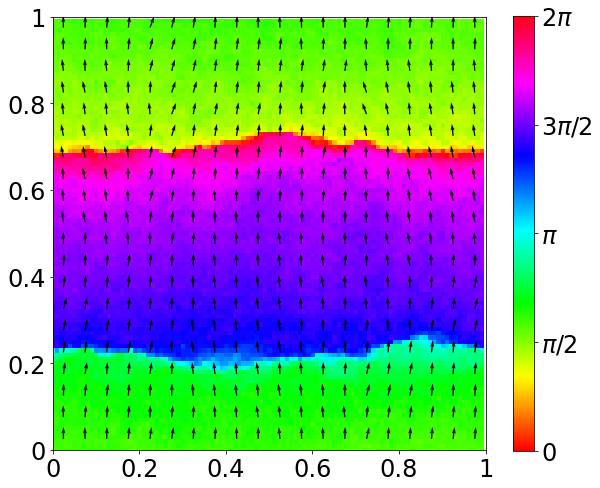}}\hspace{0.0\textwidth}
\subfloat[Density $k'_\mathrm{lo}$\label{subfig:finalNA_density_largenoise}]{\includegraphics[width=0.25\textwidth]{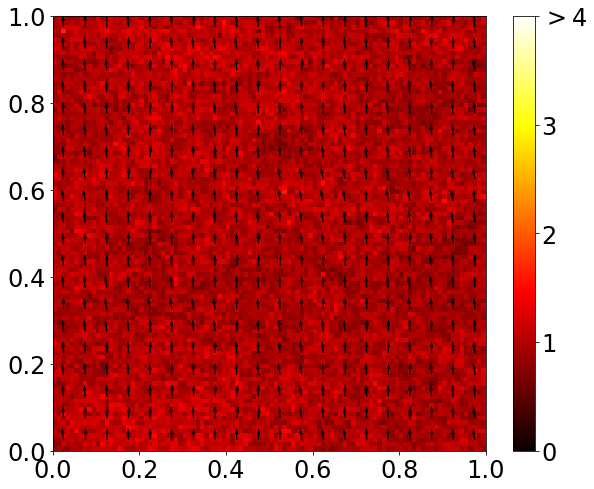}}\hspace{-0.02\textwidth}
\subfloat[Phase $k'_\mathrm{lo}$\label{subfig:finalNA_phase_largenoise}]{\includegraphics[width=0.25\textwidth]{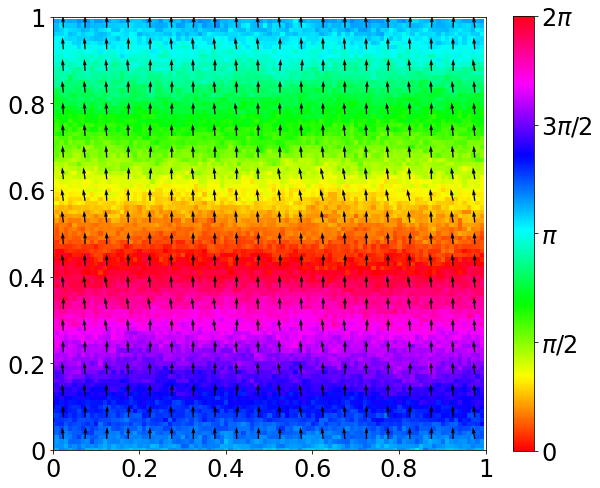}}\hspace{0.0\textwidth}
\caption{Final state after 40 units of time starting from a NA configuration for four values of $k'$. For each value of $k'$, the domain is discretized into a uniform grid with $10^4$ cells of size $0.01$. (a),(c),(e),(g) The density of particles is obtained by counting the proportion of particles in each cell. (b),(d),(f),(h) The phase in each cell is the average phase of the particles in this cell. It is arbitrarily set to 0 when the cell is empty. The corresponding videos can be found in the supplementary material: (a)-(b) Video \ref{video:verylownoise_NA} (c)-(d) Video \ref{video:lownoise_NA} (e)-(f) Video \ref{video:mediumnoise_NA} (g)-(h) Video \ref{video:largenoise_NA}. Parameters: $N=10^6$, $R=0.01$, $\tilde{\nu}=5$, $\tilde{D}=1$, $\gamma=0.2$.}
\label{fig:finalNA}
\end{figure}

\begin{enumerate} 
\item In the very-low noise case ($k'_\mathrm{hi+} = 10000$) and for any choice of the direction of the initial velocity~$v_0$, the particles immediately segregate into small regions of equal phase separated by very-low density thin boundaries. 
\begin{itemize}
\item Starting from a NA configuration, after a transition period during which many small constant-phase regions are forming, merging or expanding, the system stabilizes into a configuration where a succession of constant-phase regions with a band-like shape move in the direction $v_0 = (0,1)^{\mathrm{T}}$ (see Figs. \ref{subfig:finalNA_density_verylownoise} and~\ref{subfig:finalNA_phase_verylownoise} and Video \ref{video:verylownoise_NA}). The global direction of motion of the system $v_0$ remains constant throughout the simulation. Note that since there is no phase gradient within each region, the particles in each region are not subject to the phase attraction-repulsion force. 
\item Starting from a PA configuration, the situation is initially analogous to the NA case except that the particles are moving in the direction $(0,-1)^\mathrm{T}$. However, as times grows, unlike the NA case, the system does not reach a stable configuration within the 40 units of time of the simulation. Although we still observe the formation of band-like regions of constant phase moving at a constant speed, these regions do not have a constant shape and are perpetually subject to destruction and recombination (see Video \ref{video:verylownoise_PA}).
\item Starting from a OT configuration, unlike the NA and PA cases, the global velocity of the system does not remain constant. During the first units of time and simultaneously to the formation and recombination of constant-phase regions, the global velocity of the particles (and direction of motion) transitions from $(1,0)^\mathrm{T}$ to $(0,1)^\mathrm{T}$. After 40 units of time, the systems reaches a configuration similar to the one starting from a NA configuration but no bands stable over a long time can be clearly identified (see Video \ref{video:verylownoise_OT}). 
\end{itemize}

\item In the low noise case ($k'_\mathrm{hi} = 200$), the dynamics is similar to the very-low noise case with a few exceptions. First, the final configuration starting from the NA and OT configurations is still composed of band-like constant-phase regions moving in the direction $(0,1)^\mathrm{T}$ but the size of the regions is increased (see Figs. \ref{subfig:finalNA_density_lownoise} and \ref{subfig:finalNA_phase_lownoise} and Videos \ref{video:lownoise_NA} and \ref{video:lownoise_OT}) and the final state is more stable starting from a OT configuration. We also note that the phase in each band is not constant over time but slowly evolve. Secondly, starting from the PA configuration, the analogous configuration is much more unstable and the system finally ends up in a flocking phase with all the phases equal and an arbitrary direction of motion (see Video \ref{video:lownoise_PA}). 

\item In the medium noise case ($k'_\mathrm{med} = 10$). For all choices of $v_0$, the initial structure is preserved during a longer time (approximately 5 units of time). Then we observe the formation of thin elongated low-density regions. Unlike the previous cases, they do not clearly delimitate segregation regions and we de not observe the formation of constant-phase clusters. As time grows, the system finally ends up in a situation similar to the previous cases where band-like structures separated by thin low-density boundaries are moving along the $x_2$-axis. In the OT case, the velocity transitions from $(1,0)^\mathrm{T}$ to $(0,1)^\mathrm{T}$.  We note that the final outcome of the system is still characterized by a segregation phenomenon into band-like structures but unlike the cases where $k'$ is larger, these structures are larger and although they are clearly separated by thin low-density boundaries, there is an inner gradient of phase in each band and the phase is not preserved over time in each band as it was in the previous cases (see Figs. \ref{subfig:finalNA_density_mediumnoise} and \ref{subfig:finalNA_phase_mediumnoise} and Videos \ref{video:mediumnoise_NA} to \ref{video:mediumnoise_OT}). 

\item In the large noise case ($k'_\mathrm{lo} = 3$) and for all choices of $v_0$ the initial doubly periodic travelling wave is stable throughout the simulation (see Figs. \ref{subfig:finalNA_density_largenoise} and \ref{subfig:finalNA_phase_largenoise} and Videos~\ref{video:largenoise_NA} to~\ref{video:largenoise_OT}).
\end{enumerate} 

Finally, starting from a uniformly disordered UF state with random velocities and phases, for very-low to medium levels of noise ($k'=k'_\mathrm{hi+}$ to $k'=k'_\mathrm{med}$), the system always ends up in a flocking phase with all the phases and velocities equal. Similarly to what is observed starting from a NA, PA or OT configuration, we initially observe clusterization and phase-segregation phenomena which are all the more important that $k'$ is large (see Videos~\ref{video:verylownoise_UF}, \ref{video:lownoise_UF} and \ref{video:mediumnoise_UF}). The situation is more complex in the large noise scenario ($k'=k'_\mathrm{lo}$) as independent experiments with the same parameters can lead to different outcomes. In addition to the flocking phase (see Video \ref{video:largenoise_UF1}), we have also observed cases where the system ends up in various topologically non-trivial states. These states are still characterized by a continuous gradient of phase which is topologically constrained by the periodicity of the domain. However, unlike the doubly-periodic travelling wave solution, a wide range of complex structures can emerge characterized by non constant densities and velocities (see Videos \ref{video:largenoise_UF2} and \ref{video:largenoise_UF3}).

\subsubsection{Discussion}\label{sec:discussion_particles}

Regardless of the level of noise $k'$, all the experiments confirm that the NA configuration is the most stable one. Even though the initial doubly periodic travelling wave is not preserved, the only stable final configuration correspond to a case where the velocity is pointing in the direction $(0,1)^\mathrm{T}$ and the phase is constant along the $x_1$-axis and piecewise constant and increasing along the $x_2$ axis. It can be understood as a more general version of the NA case where in the latter the phase is linear and increasing along the $x_2$-axis. As predicted by Lemma~\ref{lem:hyperbolic}, the OT configuration is the most unstable one and except for low values of~$k'$, it never persists and quickly transitions towards a more stable NA case. Note that Lemma~\ref{lem:hyperbolic} does not make a distinction between the stability of the NA and PA cases. Numerically the former is the most stable. 

\begin{figure}[ht!]
\centering
\includegraphics[width = 0.4\textwidth]{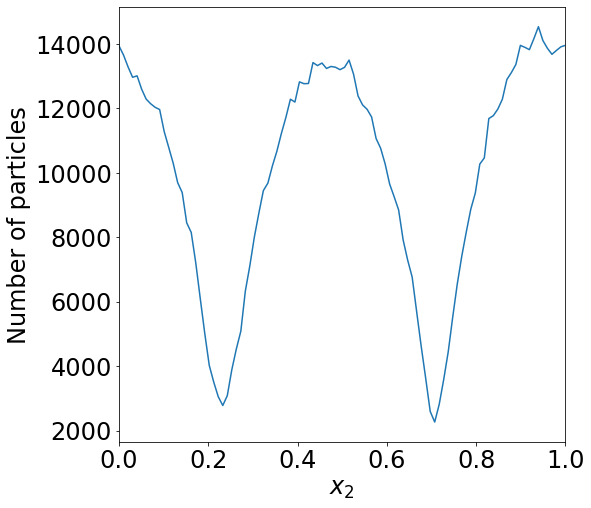}
\caption{Number of particles in the strips $[0,1]\times[\frac{k}{100},\frac{k+1}{100}]$ for $k\in\{0,\ldots,99\}$ after 40 units of time starting from the NA configuration with $k'=k'_\mathrm{med}$. Same parameters as Fig. \ref{subfig:finalNA_density_mediumnoise}.}
\label{fig:bandshape}
\end{figure}

An important observation is the ability of the particle model to produce segregation. We were not able to predict this behavior using the SH model \eqref{eq:fl_rho}-\eqref{eq:fl_al}, although we cannot exclude that it also corresponds to (possibly singular) solutions that remain to be identified. The simulation of the macroscopic model presented in the next section may also support this idea. It is also worth mentioning that the emergence of band-like structures is a well-known phenomenon in the Vicsek model \cite{chate2008collective}. However, the phenomenon observed is quite different on many aspects. First, the emergence of bands in the Vicsek model can be observed only in a very specific range of parameters and in very-low density regimes. The conditions of the presented simulations are much more general. Moreover, the bands observed here have a very different profile from the ones observed in the Vicsek model. In the Vicsek model, bands have an asymmetric profile characterized by a sharp front edge and an exponentially decaying tail whereas the bands observed here are larger and have a symmetric profile (see Fig.~\ref{fig:bandshape}). 

Regarding the doubly periodic travelling wave solution that we theoretically identified in Section \ref{subsec:explicit}, the particle simulations tend to indicate that the noise in the phase equation has a stabilizing effect. In the large noise case and for any choice of the initial velocity, we indeed do not observe any segregation but rather a stable doubly periodic travelling wave as expected, with a moving speed close to the theoretical speed predicted by the macroscopic model. 

Finally, these experiments show the robustness of topological states. Even if for lower levels of noise in the phase equation, the theoretical doubly periodic travelling wave does not persist, the system still transitions towards a state characterized by a nontrivial topology. Such scenario is never observed starting from the topologically trivial UF state. The topological robustness increases with the level of noise in the phase equation and in such regime, more general topological states can even emerge form the UF configuration, which shows that the flocking state is not the only global attractor of the system. The doubly periodic travelling wave may be seen as the simplest topological state and further investigations are needed to determine whether the topological states observed at the particle level also correspond to solutions of the hydrodynamic model.

\subsection{Simulation of the hydrodynamic NSH and SH models}\label{sec:simulationsh}

Following the methodology introduced in \cite{motsch2011numerical} for the SOH model, a finite volume discretization of the SH model is also presented. The details of the numerical method are described in Appendix \ref{sec:fv_appendix}. The results for all the simulations are shown in Videos \ref{video:lownoise_NA_SH} to~\ref{video:lownoise_OT_SH} in Appendix \ref{sec:listvideos_sh}. The main observations are summarized below.

\subsubsection{NSH case}

Simulating the NSH system is computationally easier, in particular because it does not require to discretize the term $\nabla_x\cdot(\rho\nabla_x\rho)$ in the phase equation. Starting from a doubly periodic initial condition and regardless of the direction of the initial velocity, we observe a very stable travelling wave moving in the direction of the phase gradient at the speed predicted by the model. However, starting from a slightly perturbed initial condition, we immediately observe the formation of shocks with strong local variations of the density and of the phase. Due to the CFL condition which becomes too stringent in this situation, we were not able to continue the simulation further. The same difficulty happens in the very-low noise case $k'_\mathrm{hi+}$. Note however that this observation is consistent with the particle simulations which also show this behavior.

\subsubsection{SH case}\label{sec:results_fv}

By adding the terms corresponding to a nonzero $k'\ne0$ (i.e. by adding noise in the phase equation), the situation becomes more stable even starting from perturbed initial conditions. The behavior is not immediately comparable to the behavior of the particle simulations, but shares some of the main features. For the experiments discussed below, in each scenario, the initial state is perturbed by adding a small random uniform noise for the density, phase and velocity independently for each cell. For each cell, we add to the theoretical density (equal to 1) a uniform random variable in $[-0.25,0.25]$. For the two angles which define the phase and velocity we add to the theoretical value a uniform random variable in the interval $[-0.75,0.75]$ (in radians). 

\begin{itemize}
\item In the low noise case $k'_\mathrm{hi}=200$, starting from a perturbed OT state we observe the formation of thin low-density regions very reminiscent of the ones observed in the particle simulations and a transition towards a more stable state close to a NA configuration (see Video \ref{video:lownoise_OT_SH}). Unlike the particle simulations, such behavior is not observed starting from either perturbed NA or PA states. In the PA case, even though the density does not stay uniform and we can observe the formation of thin regions of lower density, they quickly fade away and do not degenerate as in the OT case or in the particle simulations (see Video \ref{video:lownoise_PA_SH}). In the NA case, the initially perturbed density, phase and velocity are quickly restored to their theoretical un-perturbed state (see Video \ref{video:lownoise_NA_SH}), which is another numerical confirmation of the increased stability of this state. In all cases, the simulation reaches a doubly periodic travelling wave solution with a travelling wave speed correctly predicted by the model \eqref{eq:lambda_per_TW}.
\item In the medium noise and large noise cases $k'_\mathrm{med}=10$ and $k'_\mathrm{lo}=3$, regardless of the initial condition (OT, NA or PA), unlike the particle simulations, we do not observe the formation of inhomogeneities but rather a stable doubly periodic travelling wave solution (see Videos \ref{video:mediumnoise_NA_SH} to \ref{video:largenoise_OT_SH}). This behavior is similar to the one observed in the particle simulations in the large noise scenario and confirm the stabilizing effect of the noise in the phase equation. 
\end{itemize}

The simulations of the SH and particle models agree well for short times or when the phase-noise level is large enough. At long time scales when the phase-noise is small, the doubly-periodic solutions are more unstable with the particle simulations than with the SH model. In the particle simulations, we observe the formation of regions of constant phase separated by thin low density regions. By contrast, simulations of the hydrodynamic model always maintain continuous gradients in phase.

\setcounter{equation}{0}
\section{Conclusion and perspectives}
\label{sec:conclusion}

In this paper, we have presented a new swarmalator model without force reciprocity and derived its hydrodynamic limit. We have studied the hydrodynamic model in the limit of small phase noise and determined its hyperbolicity regime. Then, we have derived a class of explicit doubly-periodic travelling-wave solutions in two spatial dimensions. These solutions have non-trivial topology quantified by the index of the phase vector over a period in either dimension. Solutions with index values larger than one are possible. Then numerical simulations of these doubly-periodic travelling-wave solutions with both the particle and hydrodynamic models have been presented. They confirm that the hydrodynamic model is an accurate approximation of the particle one for short time or large phase noise. They also provide a validation of the hyperbolicity result. However, for long times and small values of the phase noise, the two models differ but both give rise to topological solutions. In forthcoming papers \cite{degond2022topological_bis, degond2022topological}, we will pursue the investigation of topological states in this swarmalator model by deriving and studying classes of travelling-wave solutions in other geometries such as strips or annulae. 

Further studies can be envisioned. For instance, numerical simulations suggest that in some parameter ranges, periodic-in-time solutions are generated. Their mathematical investigation is still open. Another direction is to explore other phase spaces that would generate solutions with more complex topologies such as higher order homotopy groups of spheres.

\newpage

\begin{center}
\bf\LARGE Appendices
\end{center}

\setcounter{equation}{0}
\section{Particle and kinetic models: additional material}
\label{sec_part_kin_add}

\subsection{Overdamped limit for potential $W$}
\label{subsec:overdamped}

System \eqref{eq:ibm_pos}-\eqref{eq:ibm_phas} is the limit as $\varepsilon \to 0$ of the following system
\begin{eqnarray}
\frac{d X_k}{dt} &=& c_0 v_k + w_k , \label{eq:ibm_pos_od} \\
d v_k &=& \frac{1}{\varepsilon} \Big\{ \varepsilon \Big( P_{v_k^\bot} \circ \big[ \big( \nu \, \bar v_k - \nabla_x V(X_k(t)) \big) \, dt + \sqrt{2D} \, dB_t^k \big] \Big) \Big\}, \label{eq:ibm_vel_od} \\
\frac{dw_k}{dt} &=& - \frac{1}{\varepsilon} \Big( w_k + \gamma \nabla_x W(X_k, \varphi_k) \Big), \label{eq:ibm_w_od} 
\end{eqnarray}
complemented with \eqref{eq:ibm_phas} (here, the parameter $\varepsilon$ has a different meaning compared to the next section). This can be interpreted as follows. Forgetting $\varepsilon$ for the time being, from~\eqref{eq:ibm_pos_od}, we see that the particle velocity is decomposed in two terms: one $c_0 v_k$ stemming from self-propulsion which has constant norm, and a complementary one $w_k$. The total force acting on the particle is thus the sum of a component acting on the self-propulsion velocity given by \eqref{eq:ibm_vel_od} and of one acting on its complement given by \eqref{eq:ibm_w_od}. The first component is unchanged from the original system (see \eqref{eq:ibm_vel}) while the second one is a relaxation force. It describes the competition between the external force $\nabla_x W$ and a friction force which hypothetically results from the surrounding medium (note that a similar friction could also be included in \eqref{eq:ibm_vel_od} but would vanish anyway because $P_{v_k^\bot} v_k =~0$). Now, $1/\varepsilon$ in factor of \eqref{eq:ibm_vel_od}, \eqref{eq:ibm_w_od} is the friction coefficient and is very large. We see that, in the limit $\varepsilon \to 0$, we recover \eqref{eq:ibm_pos}-\eqref{eq:ibm_phas} under the condition that the forces involved in \eqref{eq:ibm_vel_od} are of order $\varepsilon$ (hence the multiplication by $\varepsilon$ of all the terms involved). So, System \eqref{eq:ibm_pos}-\eqref{eq:ibm_phas} is obtained as the overdamped limit of the unrelaxed system under the assumption that the alignment force, the noise and the exterior potential are very small, of the same order as the inverse of the friction coefficient.

\subsection{Scaling of the kinetic model}
\label{sec_scaling_appendix}

We first non-dimensionalize the kinetic model \eqref{eq:kin}. We let $x_0$, $t_0$ be space and time units which we relate to each other by $x_0 = c_0 t_0$. we note that $v$, $\varphi$, $\bar v$, $\bar \varphi$ are already dimensionless. We introduce the change of variables $\tilde x = x/x_0$, $\tilde t = t/t_0$ and functions $\tilde f (\tilde x, v, \varphi, \tilde t) = x_0^n \, f (x_0 \tilde x, v, \varphi, t_0 \tilde t)$, $\tilde V(\tilde x) = t_0 x_0^{-1} V(x_0 \tilde x)$, $\tilde W(\tilde x, \varphi, \tilde t) = t_0 x_0^{-2} W(x_0 \tilde x, \varphi, t_0 \tilde t)$. We also assume that there exists $R>0$ and functions $\tilde \omega$, $\tilde \zeta$ and $\tilde \eta$ such that, for all $r \in [0,\infty)$:  
$$ \omega (r) = \frac{x_0^{n+2}}{t_0 R^n} \, \tilde \omega \Big( \frac{r}{R} \Big), \quad \zeta (r) = \frac{x_0^n}{R^n} \,  \tilde \zeta \Big( \frac{r}{R} \Big), \quad \eta (r) = \frac{x_0^n}{R^n} \, \tilde \eta \Big( \frac{r}{R} \Big). $$
We define dimensionless constants
$$ \bar D = D t_0, \quad \bar D' = D' \, t_0, \quad k = \frac{\nu}{D}, \quad k' = \frac{\nu'}{D'}, \quad \bar R = \frac{R}{x_0}. $$
In these new variables, the kinetic model reads (after dropping the tildes and bars for simplicity):
\begin{eqnarray}
&&\hspace{-1cm}
\partial_t f + \nabla_x \cdot \big[ \big( v - \gamma \nabla_x W_f(x,\varphi) \big) f \big] - \nabla_v \cdot \big[ P_{v^\bot} \nabla_x V(x) \, f \big] \nonumber \\
&&\hspace{1cm}
=  D \, \nabla_v \cdot \big[ - k P_{v^\bot} \bar v_f  \, f + \nabla_v f  \big] + D' \, \partial_\varphi \big[ - k' \sin( \bar \varphi_f - \varphi) f + \partial_\varphi f \big], \label{eq:kin_sc} \\
&&\hspace{-1cm}
W_f(x,\varphi,t) = \int_{{\mathbb R}^n \times {\mathbb S}^{n-1} \times  [0,2\pi]} \frac{1}{R^n} \omega\Big( \frac{|y-x|}{R} \Big) \, \sin(\psi - \varphi) \, f(y,w,\psi,t) \, dy \, dw \, d\psi, 
\label{eq:kin_Wf_sc} \\
&&\hspace{-1cm}
J_f(x,t) = \int_{{\mathbb R}^n \times {\mathbb S}^{n-1} \times  [0,2\pi]} \frac{1}{R^n} \zeta\Big( \frac{|y-x|}{R} \Big)  \, w \, f(y,w,\psi,t) \, dy \, dw \, d\psi. \label{eq:kin_Jf_sc} \\
&&\hspace{-1cm}
L_f(x,t) = \int_{{\mathbb R}^n \times {\mathbb S}^{n-1} \times  [0,2\pi]} \frac{1}{R^n}  \eta \Big( \frac{|y-x|}{R} \Big)  \, e^{i \psi} \, f(y,w,\psi,t) \, dy \, dw \, d\psi, \label{eq:kin_Lf_sc}  \\
&&\hspace{-1cm}
\bar v_f(x,t) =\Big( \frac{J_f}{|J_f|} \Big)(x,t), \qquad e^{i \bar \varphi_f} (x,t) = \Big( \frac{L_f}{|L_f|} \Big)(x,t) . \label{eq:kin_barv_barphi_sc}
\end{eqnarray}

We now make the following scaling assumptions: 
$$ R = \varepsilon \to 0, \quad D = {\mathcal O}\Big( \frac{1}{\varepsilon} \Big), \quad D' = {\mathcal O}\Big( \frac{1}{\varepsilon} \Big), \quad k = {\mathcal O}(1), \quad k' = {\mathcal O}(1), \quad \gamma = {\mathcal O}(1). $$
Thus, introducting  $\tilde D$ and $\tilde D'$ such that $D = \tilde D/\varepsilon$ and $D' = \tilde D'/\varepsilon$, we may assume that $\tilde D$ and $\tilde D'$ are constants. After this scaling, the problem is written (again dropping the tildes for simplicity): 
\begin{eqnarray}
&&\hspace{-1cm}
\partial_t f^\varepsilon + \nabla_x \cdot \big[ \big( v - \gamma \nabla_x W^\varepsilon_{f^\varepsilon}(x,\varphi) \big) f^\varepsilon \big] - \nabla_v \cdot \big[ P_{v^\bot} \nabla_x V(x)  \, f^\varepsilon \big] \nonumber \\
&&\hspace{-0.2cm}
=  \frac{1}{\varepsilon} \Big\{ D \, \nabla_v \cdot \big[ - k P_{v^\bot} \bar v^\varepsilon_{f^\varepsilon}  \, f^\varepsilon + \nabla_v f^\varepsilon  \big] + D' \, \partial_\varphi \big[ - k' \sin( \bar \varphi^\varepsilon_{f^\varepsilon} - \varphi) f^\varepsilon + \partial_\varphi f^\varepsilon \big] \Big\}, \label{eq:kin_eps} \\
&&\hspace{-1cm}
W^\varepsilon_f(x,\varphi,t) = \int_{{\mathbb R}^n \times {\mathbb S}^{n-1} \times  [0,2\pi]} \frac{1}{\varepsilon^n} \omega\Big( \frac{|y-x|}{\varepsilon} \Big) \, \sin(\psi - \varphi) \, f(y,w,\psi,t) \, dy \, dw \, d\psi, 
\label{eq:kin_Wf_eps} \\
&&\hspace{-1cm}
J^\varepsilon_f(x,t) = \int_{{\mathbb R}^n \times {\mathbb S}^{n-1} \times  [0,2\pi]} \frac{1}{\varepsilon^n} \zeta\Big( \frac{|y-x|}{\varepsilon} \Big)  \, w \, f(y,w,\psi,t) \, dy \, dw \, d\psi. \label{eq:kin_Jf_eps} \\
&&\hspace{-1cm}
L^\varepsilon_f(x,t) = \int_{{\mathbb R}^n \times {\mathbb S}^{n-1} \times  [0,2\pi]} \frac{1}{\varepsilon^n}  \eta \Big( \frac{|y-x|}{\varepsilon} \Big)  \, e^{i \psi} \, f(y,w,\psi,t) \, dy \, dw \, d\psi, \label{eq:kin_Lf_eps}  \\
&&\hspace{-1cm}
\bar v^\varepsilon_f(x,t) =\Big( \frac{J^\varepsilon_f}{|J^\varepsilon_f|} \Big)(x,t), \qquad e^{i \bar \varphi^\varepsilon_f} (x,t) = \Big( \frac{L^\varepsilon_f}{|L^\varepsilon_f|} \Big)(x,t) . \label{eq:kin_barv_barphi_eps}
\end{eqnarray}
Now, expanding expressions \eqref{eq:kin_Wf_eps}-\eqref{eq:kin_barv_barphi_eps} in powers of $\varepsilon$, we get: 
\begin{eqnarray} 
&&\hspace{-1cm}
W^\varepsilon_f = U_f + {\mathcal O}(\varepsilon^2), \quad U_f(x,\varphi,t) = \int_{{\mathbb S}^{n-1} \times  [0,2\pi]} \sin(\psi - \varphi) \, f(x,w,\psi,t) \, dw \, d\psi, 
\label{eq:kin_Wf_expan} \\
&&\hspace{-1cm}
J^\varepsilon_f(x,t) = \zeta_0 j_f + {\mathcal O}(\varepsilon^2), \quad j_f (x,t) = \int_{{\mathbb S}^{n-1} \times  [0,2\pi]} w \, f(x,w,\psi,t) \, dw \, d\psi, \label{eq:kin_Jf_expan} \\
&&\hspace{-1cm}
L^\varepsilon_f(x,t) = \eta_0 \ell_f + {\mathcal O}(\varepsilon^2), \quad \ell_f (x,t) = \int_{{\mathbb S}^{n-1} \times  [0,2\pi]}  e^{i \psi} \, f(x,w,\psi,t)  \, dw \, d\psi, \label{eq:kin_Lf_expan}  \\
&&\hspace{-1cm}
\bar v^\varepsilon_f = u_f + {\mathcal O}(\varepsilon^2), \quad  u_f(x,t) = \Big( \frac{j_f}{|j_f|} \Big)(x,t),  \label{eq:kin_barv_expan} \\
&&\hspace{-1cm} 
\bar \varphi^\varepsilon_f  = \alpha_f + {\mathcal O}(\varepsilon^2), \quad  e^{i \alpha_f} (x,t) = \Big( \frac{\ell_f}{|\ell_f|} \Big)(x,t) , \label{eq:kin_barphi_expan}
\end{eqnarray}
with $\zeta_0 = \int_{{\mathbb R}^n} \zeta(|x|) \, dx$, and a similar definition for $\eta_0$. Furthermore, expanding $\sin(\psi - \varphi)$ in \eqref{eq:kin_Wf_expan}, we note that $U_f$ is given by \eqref{eq:kin_U_f}.  

Finally, introducing expansions \eqref{eq:kin_Wf_expan}-\eqref{eq:kin_barphi_expan} into \eqref{eq:kin_eps} and neglecting the resulting ${\mathcal O}(\varepsilon)$ terms which will have no influence on the final result, we are led to \eqref{eq:kin_final}.

\subsection{Particle system associated with the BGK operator}
\label{sec_particle_bgk}

We use the same notations as Section \ref{sec_part}. Each particle $k \in \{1, \ldots, N\}$ is associated with an increasing sequence of random numbers $T_k^1, \, T^k_2, \ldots , T^k_n, \ldots$ which are subject to the condition that the interval between two consecutive numbers are independent random variables following a Poisson process with intensity $D$. At time $T^k_n$ the (velocity, phase) pair of the $k$-th particle jumps between from $(v_k,\varphi_k)(T^k_n-0)$ to $(v_k,\varphi_k)(T^k_n+0)$, while $X_k$ is continuous (i.e. $X_k(T^k_n+0) = X_k(T^k_n-0)$). For $t \in (T^k_n, T^k_{n+1})$, the triple $(X_k, v_k, \varphi_k)$ evolves according to the following differential system: 
\begin{eqnarray*}
\frac{d X_k}{dt} &=& c_0 v_k - \gamma \nabla_x W(X_k, \varphi_k),  \\
\frac{d v_k}{dt} &=& - P_{v_k^\bot} \nabla_x V(X_k(t)) , \\
\frac{d \varphi_k}{dt} &=& 0. 
\end{eqnarray*}
with initial condition $(X_k, v_k, \varphi_k)(T^k_n+0)$ and $(X_k,v_k,\varphi_k)(T^k_{n+1}-0)$ are the values obtained by the solution of this system at time $T^k_{n+1}$. Finally at jump time $T^k_n$, the pair $(v_k,\varphi_k)(T^k_n+0)$ is drawn according to the von Mises distribution $M_{\bar v_k(T^k_n-0)} N_{\bar \varphi_k (T^k_n-0)}$ where $\bar v_k(T^k_n-0)$ and $\bar \varphi_k (T^k_n-0)$ are computed by \eqref{eq:def_barv} and \eqref{eq:def_barphi} in which $t$ is taken equal to $T^k_n-0$. This type of jump process is known as a Piecewise Deterministic Markov Process (PDMP). 

In  \cite{diez2019propagation}, it is proved that, in the limit $N \to \infty$, the empirical measure of this process (see Section \ref{sec_kin}) converges to the following kinetic equation: 
$$
\partial_t f + \nabla_x \cdot \big[ \big( c_0 v - \gamma \nabla_x W_f(x,\varphi) \big) f \big] - \nabla_v \cdot \big[ \nu \, P_{v^\bot} \bar v_f  \, f \big] = D \big( f - M_{\bar v_f} N_{\bar \varphi_f} \big), 
$$
with $\bar v_f$ and $\bar \varphi_f$ given by \eqref{eq:kin_barv_barphi}. The scaling developed in Section \ref{sec_scaling_appendix} can be developed analogously here. In particular, it results in the localisation of $\bar v_f$ and $\bar \varphi_f$ which are then replaced by $u_f$ and $\alpha_f$ given by \eqref{eq:kin_u_f} and \eqref{eq:kin_alpha_f} respectively. After scaling and neglect of higher order terms in $\varepsilon$, the kinetic model reduces to \eqref{eq:abstractR} with the BGK type collision operator \eqref{eq:defQR}. BGK-type models of Vicsek-type dynamics have been investigated in \cite{degond2021body, degond2019phase, degond2021bulk, degond2018alignment, dimarcomotsch16}.

\setcounter{equation}{0}
\section{Limit $\varepsilon \to 0$: proofs}
\label{sec_eps_to_0_proof}

In this section, we prove Theorems \ref{thm:eps_to_0} and \ref{thm:eps_to_0_bgk}. The proofs are identical for the two theorems. We develop it for Theorem \ref{thm:eps_to_0} and only point out what is different for Theorem~\ref{thm:eps_to_0_bgk} when necessary. The proof follows a certain number of steps. 

\medskip
\noindent
\textbf{Step 1: $f$ is given by \eqref{eq:equi} with the functions $\rho$, $u$ and $\alpha$ to be determined.} Indeed, if $f^\varepsilon \to f^0$ as $\varepsilon \to 0$ smoothly, then $f^0$ satisfies $Q(f^0) = 0$, which in view of Lemma \ref{lem:equi}~(iii) means that, at any given point $(x,t)$, $f$ is given by \eqref{eq:equi_homo}. At a different point $(x',t')$, the equilibrium \eqref{eq:equi_homo} may be different. This means that $\rho$, $u$ and $\alpha$ are functions of $(x,t)$, still to be determined, and that $f$ is given by \eqref{eq:equi}. 

\medskip
\noindent
\textbf{Step 2: derivation of the mass conservation equation \eqref{eq:fl_rho}.} This is simply done by integrating \eqref{eq:abstract} with respect to $(v,\varphi)$ and using that for any smooth functions $f$, 
$$ \int Q(f) \, dv \, d \varphi = 0, $$ 
(in this discussion, we omit the integration domain ${\mathbb S}^{n-1} \times [0,2\pi)$ any time the context is clear). This cancels the $1/\varepsilon$ singularity and leads to 
$$ \int T(f^\varepsilon) \, dv \, d \varphi = 0.  $$
Letting $\varepsilon \to 0$, we finally get 
$$ \int T(f^0) \, dv \, d \varphi = 0.  $$
We note that the $\nabla_v$ term in the expression \eqref{eq:expressT} of $T$ cancels in the integration with respect to $v$. 
The time and space derivatives commute with the integrals in $v$ and $\varphi$ and we get
\begin{equation}
\partial_t \rho + \nabla_x \cdot \Big( \rho \, \int \big( v - \gamma \nabla_x U_{f^0}(x,\varphi) \big) \, M_u(v) \, N_\alpha(\varphi) \, dv \, d\varphi \Big)=0. 
\label{eq:partial_t_rho}
\end{equation}
Now, we have from \eqref{eq:kin_U_f}:  
$$ U_{f^0}(x,\varphi) = c'_1 \, \rho \sin(\alpha - \varphi), $$
with 
\begin{equation}
 c'_1 = \frac{\displaystyle \int_0^{2 \pi} \cos \varphi \,  e^{k' \cos \varphi} \, d\varphi}{\displaystyle \int_0^{2 \pi} e^{k' \cos \varphi} \, d\varphi}. 
\label{eq:express_c'1}
\end{equation}
So, we get
\begin{equation} 
\nabla_x U_{f^0}(x,\varphi)  = c'_1 \, \big( \nabla_x \rho \, \sin(\alpha - \varphi) + \rho \, \cos (\alpha - \varphi) \nabla_x \alpha \big). 
\label{eq:nablaxU}
\end{equation}
Inserting \eqref{eq:nablaxU} into \eqref{eq:partial_t_rho} leads to \eqref{eq:fl_rho} with 
\begin{equation}
 c_1 = \frac{\displaystyle \int_0^\pi \cos \theta \,  e^{k \cos \theta} \, \sin^{n-2} \theta \, d\theta}{\displaystyle \int_0^\pi e^{k \cos \theta} \, \sin^{n-2} \theta \, d\theta}. 
\label{eq:express_c1}
\end{equation}
and 
\begin{equation}
b = - \gamma \, {c'_1}^2. 
\label{eq:express_b}
\end{equation}

\medskip
\noindent
\textbf{Step 3: Computation of the generalized collision invariants (GCI).} To find equations for $u$ and $\alpha$ is a not as straightforward, as there are no collision invariants, i.e. function $\chi(v,\varphi)$ such that for all smooth functions $f$
$$ \int Q(f) \, \chi \, dv \, d\varphi  = 0, $$
other than constant functions. In \cite{degond2008continuum}, it was shown that this difficulty can be solved by the concept of generalized collision invariant (GCI). We summarize the approach here and refer to \cite{degond2008continuum, frouvelle2012continuum} for details. To define the GCI concept, we first introduce the following operators 
\begin{itemize}
\item if the collision operator is $Q$ given by \eqref{eq:defQ}, then ${\mathcal Q}(f;u,\alpha)$ is defined for any $(u,\alpha) \in {\mathbb S}^{n-1} \times {\mathbb R}/(2 \pi {\mathbb Z})$ by 
\begin{equation}
{\mathcal Q}(f;u,\alpha) =D \, \nabla_v \cdot \Big[ M_u \nabla_v \Big( \frac{f}{M_u} \Big) \Big] + D' \, \partial_\varphi \Big[ N_\alpha \partial_\varphi \Big( \frac{f}{N_\alpha} \Big) \Big]. 
\label{eq:def_calQ}
\end{equation}
\item if the collision operator is $Q_R$ given by \eqref{eq:defQR}, then 
\begin{equation}
{\mathcal Q}_R(f;u,\alpha) = D \, \big( \rho_f M_u \, N_\alpha - f \big). 
\label{eq:def_calQR}
\end{equation}
\end{itemize}
We note that 
\begin{equation}
Q(f) = {\mathcal Q}(f;u_f,\alpha_f),
\label{eq:rel_Q_calQ}
\end{equation}
and we have a similar relation between $Q_R$ and ${\mathcal Q}_R$. Then, we have the: 

\begin{definition}
Given $(u,\alpha) \in {\mathbb S}^{n-1} \times {\mathbb R}/(2 \pi {\mathbb Z})$, the function $\chi_{u,\alpha}$: ${\mathbb S}^{n-1} \times {\mathbb R}/(2 \pi {\mathbb Z}) \to {\mathbb R}$, $(v,\varphi) \mapsto \chi_{u,\alpha}(v,\varphi)$ is a GCI for $Q$ associated to $(u,\alpha)$ if and only if the following holds:
\begin{eqnarray}
&&\hspace{-1cm}
\int {\mathcal Q}(f;u,\alpha) \, \chi_{u,\alpha} \, dv \, d\varphi  = 0, \quad \forall f \textrm{ such that } P_{u^\bot} \int f \, v \, dv \, d\varphi =0 \nonumber \\
&&\hspace{6cm}
\textrm{ and } \int f \, \sin (\varphi - \alpha) \, dv \, d\varphi = 0. 
\label{eq:def_GCI}
\end{eqnarray}
We have a similar definition of a GCI for $Q_R$ by replacing ${\mathcal Q}$ by ${\mathcal Q}_R$. 
\label{def:gci}
\end{definition}

We have the 

\begin{proposition}
The set ${\mathcal G}_{u,\alpha}$ of GCI $\chi_{u,\alpha}$ is a vector space given by 
\begin{equation}
{\mathcal G}_{u,\alpha} = \big\{ A \, \tilde h \big(\cos (\varphi - \alpha) \big) \, \sin (\varphi - \alpha) + h(u \cdot v ) \, P_{u^\bot} v \cdot B + C \, \, | \, \, A, C \in {\mathbb R}, \, B \in \{u\}^\bot \big\}, 
\label{eq:gci_set}
\end{equation}
where the functions $h$ and $\tilde h$: $(-1,1) \to {\mathbb R}$ are given as follows:  
\begin{itemize}
\item Case of operator $Q$: $h$ and $\tilde h$ are such that the functions
\begin{equation}
\tilde g(\varphi) = \tilde h \big(\cos \varphi \big) \, \sin \varphi,  \quad  g(\theta) = h(\cos \theta) \, \sin \theta, 
\label{eq:express_h_htilde}
\end{equation}
are the solutions of the following equations: 
\begin{equation}
- \partial_\varphi \big( e^{k' \cos \varphi} \, \partial_\varphi \tilde g\big) = \sin \varphi \, e^{k' \cos \varphi}, 
\label{eq:eq_gtilde}
\end{equation}
for $\tilde g$ and 
\begin{equation}
- \partial_\theta \big( \sin^{n-2} \theta \, e^{k \cos \theta} \, \partial_\theta  g \big) + \frac{n-2}{\sin^2 \theta} \,  \sin^{n-2} \theta \, e^{k \cos \theta} \, g = \sin \theta \, \sin^{n-2} \theta \, e^{k \cos \theta} 
\label{eq:eq_g}
\end{equation}
for $g$; these solutions are unique in the spaces $H^1(0,\pi)$ for $\tilde g$ and 
$$ \big\{ g \, \, | \, \, \sin^{\frac{n}{2} - 2} \theta \, g \in L^2(0,\pi), \, \, \sin^{\frac{n}{2} - 1} \theta \, g \in H^1(0,\pi) \big\}, $$
for $g$; the functions $g$ and $\tilde g$ are nonnegative on $[0,\pi]$ and $\tilde g$ can be extended into a smooth odd function on $[-\pi,\pi]$, 
\item Case of operator $Q_R$: $h$ and $\tilde h$ are given by 
\begin{equation}
\tilde h = 1, \qquad h=1. \label{eq:htilhbgk}
\end{equation}
\end{itemize}
\label{prop:gci}
\end{proposition}

\noindent
\textbf{Proof of Proposition \ref{prop:gci}.} \textbf{Case 1: collision operator $Q$}. In \cite{degond2008continuum, frouvelle2012continuum}, it is shown that \eqref{eq:def_GCI} is equivalent to saying that 
\begin{equation}
\exists A \in {\mathbb R}, \, \, \exists B \in \{u\}^\bot \textrm{ such that } {\mathcal Q}^*(\chi_{u,\alpha};u,\alpha) = A \sin(\varphi - \alpha) + B \cdot v, 
\label{eq:prop_GCI} 
\end{equation}
where ${\mathcal Q}^*(\cdot;u,\alpha)$ is the formal $L^2$-adjoint of ${\mathcal Q}(\cdot;u,\alpha)$. Computing this adjoint, we find that $\chi = \chi_{u,\alpha}$ is a GCI if and only if it satisfies the following problem
\begin{equation}
D \frac{1}{M_u} \nabla_v \cdot \big( M_u \nabla_v \chi \big) + D' \frac{1}{N_\alpha} \partial_\varphi \big( N_\alpha \partial_\varphi \chi \big) = A \sin(\varphi - \alpha) + B \cdot v. 
\label{eq:GCI_eq}
\end{equation}
For given $A$ and $B$, by Lax-Milgram theorem, it can be shown that this problem has a unique solution in the subspace $\dot H^1$ of $H^1({\mathbb S}^{n-1} \times {\mathbb R}/(2 \pi {\mathbb Z}))$ consisting of functions $g$ satisfying $\int g \, dv \, d \varphi = 0$. Furthermore, any solution to \eqref{eq:GCI_eq} in $H^1$ is equal to this special solution up to an additive constant. Denoting by $\chi^{A,B}$ the unique solution of \eqref{eq:GCI_eq} in $\dot H^1$ corresponding to the pair $(A,B)$, we have $\chi^{A,B} = \chi^{A,0} + \chi^{0,B}$ with $\chi^{A,0}$ only depending on $\varphi$ and solution to 
\begin{equation}
D' \frac{1}{N_\alpha} \partial_\varphi \big( N_\alpha \partial_\varphi \chi^{A,0} \big) = A \sin(\varphi - \alpha) , 
\label{eq:chiA}
\end{equation}
and $\chi^{0,B}$ only depending on $v$ and solution to 
\begin{equation}
D \frac{1}{M_u} \nabla_v \cdot \big( M_u \nabla_v \chi \big) =  B \cdot v. 
\label{eq:chiB}
\end{equation}
These two problems have been solved in \cite{degond2008continuum, frouvelle2012continuum} (note that \eqref{eq:chiA} is a special case of \eqref{eq:chiB} corresponding to the dimension $n=2$). Their solutions are given as follows:
\begin{equation}
\chi^{A,0} = - A \, \tilde h \big(\cos (\varphi - \alpha)\big) \, \sin (\varphi - \alpha), \quad  \chi^{0,B} = - h(u \cdot v) \, B \cdot v,  
\label{eq:express_chiA_chiB}
\end{equation}
with $h$ and $\tilde h$ given by \eqref{eq:express_h_htilde}. This leads to \eqref{eq:gci_set}. 

\noindent
\textbf{Case 2: collision operator $Q_R$}. Eq. \eqref{eq:prop_GCI} stands but now \eqref{eq:GCI_eq} is changed into
\begin{equation}
D \Big( \int \chi(w,\psi) \, M_u(w) \, N_\alpha(\psi) \, dw \, d \psi - \psi(v,\varphi) \Big) = A \sin(\varphi - \alpha) + B \cdot v. 
\label{eq:GCIR_eq}
\end{equation}
It is clear that any two solutions of this equation differ by a constant. So, we can single out a solution by requiring that $\int \chi(w,\psi) \, M_u(w) \, N_\alpha(\psi) \, dw \, d \psi =0$. It follows that 
${\mathcal G}_{u,\alpha}$ is given by \eqref{eq:gci_set} with $h=1$ and $\tilde h = 1$.
\endproof

We will now write 
$$ \chi_\alpha (\varphi) = \tilde h \big(\cos (\varphi - \alpha) \big) \, \sin (\varphi - \alpha), \quad \vec \chi_u(v) = h(u \cdot v ) \, P_{u^\bot} v, $$
so that 
\begin{equation}
{\mathcal G}_{u,\alpha} = \big\{ A \, \chi_\alpha(\varphi) + \vec \chi_u(v) \cdot B + C \, \, | \, \, A, C \in {\mathbb R}, \, B \in \{u\}^\bot \big\}. 
\label{eq:gci_set2}
\end{equation}
We remark that $\vec \chi_u$ is a vector (perpendicular to $u$) further referred to as the vector GCI. 

The main use of the GCI is as follows. We first note that 
$$ P_{u_f^\bot} \int f \, v \, dv \, d\varphi = P_{u_f^\bot} (|j_f| u_f) =  0,  \quad \int f \, \sin (\varphi - \alpha_f) \, dv \, d\varphi = |\ell_f| \sin (\alpha_f - \alpha_f) = 0. $$
Therefore, $f$ satisfies the conditions of \eqref{eq:def_GCI} for $(u,\alpha) = (u_f, \alpha_f)$. Thanks to \eqref{eq:rel_Q_calQ}, we deduce that 
$$ \int Q(f) \, \chi_{u_f, \alpha_f} \, dv \, d\varphi = \int {\mathcal Q}(f;u_f,\alpha_f) \, \chi_{u_f, \alpha_f} \, dv \, d\varphi = 0. $$
From this, we obtain: 
$$ \int T(f^\varepsilon) \chi_{u_{f^\varepsilon}, \alpha_{f^\varepsilon}} \, dv \, d\varphi = \frac{1}{\varepsilon} \int Q(f^\varepsilon) \chi_{u_{f^\varepsilon}, \alpha_{f^\varepsilon}} \, dv \, d\varphi = 0. $$
In particular, we get 
$$ \int T(f^\varepsilon) \vec \chi_{u_{f^\varepsilon}} \, dv \, d\varphi = 0 , \qquad  \int T(f^\varepsilon) \chi_{\alpha_{f^\varepsilon}} \, dv \, d\varphi = 0. $$
But it is clear that $\chi_\alpha$ is continuous with respect to $\alpha$ and  $\vec \chi_u$ with respect to $u$. We can then let $\varepsilon \to 0$ and obtain
\begin{equation}
 \int T(f^0) \vec \chi_{u_{f^0}} \, dv \, d\varphi = 0 , \qquad  \int T(f^0) \chi_{\alpha_{f^0}} \, dv \, d\varphi = 0. 
\label{eq:fluid_abstract}
\end{equation}
These are the two missing equations of the fluid model as shown in the next step.

\medskip
\noindent
\textbf{Step 4: Explicit forms of the equations for $u$ and $\alpha$.} In this step, we are making the two equations \eqref{eq:fluid_abstract} explicit. We have
\begin{eqnarray}
T(f_0) &=&  (\partial_t + v \cdot \nabla_x) (\rho M_u N_\alpha) - \gamma \nabla_x \cdot \big( \nabla_x U_{\rho M_u N_\alpha} \, \rho M_u N_\alpha \big) 
\nonumber \\
&& \hspace{5.5cm}
- \nabla_v \cdot \big[ P_{v^\bot} \nabla_x V(x) \, \rho M_u N_\alpha \big] 
. \label{eq:Tf0_1}
\end{eqnarray}
Using the notations $D = \partial_t + v \cdot \nabla_x$ and $v_\bot = P_{u^\bot}v$ as well as \eqref{eq:nablaxU}, computations give
\begin{eqnarray}
&&\hspace{-1cm}
D (\rho M_u N_\alpha) = M_u N_\alpha \big[ D \rho + k \rho  v_\bot \cdot Du + k' \rho \sin(\varphi- \alpha) D \alpha \big], \label{eq:Tf0_comput1}\\
&&\hspace{-1cm}
\nabla_x \cdot \big( \nabla_x U_{\rho M_u N_\alpha} \, \rho M_u N_\alpha \big) = -  c'_1 \Big\{ \big( \rho \Delta_x \rho - \rho^2 |\nabla_x \alpha|^2 + |\nabla_x \rho|^2 \big) \sin(\varphi- \alpha) \nonumber \\
&&\hspace{0cm}
 - \big( 3 \rho \nabla_x \rho \cdot \nabla_x \alpha + \rho^2 \Delta_x \alpha  \big) \cos(\varphi- \alpha) \nonumber \\
&&\hspace{0cm}
+ k \rho v_\bot \cdot \Big( \big[ \nabla_x \rho \sin (\varphi - \alpha) - \rho \nabla_x \alpha \cos(\varphi - \alpha) \big] \cdot \nabla_x \Big) u \nonumber \\
&&\hspace{0cm}
+ k' \rho \sin(\varphi - \alpha) \big( \nabla_x \rho \sin (\varphi - \alpha) - \rho \nabla_x \alpha \cos(\varphi - \alpha) \big) \cdot \nabla_x \alpha \Big\} M_u \, N_\alpha , \label{eq:Tf0_comput2}\\
&&\hspace{-1cm}
- \nabla_v \cdot \big[ P_{v^\bot} \nabla_x V (\rho M_u N_\alpha) \big] = \rho M_u N_\alpha \big( (n-1) \,  \nabla_x V \cdot v - k \, P_{v^\bot} \nabla_x V \cdot P_{v^\bot} u \big). 
\label{eq:Tf0_comput3}
\end{eqnarray}
Using the decomposition $v = (v \cdot u) u + v_\bot$, we can write 
$$ T(\rho \, M_u \, N_\alpha) = M_u \, N_\alpha \big( {\mathcal T}_{ee} + {\mathcal T}_{eo} +{\mathcal T}_{oe} + {\mathcal T}_{oo} \big) , $$
where ${\mathcal T}_{ee}$ is even with respect to both $v_\bot$ and $\sin(\varphi - \alpha)$, ${\mathcal T}_{eo}$ is even with respect to $v_\bot$ and odd with respect to $\sin(\varphi - \alpha)$, ${\mathcal T}_{oe}$ is odd with respect to $v_\bot$ and even with respect to $\sin(\varphi - \alpha)$ and finally ${\mathcal T}_{oo}$ is odd with respect to both arguments. We have 
\begin{eqnarray}
&&\hspace{-1cm}
{\mathcal T}_{eo} = k' \rho \sin(\varphi - \alpha) \, \big( \partial_t + (v \cdot u) \, u \cdot \nabla_x \big) \alpha + \gamma c'_1 \Big[ \big(\rho \Delta_x \rho - \rho^2 |\nabla_x \alpha|^2  \nonumber \\
&& \hspace{2cm}
+ |\nabla_x \rho|^2 \big) \sin(\varphi-\alpha) - k' \rho^2 |\nabla_x \alpha|^2 \sin(\varphi-\alpha) \cos(\varphi-\alpha) \Big]  
, \label{eq:Teo}\\
&&\hspace{-1cm}
{\mathcal T}_{oe} = v_\bot \cdot \nabla_x \rho + k \rho v_\bot \cdot \big( \partial_t u + (v \cdot u) \, (u \cdot \nabla_x) u \big) + (n-1) \rho v_\bot \cdot \nabla_x V \nonumber \\
&&\hspace{2cm}
+ \rho k (u \cdot v) v_\bot \cdot \nabla_x V  - \gamma c'_1 k \rho^2 \cos(\varphi-\alpha) \, v_\bot \cdot \big( (\nabla_x \alpha \cdot \nabla_x) u \big)
, \label{eq:Toe}
\end{eqnarray}
while the other terms will not be needed in the forthcoming computations. 

We now consider the first Eq. \eqref{eq:fluid_abstract}. Since $\chi_{u_{f_0}}$ is odd with respect to $v$ and even with respect to $\alpha$, the only term in $T(\rho \, M_u \, N_\alpha)$ which will not vanisy in the integration by imparity will be that corresponding to ${\mathcal T}_{oe}$. By Lemma 4.1 of \cite{degond2020nematic}, we have, for any function $k(v \cdot u)$, 
$$ \int_{{\mathbb S}^{n-1}} k(v \cdot u) \, v_\bot \otimes v_\bot \, dv = \frac{1}{n-1} \int_{{\mathbb S}^{n-1}} k(v \cdot u) \, (1 - (v \cdot u)^2) \, dv \, P_{u^\bot}. $$
Using this, the first Eq. \eqref{eq:fluid_abstract} leads to
\begin{eqnarray}
&&\hspace{-1cm}
C_1 P_{u^\bot} \nabla_x \rho + k C_1  \rho \partial_t u + k C_2 \rho (u \cdot \nabla_x) u + \big((n-1) C_1  + k C_2 \big) \rho P_{u^\bot} \nabla_x V \nonumber \\ 
&&\hspace{7cm}
- \gamma k {c'_1}^2 C_1 \rho^2 (\nabla_x \alpha \cdot \nabla_x) u =0,
\label{eq:fl_first}
\end{eqnarray}
with
\begin{equation}
C_i = \frac{1}{n-1} \frac{\displaystyle \int_0^\pi \cos^{i-1} \theta \,  h(\cos \theta) \, e^{k \cos \theta}  \, \sin^n \theta \, d\theta}{\displaystyle \int_0^\pi e^{k \cos \theta} \, \sin^{n-2} \theta \, d\theta}, \quad i=1,2. \label{eq:express_C12} 
\end{equation}
Dividing \eqref{eq:fl_first} by $k C_1$, we get \eqref{eq:fl_u} with
\begin{eqnarray}
c_2 &=& \frac{C_2}{C_1} = \frac{\displaystyle \int_0^\pi \cos \theta \,  h(\cos \theta) \, e^{k \cos \theta}  \, \sin^n \theta \, d\theta}{\displaystyle \int_0^\pi h(\cos \theta) \, e^{k \cos \theta} \, \sin^n \theta \, d\theta}, 
\label{eq:express_c2} \\
\Theta &=&  \frac{1}{k}, \qquad \kappa =  \frac{n-1}{k} + c_2. \label{eq:express_Thetkap} 
\end{eqnarray}
and $b$ is given by \eqref{eq:express_b}. 

We proceed similarly for the second Eq. \eqref{eq:fluid_abstract}. In this case, $\chi_{\alpha_{f_0}}$ is even with respect to $v$ and odd with respect to $\varphi$. So, the only term of $T(\rho \, M_u \, N_\alpha)$ which remains is that corresponding to ${\mathcal T}_{eo}$. This leads to
\begin{eqnarray}
&&\hspace{-1cm}
k' C'_1 \rho \partial_t \alpha + c_1 C'_1 k' \rho (u \cdot \nabla_x) \alpha + \gamma c'_1 C'_1 ( \rho \Delta_x \rho- \rho^2 |\nabla_x \alpha|^2 
\nonumber \\ 
&&\hspace{6cm}
+ |\nabla_x \rho|^2 ) - \gamma c'_1 k' C'_2 \rho^2 |\nabla_x \alpha|^2 =0,
\label{eq:fl_second}
\end{eqnarray}
with
\begin{equation}
 C'_i = \frac{\displaystyle \int_0^{2 \pi} \cos^{i-1} \varphi \, \tilde h(\cos \varphi) \,  e^{k' \cos \varphi} \, \sin^2 \varphi \, d\varphi}{\displaystyle \int_0^{2 \pi} e^{k' \cos \varphi} \, d\varphi} \quad i=1,2. 
\label{eq:express_C34}
\end{equation}
Dividing by $C'_1 k'$, we get \eqref{eq:fl_al} with
\begin{eqnarray}
b'&=& - \gamma c'_1 \Big( \frac{1}{k'} + c'_2 \Big)  ,
\label{eq:express_b'} \\
c'_2&=& \frac{C'_2}{C'_1} = \frac{\displaystyle \int_0^{2 \pi} \cos \varphi \, \tilde h(\cos \varphi) \,  e^{k' \cos \varphi} \, \sin^2 \varphi \, d\varphi}{\displaystyle \int_0^{2 \pi} \tilde h(\cos \varphi) \,  e^{k' \cos \varphi} \, \sin^2 \varphi \, d\varphi} ,
\label{eq:express_c'2} \\
\Theta' &=& -  \frac{\gamma c'_1 }{k'}.  \label{eq:express_Thetapr} 
\end{eqnarray}

Finally, the properties of the coefficients listed at the end of the theorem statement are direct consequences of \cite{frouvelle2012continuum}. This ends the proof of Theorem \ref{thm:eps_to_0}. \endproof

\setcounter{equation}{0}
\section{Small noise limit in the phase equation}
\label{sec_small_noise_proofs}

\subsection{Expression of the system: proof of Lemma \ref{lem:small_noise_limit}}
\label{sec_small_noise_lemma_proof}

We recall that $b$, $b'$ and $\Theta'$ are given by \eqref{eq:express_b}, \eqref{eq:express_b'}, \eqref{eq:express_Thetapr} respectively. Now, it is proved in \cite{frouvelle2012continuum} that $c'_1 \to 1$, $c'_2 \to 1$ as $k' \to \infty$. Thus, \eqref{eq:small_noise_coef} follows immediately. \endproof


\subsection{Hyperbolicity: proof of Lemma \ref{lem:hyperbolic}}
\label{sec_hyperbolic_proof}

Let $(\rho_0, u_0, z_0) \in (0,\infty) \times {\mathbb S}^{n-1} \times {\mathbb R}^n$ be given. Then, $(\rho, u, z) = (\rho_0, u_0, z_0)$ is a spatially uniform stationary solution of System \eqref{eq:fl_rho_nl}-\eqref{eq:fl_const_nl}. The linearization of System \eqref{eq:fl_rho_nl}-\eqref{eq:fl_const_nl} about this equilibrium solution is then given by 
\begin{eqnarray}
&&\hspace{-1cm}
\partial_t \rho + (c_1 u_0 + 2 b \rho_0 z_0) \cdot \nabla_x \rho + c_1 \rho_0  \, \nabla_x \cdot u + b \rho_0^2 \, \nabla_x \cdot z = 0, \label{eq:fl_rho_lin} \\
&&\hspace{-1cm}
\partial_t u + \Theta \rho_0^{-1} \, P_{u_0^\bot} \nabla_x\rho + \big( (c_2 u_0 + b \rho_0 z_0) \cdot \nabla_x \big) u = 0, \label{eq:fl_u_lin} \\
&&\hspace{-1cm}
\partial_t z + b |z_0|^2 \, \nabla_x \rho + \nabla_x u \, z_0 + \nabla_x z \, (c_1 u_0 + 2 b \rho_0 z_0)  = 0, \label{eq:fl_al_lin} \\
&&\hspace{-1cm}
\nabla_x \wedge z = 0. \label{eq:fl_const_lin} \\
&&\hspace{-1cm}
u \cdot u_0 = 0. \label{eq:fl_const_ortho}
\end{eqnarray}
In \eqref{eq:fl_al_lin}, $\nabla_x u$ denotes the gradient matrix of $u$ i.e. $(\nabla_x u)_{ij} = \partial_{x_i} u_j$, for all $i, \, j \in \{1, \ldots, n \}$. The expression $\nabla_x u \, z_0$ refers to the multiplication of the matrix $\nabla_x u$ and the vector $z_0$. Similar definitions apply to $\nabla_x z$. The constraint \eqref{eq:fl_const_ortho} expresses that the first order variation of a normalized vector is orthogonal to that vector. 

We take the partial Fourier transform of this system with respect to $x$ and denote the resulting unknown by  $(\hat \rho, \hat u, \hat z)(\xi,t)$, with $\xi$ the Fourier dual variable to $x$. We assume $\xi \not = 0$ (the case $\xi = 0$ corresponds to constants, which we already know are solutions of the linearized system). We recall that $\tau = \xi/|\xi|$. From \eqref{eq:fl_const_lin} we deduce that $\hat z = \tilde z \tau$, where $\tilde z \in {\mathbb R}$. Also, due to \eqref{eq:fl_const_ortho}, we can project \eqref{eq:fl_u_lin} on $\{u_0\}^\bot$, and the component $u_1$ of $u$ on $e_1$ is always zero. 

Using these remarks, the Fourier transform of System \eqref{eq:fl_rho_lin}-\eqref{eq:fl_const_lin} leads to
\begin{eqnarray*}
&&\hspace{-1cm}
\frac{1}{i |\xi|} \partial_t \hat \rho + \tau \cdot (c_1 u_0 + 2 b \rho_0 z_0) \hat \rho + c_1 \rho_0  \, \tau \cdot \hat u + b \rho_0^2 \, \tilde z = 0,  \\
&&\hspace{-1cm}
\frac{1}{i |\xi|} \partial_t \hat u + \Theta \rho_0^{-1} \, P_{u_0^\bot} \tau \, \hat \rho + (c_2 u_0 + b \rho_0 z_0) \cdot \tau \, \hat u = 0,  \\
&&\hspace{-1cm}
\frac{1}{i |\xi|} \partial_t \tilde z + b |z_0|^2 \, \hat \rho + \hat u \cdot z_0 + (c_1 u_0 + 2 b \rho_0 z_0) \cdot \tau \, \hat z  = 0. 
\end{eqnarray*}
Choosing a reference frame such that $u_0$, $z_0$ and $\tau$ are expressed by \eqref{eq:frame}, and denoting by $V = (\hat \rho , \hat u_2 , \hat u_3 , \hat u_4 ,  \ldots ,  \hat u_n , \tilde z )^T$, we can write 
$$ \frac{1}{i |\xi|} \partial_t V + {\mathbb A} V = 0, $$
where ${\mathbb A}$ is given in dimension $n \geq 3$ by:
\begin{equation}
{\mathbb A} = \left( \begin{array}{ccccccc} X_1  & c_1 \rho_0 \sin \theta \sin \phi & c_1 \rho_0 \cos \theta & 0 & \ldots & 0 & b \rho_0^2 \\
\Theta \rho_0^{-1} \sin \theta \sin \phi& X_2 & 0 & 0 & \ldots & 0 & 0 \\
\Theta \rho_0^{-1} \cos \theta & 0 & X_2 & 0 & \ddots & 0 & 0 \\
0 & 0 & 0 & X_2 & \ldots & 0 & 0 \\
\vdots & \vdots & \vdots & \ddots & \ddots & \vdots & \vdots \\
0 & 0 & 0 & \ldots & \ldots & X_2 & 0 \\
b |z_0|^2 & |z_0| \sin \delta & 0 & \ldots & \ldots & 0 & X_1
\end{array} \right), 
\label{eq:defbbA}
\end{equation}
with
$$ X_1 = \sin \theta \, \big( c_1 \cos \phi + 2 b \rho_0 |z_0| \cos (\phi-\delta) \big), \qquad X_2 = \sin \theta \, \big( c_2 \cos \phi +  b \rho_0 |z_0| \cos (\phi-\delta) \big). $$
In dimension $n=2$, we can choose $\theta = \pi/2$ and  ${\mathbb A}$ is given by:
$$ {\mathbb A} = \left( \begin{array}{ccc} X_1  & c_1 \rho_0 \sin \phi & b \rho_0^2 \\
\Theta \rho_0^{-1} \sin \phi& X_2 &  0 \\
b |z_0|^2 & |z_0| \sin \delta & X_1
\end{array} \right).  $$
We recall that the system is hyperbolic about $(\rho_0, u_0, z_0)$ if and only if ${\mathbb A}$ is diagonalizable with real eigenvalues for all values of $\theta$ and $\phi$.

In the case $n \geq 3$, developing the determinant with respect to the last column, we find:
\begin{eqnarray}
&& \hspace{-1cm}
\det({\mathbb A}-\lambda \textrm{Id}) = (X_1 - \lambda) (X_2 - \lambda)^{n-3} \left| \begin{array}{ccc} X_1 - \lambda & c_1 \rho_0 \sin \theta \sin \phi & c_1 \rho_0 \cos \theta \\
\Theta \rho_0^{-1} \sin \theta \sin \phi& X_2 - \lambda &  0 \\
\Theta \rho_0^{-1} \cos \theta & 0 & X_2 - \lambda 
\end{array} \right| \nonumber  \\
&& \hspace{2cm}
+ (-1)^{n+2} b \rho_0^2 \, \prod_{k=3}^n \Big( (-1)^{2k-1} (X_2 -  \lambda) \Big) \, \left| \begin{array}{cc} \Theta \rho_0^{-1} \sin \theta \sin \phi& X_2 - \lambda \\
b |z_0|^2 & |z_0| \sin \delta 
\end{array} \right| \nonumber  \\
&& \hspace{0cm} = (X_2-\lambda)^{n-2} \Big\{ (X_1-\lambda) \big[ (X_1-\lambda) (X_2 - \lambda) - L  \big] + MR - M^2 (X_2-\lambda) \Big\} \nonumber \\
&& \hspace{0cm} =: (X_2-\lambda)^{n-2} P(\lambda),
\label{eq:detA-lamId}
\end{eqnarray}
with
\begin{equation}
 L = c_1 \Theta ( \cos^2 \theta + \sin^2 \theta \sin^2 \phi), \quad M= b \rho_0 |z_0|, \quad R = \Theta \sin \theta \sin \phi \sin \delta.
\label{eq:LMR}
\end{equation}
After rearranging, we get 
\begin{equation}
P(\lambda) = A \lambda^3 + B \lambda^2 + C \lambda + D, 
\label{eq:defP}
\end{equation}
with 
\begin{eqnarray}
A &=&-1, \quad B= 2 X_1 + X_2, \quad C = -2X_1 X_2 -X_1^2 +L+M^2, 
\label{eq:defABC} \\
D &=& X_1^2 X_2 - L X_1 + MR - M^2 X_2. 
\label{eq:deD} 
\end{eqnarray}
In the case $n=2$, a direct computation shows that \eqref{eq:detA-lamId} with $P$ given by \eqref{eq:defP}, \eqref {eq:defABC} \eqref{eq:deD} is still true provided we make $\theta = \pi/2$ in \eqref{eq:LMR}. 

In the case $n \geq 3$, we see that $X_2$ is an eigenvalue of ${\mathbb A}$ with multiplicity at least $n-2$. We show that the associated eigenspace has dimension at least $n-2$. First, by inspection of \eqref{eq:defbbA}, it is clear that the space ${\mathbb V}= \{ (0,0,0, v_4, \ldots, v_n, 0) \, \, | \, \,  (v_4, \ldots, v_n) \in {\mathbb R}^{n-3} \}$ is a subspace of $\ker({\mathbb A}-X_2 \textrm{Id})$. Now, we may remove the lines and columns of ${\mathbb A}-X_2 \textrm{Id}$ of indices comprised between $4$ and $n$ and check the following determinant 
$$ \left| \begin{array}{cccc} Y  & c_1 \rho_0 \sin \theta \sin \phi & c_1 \rho_0 \cos \theta & b \rho_0^2 \\
\Theta \rho_0^{-1} \sin \theta \sin \phi& 0 & 0 & 0 \\
\Theta \rho_0^{-1} \cos \theta & 0 & 0 & 0 \\
b |z_0|^2 & |z_0| \sin \delta & 0 & Y
\end{array} \right|, $$
with 
$$ 
Y=X_1-X_2 = \sin \theta \big( (c_1-c_2) \cos \phi + M \cos (\phi-\delta) \big). 
$$
Developing with respect to the last column, we easily realize that this determinant is equal to $0$, showing that there is a one-dimensional complement space to ${\mathbb V}$ in $\ker({\mathbb A}-\lambda \textrm{Id})$ and consequently, that this eigenspace  is at least of dimension $n-2$. 

According to classical results about the cubic equation \cite{jacobson2012basic}, $P$ has three distinct real roots if and only if its discriminant $\Delta$ is positive, i.e. 
$$ \Delta = : \frac{1}{27} \big[ 4 (B^2 + 3C)^3 - (2 B^3 + 9BC + 27 D)^2 \big] >0. $$
Some algebra leads to 
$$ B^2 + 3C = Y^2 + 3 (L + M^2), \quad 2 B^3 + 9BC + 27 D = -2 Y^3 -9Y (L-2M^2) + 27 MR, $$
and finally we get $\Delta$ as a polynomial of $Y$: 
\begin{eqnarray}
\Delta &=& 4 Y^4 M^2 + 4 Y^3 MR + Y^2 (L^2 + 20 L M^2 - 8 M^4)  \nonumber \\
&&\hspace{4cm}  + 18 YMR (L - 2 M^2) + 4 (L+M^2)^3 - 27 M^2 R^2. \label{eq:defDelta}
\end{eqnarray}

\medskip
\noindent
\textbf{Proof of (i).} We first consider the case where $\delta = 0$ or $\delta = \pi$. In this case \eqref{eq:LMR} shows that $R=0$ and $\Delta$ reduces to 
\begin{eqnarray*}
\Delta &=& 4 Y^4 M^2 + Y^2 (L^2 + 20 L M^2 - 8 M^4) + 4 (L+M^2)^3 \\
&=& 4 \big( M^2 (Y^2-M^2)^2 + L^3 + 3 L^2 M^2 + 3 L M^4 \big) + Y^2 (L^2 + 20 L M^2). 
\end{eqnarray*}
Since $L \geq 0$, all terms in the last expression are nonnegative so we get $\Delta \geq 0$. If $L \not = 0$, we see that $\Delta >0$ and so, the three roots of $P$ are real and distinct. If $L=0$, we have $ \Delta = 4 M^2 (Y-M)^2 (Y+M)^2$, so that
$$ \Delta =0 \quad \Longleftrightarrow \quad Y = \epsilon_1 M, $$
with $\epsilon_1 = \pm 1$. On the other hand, from \eqref{eq:LMR}, we have  
\begin{equation} 
L=0 \quad \Longleftrightarrow \quad \theta = \frac{\pi}{2} \, \, \textrm{and} \, \, ( \phi = 0 \, \, \textrm{or} \, \, \phi = \pi). 
\label{eq:L=0}
\end{equation}
In this case, we have $ Y = \epsilon_2 \big( (c_1-c_2) + \epsilon_3 M \big)$ with $\epsilon_2 = \cos \phi = \pm1$ and $\epsilon_3 = \cos \delta = \pm1$. Therefore, we have $\Delta = 0$ if and only if 
$$ (1- \epsilon_1 \epsilon_2 \epsilon_3) M = \epsilon_1 \epsilon_2 (c_1-c_2) .  $$
Since $c_1-c_2 >0$ (see Theorem \ref{thm:eps_to_0}), we must have $(1- \epsilon_1 \epsilon_2 \epsilon_3)=2$. Hence $\Delta$ can only be zero if $2 |M| = c_1 - c_2$, which is ruled out in the assumptions of the theorem. Thus, it follows that $P$ has three real and distinct roots. 

Now, we study whether one of these roots coincides with the root $X_2$ found earlier. Indeed, in this case, we have $P(X_2)=0$. From \eqref{eq:defP}, \eqref{eq:defABC}, \eqref{eq:deD} we readily get that $P(X_2) = LY$. So, $P(X_2) = 0$ implies $L=0$ or $Y=0$. 
\begin{itemize}
\item  If $L=0$, injecting \eqref{eq:L=0} into \eqref{eq:defbbA} shows that the eigenspace associated to the eigenvalue $X_2$ contains the subspace ${\mathbb W}= \{ (0,v_2, \ldots, v_n, 0) \, \, | \, \,  (v_2, \ldots, v_n) \in {\mathbb R}^{n-1} \}$. Since $Y^2 - M^2 = 0$ is not allowed by the assumptions of the theorem, we see that ${\mathbb A}-X_2 \textrm{Id}$ has rank $2$ which shows that $\ker({\mathbb A}-X_2 \textrm{Id})$ has dimension exactly equal to $n-1$. Since the roots of $P$ are distinct, $X_2$ is a simple root of $P$ so that, as an eigenvalue of ${\mathbb A}$, $X_2$ has multiplicity $n-1$. So, in this case, ${\mathbb A}$ is diagonalizable with real eigenvalues. 
\item If $Y=0$, injecting it into \eqref{eq:defbbA} readily shows that ${\mathbb A}- X_2 \textrm{Id}$ has rank $2$. We conclude similarly as in the previous case. 
\end{itemize}

Now, in the general case where no root of $P$ coincides with $X_2$, the eigenspaces associated with the roots of $P$ are one-dimensional. Then, $X_2$ is an eigenvalue of multiplicity exactly equal to $n-2$ and since the associated eigenspace has dimension at least $n-2$ as previously shown, its dimension is exactly $n-2$. It follows that ${\mathbb A}$ is diagonalizable with real eigenvalues. 

This shows that the model is hyperbolic about $(\rho_0, u_0, z_0)$ when $z_0 \parallel u_0$. The case where $z_0 = 0$ is obvious and left to the reader.  

\medskip
\noindent
\textbf{Proof of (ii), case $|M|$ large.} We notice that the condition that $\rho_0 |b| |z_0|$ large just means that $|M|$ is large. If we introduce $T$ and $Z$ such that 
$$ Y = T+MZ, \quad T=(c_1-c_2) \sin \theta \cos \phi, \quad Z = \sin \theta \cos (\phi-\delta), $$
and insert it in \eqref{eq:defDelta}, we get that $\Delta$ is a polynomial in $M$ of degree $6$ which is written:
\begin{eqnarray*}
\Delta &=& 4 (Z^2-1)^2 M^6 + 16 T Z (Z^2-1) M^5 \\
&+& \big( 8 T^2 (3 Z^2-1) + 12 L + 20 L Z^2 + 4 Z^3 R - 36 Z R \big) M^4 + \textrm{l.o.t}, 
\end{eqnarray*}
where ``l.o.t'' stand for ``lower order terms''. Suppose $Z= \pm 1=: \epsilon$. Then, $\sin \theta = 1$ and $\cos(\phi- \delta) = \epsilon$. This means that $\theta = \pi/2$ and $\phi = \delta$ (if $\epsilon = 1$) or $\phi = \delta + \pi$ (if $\epsilon = -1$). This implies 
$$ L = c_1 \Theta \sin^2 \delta, \quad R = \epsilon \Theta \sin^2 \delta, \quad T = \epsilon (c_1-c_2) \cos \delta. $$
In this case, $\Delta$ reduces to 
\begin{eqnarray*}
\Delta &=& 16 \big( T^2 + 2 (L-\epsilon R) \big) M^4 + \textrm{l.o.t} \\
&=& 16 \big( (c_1-c_2)^2 \cos^2 \delta - 2 (1-c_1) \Theta \sin^2 \delta \big) M^4 + \textrm{l.o.t} 
\end{eqnarray*}
Thus, whenever $\delta$ is such that 
$$ 0 \leq \cot^2 \delta < E, \qquad E= \frac{2 \Theta (1-c_1)}{(c_1-c_2)^2} ,$$ 
(we notice that $E>0$ thanks to Theorem \ref{thm:eps_to_0}), the leading order of $\Delta$ as a polynomial in $M$ is a negative coefficient times $M^4$. Thus, $\Delta$ is negative for large enough $M$. The coefficients of the lower order terms are bounded by constants that only depend on $c_1$, $c_2$ and $\Theta$. Thus, there exists a constant $C_2>0$ which only depends on $c_1$, $c_2$ and $\Theta$ such that for $|M| > C_2$ and $\delta$ such that $\cot^2 \delta < E/2$, then $\Delta <0$, showing that the System is not hyperbolic about a state of corresponding $(\rho_0,u_0,z_0)$.

\medskip
\noindent
\textbf{Proof of (ii), case $|M|$ small.} Suppose now $\theta = \pi/2$ and $\phi$ such that $L = c_1 \Theta \sin^2 \phi = e \ll 1$. Let us also choose $\delta = \phi+\pi/2$ so that, for instance: 
$$ \sin \phi = \epsilon \sqrt{\frac{e}{c_1 \Theta}}, \quad \cos \phi = \sin \delta = \sqrt{1- \frac{e}{c_1 \Theta}} = 1 + {\mathcal O}(e), $$
with $\epsilon = \pm 1$. 
Then, we get 
$$ R = \epsilon \sqrt{\frac{e \Theta}{c_1}} \, \big( 1 + {\mathcal O}(e) \big), \quad  Y = c_1 - c_2 + {\mathcal O}(e). $$
Finally, let us choose $M = \alpha e$ with $\alpha$ a constant to be chosen later. Inserting these assumptions into \eqref{eq:defDelta}, we check that the leading order term when $e \to 0$ is coming from the second term of \eqref{eq:defDelta}, so that 
$$ \Delta = 4 \epsilon (c_1-c_2)^3 \sqrt{\frac{\Theta}{c_1}} \alpha e^{3/2} \big( 1 + o(1) \big), $$
as $e \to 0$. Taking $\alpha = - \epsilon$, we see that this leading order term is negative. By the same arguments as in the previous case, this shows that there exists a constant $C_1>0$ which only depends on $c_1$, $c_2$ and $\Theta$ such that for all $|M| < C_1$, we can find $\delta$ (depending on $|M|$) such that $\Delta <0$. This proves that the System is not hyperbolic about a state of corresponding $(\rho_0,u_0,z_0)$ either. \endproof

\setcounter{equation}{0}
\section{Doubly periodic travelling-wave solutions: proofs}
\label{sec_TW_stat}

We look for a solution of the form \eqref{eq:rho_per_TW}, \eqref{eq:u_per_TW}, \eqref{eq:alp_per_TW}. 
Since $\rho $, $u$ and $\nabla_x \alpha = 2 \pi (p e_1 + m e_2)$ are constant in space and time, Eqs \eqref{eq:fl_rho_sm} and \eqref{eq:fl_u_sm} (with $V=0$) are trivially satisfied. The only equation left to verify is \eqref{eq:fl_al_sm}; It leads to 
$$ - \lambda + (c_1 U_1 + 2 \pi b p) 2 \pi p + (c_1 U_2 + 2 \pi b m) 2 \pi m = 0, $$
which, after rearrangement, is nothing but \eqref{eq:lambda_per_TW}. This gives the travelling-wave solutions. 

Now, such solutions are stationary if $\lambda = 0$. In such a case, supposing $(p,m) \not = (0,0)$, $U$ must satisfy \eqref{eq:TW_constraint2}. Letting $X = p e_1 + m e_2$, this equation is written 
\begin{equation} 
U \cdot \frac{X}{|X|^2} = - \frac{2 \pi b}{c_1}.  
\label{eq:eq_for_U}
\end{equation}
Using Cauchy-Schwarz inequality and the fact that $|U|=1$, we find that 
\begin{equation}
|X| \leq \frac{c_1}{2 \pi |b|},
\label{eq:cond_exist_U}
\end{equation}
is a necessary condition for the existence of $U$. But this condition is exactly \eqref{eq:TW_constraint}. It is also a sufficient condition. If the inequality \eqref{eq:cond_exist_U} is satisfied, we can write 
\begin{eqnarray} 
U &=& - \frac{2 \pi b |X|}{c_1} \frac{X}{|X|} + \sigma \Big( 1 - \Big( \frac{2 \pi b |X|}{c_1} \Big)^2 \Big)^{1/2} \Big( \frac{X}{|X|} \Big)^\bot, \nonumber \\
&=& - \frac{2 \pi b}{c_1} \left( \begin{array}{c} p \\ m \end{array} \right) + \sigma \Big( \frac{1}{p^2 + m^2} - \Big( \frac{2 \pi b}{c_1} \Big)^2 \Big)^{1/2} \left( \begin{array}{c}  - m \\ p \end{array} \right) \label{eq:U_stat}
\end{eqnarray}
where $(\frac{X}{|X|})^\bot$ is the vector obtained by rotating $\frac{X}{|X|}$ by an angle of $\pi/2$ and $\sigma = \pm 1$. This gives two solutions $U$ except if the factor of $\sigma$ is zero, which is the case where \eqref{eq:cond_exist_U} is an equality. In this case, the solution is given by 
$$U = - \textrm{Sign}(b) X/|X| = - \textrm{Sign}(b) \frac{1}{\sqrt{p^2 + m^2}} \left( \begin{array}{c} p \\ m \end{array} \right).$$ 
Finally, the case $(p,m) = (0,0)$ is obvious, which ends the proof. 
\endproof


\section{Numerical methods}\label{sec:numericalmethods}

In this section, we give additional details on the numerical methods used to produce the simulations shown in Section \ref{sec:numerics}. The code is freely available on the \texttt{GitHub} page of the second author at 
\begin{center}
\url{https://github.com/antoinediez/Swarmalators}
\end{center}
The particle scheme is written in \texttt{Python} and the finite volume scheme in \texttt{Julia}. 

\subsection{Particle scheme}

Simulating mean-field particle systems is relatively easy though computationally expensive when the number of particles becomes large. In order to simulate the particle system with up to 3.5 millions particles, we rely on the highly-efficient GPU framework introduced in the \texttt{SiSyPHE} library \cite{diez2021sisyphe} which is based on the \texttt{KeOps} library \cite{charlier2021kernel}. The \texttt{SiSyPHE} library is a versatile \texttt{Python} library designed for the simulation of collective dynamics models which already includes classical models such as the Vicsek model. Thanks to the object-oriented implementation of the library and since the present model is an elaboration of the Vicsek model, only a simple extension of the base class \texttt{Vicsek} of the \texttt{SiSyPHE} library is needed in order to incorporate the new phase variable and its contribution to the dynamics. From a methodological point of view, the (stochastic) particle system \eqref{eq:ibm_pos}-\eqref{eq:ibm_phas} is discretized using a first-order Euler-Maruyama scheme. For a given interaction radius $R$ and with the notations of Section \ref{sec:scaling_numerics}, the time-step is taken equal to $\Delta t = 10^{-2}/ \max(C\gamma,\nu,D,\nu',D')$. All the
particle simulations have been run using an Nvidia GTX 2080 Ti GPU chip on the GPU cluster of the Department of Mathematics at Imperial College London.

\subsection{Finite volume scheme}\label{sec:fv_appendix}

Following the methodology introduced in \cite{motsch2011numerical} for the SOH model, the finite volume scheme is based on the following formulation of the system \eqref{eq:fl_rho}-\eqref{eq:fl_al} as the relaxation limit $\varepsilon\to0$ of a system written in conservative form. 
\begin{align}
& \partial_t \rho^\varepsilon + \nabla_x \cdot [\rho(c_1 u^\varepsilon + b\rho^\varepsilon\nabla_x\alpha^\varepsilon)] = 0,  \\ 
& \partial_t(\rho^\varepsilon u^\varepsilon) + \nabla_x\cdot [ \rho^\varepsilon u^\varepsilon \otimes (c_2 u^\varepsilon + b\rho^\varepsilon\nabla_x\alpha^\varepsilon) ] + \Theta \nabla_x\rho^\varepsilon = \frac{\rho^\varepsilon}{\varepsilon}(1-|u^\varepsilon|^2)u^\varepsilon, \\ 
& \partial_t(\rho^\varepsilon \cos\alpha^\varepsilon) + \nabla_x \cdot [\rho^\varepsilon\cos\alpha^\varepsilon  (c_1 u^\varepsilon + b\rho^\varepsilon\nabla_x\alpha^\varepsilon)] \nonumber\\
&\qquad\qquad\qquad\qquad= -\sin\alpha^\varepsilon \big[(b-b')|\rho^\varepsilon\nabla_x \alpha^\varepsilon|^2 +  \Theta' \nabla_x\cdot (\rho^\varepsilon\nabla_x\rho^\varepsilon)\big], \\ 
& \partial_t(\rho^\varepsilon \sin\alpha^\varepsilon) + \nabla_x \cdot [\rho^\varepsilon\sin\alpha^\varepsilon  (c_1 u^\varepsilon + b\rho^\varepsilon\nabla_x\alpha^\varepsilon)] \nonumber\\
&\qquad\qquad\qquad\qquad= \cos\alpha^\varepsilon \big[(b-b')|\rho^\varepsilon \nabla_x \alpha^\varepsilon|^2 + \Theta' \nabla_x\cdot (\rho^\varepsilon\nabla_x\rho^\varepsilon)\big],
\end{align}
where for numerical stability reasons, we use the variable $(\cos\alpha,\sin\alpha)$ instead of just $\alpha$. We first solve the conservative part using a custom HLLE scheme \cite{leveque2002finite,einfeldt1991godunov} and we use a splitting method for the source terms as outlined below. 
\begin{enumerate}
\item Using a dimensional splitting, solving the conservative part 
\begin{align*}
\partial_t \rho + \nabla_x \cdot [\rho(c_1 u + b\rho\nabla_x\alpha)] &= 0,  \\ 
 \partial_t(\rho u) + \nabla_x\cdot [ \rho u \otimes (c_2 u + b\rho\nabla_x\alpha) ] + \Theta \nabla_x\rho &= 0, \\ 
 \partial_t(\rho \cos\alpha) + \nabla_x \cdot [\rho\cos\alpha  (c_1 u + b\rho\nabla_x\alpha)] &=0\\
 \partial_t(\rho \sin\alpha) + \nabla_x \cdot [\rho\sin\alpha  (c_1 u + b\rho\nabla_x\alpha)] &=0,
\end{align*}
reduces to solving two 1D equations. In order to compute the numerical flux between two datas $(\rho_\ell, u_{1\ell}, u_{2\ell}, \cos\alpha_\ell, \sin\alpha_\ell)$ and  $(\rho_r, u_{1r}, u_{2r}, \cos\alpha_r, \sin\alpha_r)$ we first need to approximate the phase gradient (in the $x_1$-direction). A finite difference approximation can be computed from $(\cos \alpha_\ell, \sin\alpha_\ell)$ and $(\cos \alpha_r, \sin\alpha_r)$ only by taking the argument of the following complex number:
\begin{equation}\label{eq:discretegradalpha} \partial_{x_1} \alpha \simeq z_{\ell r} := \mathrm{arg}\big(\cos\alpha_r\cos\alpha_\ell + \sin\alpha_r\sin\alpha_\ell + i(\sin\alpha_r\cos\alpha_\ell - \sin\alpha_\ell\cos\alpha_r)\big)/\Delta x,\end{equation}
where $\Delta x$ is the space discretization step. Then, using the change of variable $(q_0,q_1,q_2,q_3,q_4) = (\rho,\rho u_1,\rho u_2,\rho\cos\alpha,\rho\sin\alpha)$ we are led to the computation of the Jacobian matrix of the flux function
\[f(q_0,q_1,q_2,q_3,q_4) := \left(
\begin{array}{c}
c_1 q_1 +  bz_{\ell r} q^2_0\\ 
c_2 q_1^2/q_0 + \Theta q_0 +  bz_{\ell r}q_0q_1 \\ 
c_2 q_1q_2/q_0 + b z_{\ell r} q_0q_2 \\ 
c_1 q_3q_1/q_0 + bz_{\ell r} q_0q_3\\
c_1 q_4 q_1/q_0 + bz_{\ell r} q_0q_4
\end{array}
\right).\]
A direct computation shows that the Jacobian matrix of $f$ has four eigenvalues, one with multiplicity 2:
\[\nu_1 = c_1 q_1/q_0  + b z_{\ell r}q_0,\]
and three with multiplicity 1:
\begin{align*}
\nu_2 &= c_2 q_1/q_0  + b z_{\ell r} q_0,\\
\nu_+ &= \frac{1}{2}\big(2c_2 q_1/q_0  + 3b z_{\ell r}q_0 + \sqrt{\Delta}\big),\\
\nu_-  &= \frac{1}{2}\big(2c_2 q_1/q_0  + 3b z_{\ell r}q_0 - \sqrt{\Delta}\big),
 \end{align*}
where $\Delta = 4c_2(c_2 - c_1)(q_1/q_0)^2 + 4c_1\Theta + 4bz_{\ell r}(c_1-c_2)q_1 + (bz_{\ell r} q_0)^2$. Using these values, the computation of the numerical flux using a HLLE scheme is explained in \cite{leveque2002finite,einfeldt1991godunov}. Note that since the eigenvalues depend on $z_{\ell r}$ and thus on $\rho$ and $\nabla_x\alpha$ (which are not ensured to be bounded), we have to use an adaptive time step~$\Delta t$ in order to guarantee the CFL condition $\frac{\Delta t}{\Delta x} \max(|\nu_1|,|\nu_2|,|\nu_+|,|\nu_-|) \leq 1$ at each iteration. For better stability, in the experiments, $\Delta t$ is chosen so that the CFL number does not exceed 0.1. 
\item The relaxation part reads 
\begin{align*}
& \partial_t \rho^\varepsilon = 0,  \\ 
& \partial_t(\rho^\varepsilon u^\varepsilon)  = \frac{\rho^\varepsilon}{\varepsilon}(1-|u^\varepsilon|^2)u^\varepsilon, \\ 
& \partial_t(\rho^\varepsilon \cos\alpha^\varepsilon) = 0, \\ 
& \partial_t(\rho^\varepsilon \sin\alpha^\varepsilon)  =0.
\end{align*}
It can be solved explicitly but as shown in \cite{motsch2011numerical}, when $\varepsilon \to0$ it reduces to a mere normalization of the velocity. 

\item The other source terms reads
\begin{align*}
& \partial_t \rho = 0,  \\ 
& \partial_t u = 0 \\ 
& \partial_t \cos\alpha = -\sin\alpha \big[(b-b')\rho|\nabla_x \alpha|^2 + \Theta'(\Delta \rho + |\nabla_x\sqrt{\rho}|^2)\big], \\ 
& \partial_t \sin\alpha = \cos\alpha \big[(b-b')\rho|\nabla_x \alpha|^2 + \Theta'(\Delta \rho + |\nabla_x\sqrt{\rho}|^2)\big].
\end{align*}
We solve this part using an explicit Euler scheme. We use a finite difference approximation of the spatial derivatives on the right-hand side, with a classical five-point discretisation of the Laplacian term and using the same method as before \eqref{eq:discretegradalpha} for the gradient in $\alpha$.  
\end{enumerate}

\section{List of supplementary videos}\label{sec:listvideos}

The videos can be found using the following link: \\
\url{https://figshare.com/projects/Topological_states_and_continuum_model_for_swarmalators_without_force_reciprocity/139912}

\medskip
\noindent
The videos show the outcome of the simulations discussed in Sections~\ref{sec:segregationparticles} and~\ref{sec:simulationsh}. For each video, the left panel shows the spatial density and the right panel the average phase. These quantities are computed using a spatial discretization on a uniform grid with constant step $\Delta x = 0.01$. For the particle simulations, the value of the density in a cell is computed as the proportion of particles in this cell and the phase is their average phase (it is arbitrarily set to 0 if the cell is empty). Similarly the average velocity is computed on a spatial grid with step size $\Delta x = 0.05$ and depicted by black arrows. For the particle simulations, three particles are represented on the right panel by three disks (of arbitrary radius) colored according to the phases of these particles. 

\subsection{Particle simulations}\label{sec:listvideos_particles}

The following videos supplement the results presented in Section \ref{sec:results_particles} and discussed in Section \ref{sec:discussion_particles}. Note that the phases of the particles range from 0 to $2\pi$ but they have been rescaled so that the range of the colorbar of the right-panel is between 0 and 1. 

\begin{video}[Very low noise NA]\label{video:verylownoise_NA}
Particle simulation starting form a NA configuration with $k'=k'_\mathrm{hi+}=10000$ and the other parameters given in Section \ref{sec:parameters_particles}. The particles segregate into small regions of equal phase separated by thin low-density regions. The systems eventually reaches a stable state where well-separated band-like structures characterized by a constant phase travel at a constant speed in the direction opposite to the phase gradient. See Figs. \ref{subfig:finalNA_density_verylownoise} and \ref{subfig:finalNA_phase_verylownoise}. 
\end{video}

\begin{video}[Very low noise PA]\label{video:verylownoise_PA}
Particle simulation starting form a PA configuration with $k'=k'_\mathrm{hi+}=10000$ and the other parameters given in Section \ref{sec:parameters_particles}. The particles segregate into small regions of equal phase separated by thin low-density regions. After 40 units of time, well-separated band-like structures characterized by a constant phase can be identified. They travel at a constant speed in the same direction as the phase gradient but their shapes are not stable. 
\end{video}

\begin{video}[Very low noise OT]\label{video:verylownoise_OT}
Particle simulation starting form a OT configuration with $k'=k'_\mathrm{hi+}=10000$ and the other parameters given in Section \ref{sec:parameters_particles}. The particles segregate into small regions of equal phase separated by thin low-density regions. The global velocity of the particles transitions from $(1,0)^\mathrm{T}$ to $(0,1)^\mathrm{T}$. Although band-like structures can be identified, they are not as stable as in the NA case. 
\end{video}

\begin{video}[Very low noise UF]\label{video:verylownoise_UF}
Particle simulation starting form a UF configuration with $k'=k'_\mathrm{hi+}=10000$ and the other parameters given in Section \ref{sec:parameters_particles}. The particles segregate into small regions of equal phase separated by thin low-density regions. These regions are very dynamic with a lot of merging and mixing. After about 10 units of time, the system reaches a flocking state in which all the phases and velocity equal. 
\end{video}

\begin{video}[Low noise NA]\label{video:lownoise_NA}
Particle simulation starting form a NA configuration with $k'=k'_\mathrm{hi}=200$ and the other parameters given in Section \ref{sec:parameters_particles}. The behavior is the same as the one described  in the caption of Video \ref{video:verylownoise_NA} but the bands are larger and their phase slowly varies in time. See Figs. \ref{subfig:finalNA_density_lownoise} and \ref{subfig:finalNA_phase_lownoise}. 
\end{video}

\begin{video}[Low noise PA]\label{video:lownoise_PA}
Particle simulation starting form a PA configuration with $k'=k'_\mathrm{hi}=200$ and the other parameters given in Section \ref{sec:parameters_particles}. The behavior is initially the same as the one described  in the caption of Video \ref{video:verylownoise_PA} but the bands are not stable and the systems finally reaches a flocking phase.  

\end{video}

\begin{video}[Low noise OT]\label{video:lownoise_OT}
Particle simulation starting form a OT configuration with $k'=k'_\mathrm{hi}=200$ and the other parameters given in Section \ref{sec:parameters_particles}. The behavior is initially the same as the one described  in the caption of Video \ref{video:verylownoise_OT} but the systems finally reaches a stable state which is the same state as the one starting from the NA configuration. 

\end{video}

\begin{video}[Low noise UF]\label{video:lownoise_UF}
Particle simulation starting form a UF configuration with $k'=k'_\mathrm{hi}=200$ and the other parameters given in Section \ref{sec:parameters_particles}. The behavior is the same as the one described in the caption of Video \ref{video:verylownoise_UF} but the segregation regions are larger. 
\end{video}

\begin{video}[Medium noise NA]\label{video:mediumnoise_NA}
Particle simulation starting form a NA configuration with $k'=k'_\mathrm{med}=10$ and the other parameters given in Section \ref{sec:parameters_particles}. The initial doubly periodic travelling wave persists during about 5 units of time. Then thin low-density regions emerge and delimitate band-like structures. Unlike the cases with less noise in the phase equation, the bands are larger and an inner gradient in phase can be identified inside each band. 
\end{video}

\begin{video}[Medium noise PA]\label{video:mediumnoise_PA}
Particle simulation starting form a PA configuration with $k'=k'_\mathrm{med}=10$ and the other parameters given in Section \ref{sec:parameters_particles}. The behavior is similar to the one presented in the caption of Video \ref{video:mediumnoise_NA} except that the bands are moving in the opposite direction. Moreover, the bands shape is less stable. 
\end{video}

\begin{video}[Medium noise OT]\label{video:mediumnoise_OT}
Particle simulation starting form a OT configuration with $k'=k'_\mathrm{med}=10$ and the other parameters given in Section \ref{sec:parameters_particles}. The behavior is initially similar to the one presented in the caption of Video \ref{video:mediumnoise_NA} except that the global velocity of the particles transition from $(1,0)^\mathrm{T}$ to $(0,1)^\mathrm{T}$ and the systems finally reaches the same stable state as the one starting from a NA configuration. 
\end{video}

\begin{video}[Medium noise UF]\label{video:mediumnoise_UF}
Particle simulation starting form a UF configuration with $k'=k'_\mathrm{med}=10$ and the other parameters given in Section \ref{sec:parameters_particles}. The behavior is the same as the one described in the caption of Video \ref{video:lownoise_UF}. 
\end{video}

\begin{video}[Large noise NA]\label{video:largenoise_NA}
Particle simulation starting form a NA configuration with $k'=k'_\mathrm{lo}=3$ and the other parameters given in Section \ref{sec:parameters_particles}. The initial doubly periodic travelling wave is stable and persists during the 40 units of time of the simulation. The theoretical travelling wave speed is $\lambda\simeq0.03$. The measured speed is approximately equal to $0.05$.  
\end{video}

\begin{video}[Large noise PA]\label{video:largenoise_PA}
Particle simulation starting form a PA configuration with $k'=k'_\mathrm{lo}=3$ and the other parameters given in Section \ref{sec:parameters_particles}. The initial doubly periodic travelling wave is stable and persists during the 40 units of time of the simulation. The theoretical travelling wave speed is $\lambda\simeq-1.75$. The measured speed is approximately equal to $-1.72$.  
\end{video}

\begin{video}[Large noise OT]\label{video:largenoise_OT}
Particle simulation starting form a OT configuration with $k'=k'_\mathrm{lo}=3$ and the other parameters given in Section \ref{sec:parameters_particles}. The initial doubly periodic travelling wave is stable and persists during the 40 units of time of the simulation. The theoretical travelling wave speed is $\lambda\simeq-0.86$. The measured speed is approximately equal to $-0.84$.  
\end{video}

\begin{video}[Large noise UF 1]\label{video:largenoise_UF1}
Particle simulation starting form a UF configuration with $k'=k'_\mathrm{lo}=3$ and the other parameters given in Section \ref{sec:parameters_particles}. The system reaches a flocking state (although two phases can be identified)
\end{video}

\begin{video}[Large noise UF 2]\label{video:largenoise_UF2}
Particle simulation starting form a UF configuration with $k'=k'_\mathrm{lo}=3$ and the other parameters given in Section \ref{sec:parameters_particles}. The system reaches a state close to a doubly periodic travelling wave but there are also waves of velocities. 
\end{video}

\begin{video}[Large noise UF 3]\label{video:largenoise_UF3}
Particle simulation starting form a UF configuration with $k'=k'_\mathrm{lo}=3$ and the other parameters given in Section \ref{sec:parameters_particles}. The system reaches a kind of doubly periodic travelling wave solution but with a much more complex shape. 
\end{video}

\subsection{Simulations of the SH system}\label{sec:listvideos_sh}

The following videos supplement the results presented in Section \ref{sec:results_fv}. 

\begin{video}[Low noise NA]\label{video:lownoise_NA_SH}
Simulation of the SH system starting from a perturbed NA configuration with $k'=k'_\mathrm{hi}=200$ and the other parameters given in Section \ref{sec:parameters_particles}. The initial noise quickly resorbs and the system finally reaches a stable doubly periodic travelling wave with no perceptible inhomogeneities. 
\end{video}

\begin{video}[Low noise PA]\label{video:lownoise_PA_SH}
Simulation of the SH system starting from a perturbed PA configuration with $k'=k'_\mathrm{hi}=200$ and the other parameters given in Section \ref{sec:parameters_particles}. The initial noise quickly resorbs and the system finally reaches a stable doubly periodic travelling wave although small inhomogeneities in the density are perceptible. 
\end{video}

\begin{video}[Low noise OT]\label{video:lownoise_OT_SH}
Simulation of the SH system starting from a perturbed OT configuration with $k'=k'_\mathrm{hi}=200$ and the other parameters given in Section \ref{sec:parameters_particles}. The initial noise quickly resorbs but small inhomogeneities in the density are perceptible and eventually degenerate into thin very-low density regions. After a transition period during which the global velocity transitions towards a limit value close to $(0,1)^\mathrm{T}$, these structures disappear and the system reaches the corresponding stable doubly periodic travelling wave solution (although although small inhomogeneities in the density remain perceptible).\end{video}

\begin{video}[Medium noise NA]\label{video:mediumnoise_NA_SH}
Simulation of the SH system starting from a perturbed NA configuration with $k'=k'_\mathrm{lo}=10$ and the other parameters given in Section \ref{sec:parameters_particles}. The initial noise quickly resorbs and the system finally reaches a stable doubly periodic travelling wave with no perceptible inhomogeneities.\end{video}

\begin{video}[Medium noise PA]\label{video:mediumnoise_PA_SH}
Simulation of the SH system starting from a perturbed PA configuration with $k'=k'_\mathrm{lo}=10$ and the other parameters given in Section \ref{sec:parameters_particles}. Same observations as Video \ref{video:mediumnoise_NA_SH}.\end{video}

\begin{video}[Medium noise OT]\label{video:mediumnoise_OT_SH}
Simulation of the SH system starting from a perturbed OT configuration with $k'=k'_\mathrm{lo}=10$ and the other parameters given in Section \ref{sec:parameters_particles}. Same observations as Video \ref{video:mediumnoise_NA_SH}.
\end{video}

\begin{video}[Large noise NA]\label{video:largenoise_NA_SH}
Simulation of the SH system starting from a perturbed NA configuration with $k'=k'_\mathrm{lo}=3$ and the other parameters given in Section \ref{sec:parameters_particles}. Same observations as Video \ref{video:mediumnoise_NA_SH}.
\end{video}

\begin{video}[Large noise PA]\label{video:largenoise_PA_SH}
Simulation of the SH system starting from a perturbed PA configuration with $k'=k'_\mathrm{lo}=3$ and the other parameters given in Section \ref{sec:parameters_particles}. Same observations as Video \ref{video:mediumnoise_NA_SH}.
\end{video}

\begin{video}[Large noise OT]\label{video:largenoise_OT_SH}
Simulation of the SH system starting from a perturbed OT configuration with $k'=k'_\mathrm{lo}=3$ and the other parameters given in Section \ref{sec:parameters_particles}. Same observations as Video \ref{video:mediumnoise_NA_SH}.
\end{video}


\end{document}